\documentclass[aps,pra,twocolumn,showkeys,amssymb,floatfix,longbibliography,superscriptaddress]{revtex4-1}

\usepackage{hyperref}
\usepackage{braket}
\usepackage{xcolor}
\usepackage{mathtools}
\usepackage{multirow}
\usepackage{bm}
\usepackage{placeins}

\DeclareRobustCommand\id{\leavevmode\hbox{\small1\normalsize\kern-.33em1}}

\let\baraccent=\=
\renewcommand{\=}[1]{\stackrel{#1}{=}}

\newlength\stextwidth

\allowdisplaybreaks

\begin{document}

\title{Real-time simulation of flux qubits used for quantum annealing}

\author{Madita Willsch}
\affiliation{Institute for Advanced Simulation, J\"ulich Supercomputing Centre,\\
Forschungszentrum J\"ulich, D-52425 J\"ulich, Germany}
\affiliation{RWTH Aachen University, D-52056 Aachen, Germany}
\author{Dennis Willsch}
\affiliation{Institute for Advanced Simulation, J\"ulich Supercomputing Centre,\\
Forschungszentrum J\"ulich, D-52425 J\"ulich, Germany}
\affiliation{RWTH Aachen University, D-52056 Aachen, Germany}
\author{Fengping Jin}
\affiliation{Institute for Advanced Simulation, J\"ulich Supercomputing Centre,\\
Forschungszentrum J\"ulich, D-52425 J\"ulich, Germany}
\author{Hans De Raedt}
\affiliation{Zernike Institute for Advanced Materials,\\
University of Groningen, Nijenborgh 4, NL-9747 AG Groningen, The Netherlands}
\author{Kristel Michielsen}
\affiliation{Institute for Advanced Simulation, J\"ulich Supercomputing Centre,\\
Forschungszentrum J\"ulich, D-52425 J\"ulich, Germany}
\affiliation{RWTH Aachen University, D-52056 Aachen, Germany}

\date{\today}

\begin{abstract}
The real-time flux dynamics of up to three superconducting quantum interference devices (SQUIDs) are studied by numerically solving the time-dependent Schr\"odinger equation.
The numerical results are used to scrutinize the mapping of the flux degrees of freedom onto two-level systems (the qubits) as well as the performance of the intermediate SQUID as a tunable coupling element.
It is shown that the qubit representation yields a good description of the flux dynamics during quantum annealing and the presence of the tunable coupling element does not have negative effects on the overall performance.
Additionally, data obtained from a simulation of the dynamics of two-level systems during quantum annealing are compared to experimental data produced by the D-Wave 2000Q quantum annealer.
The effects of finite temperature are incorporated in the simulation by coupling the qubit system to a bath of two-level systems.
It is shown that an environment modeled as non-interacting two-level systems coupled to the qubits can produce data which matches the experimental data much better than the simulation data of the qubits without coupling to an environment and better than data obtained from a simulation of an environment modeled as interacting two-level systems coupling to the qubits.
\end{abstract}

\keywords{quantum computation; superconducting qubits; quantum annealing; SQUID; product-formula algorithm; optimization problems}

\maketitle

\section{Introduction}\label{sec:introduction}

Theoretically, an ideal quantum computer is described in terms of two-level systems, the qubits~\cite{nielsen_chuang}.
However, almost all currently popular technologies such as ion traps~\cite{cirac95,monroe95,haeffner03,Hanneke2009,Schindler2013,ballance16}, quantum dots~\cite{loss98,levy02}, or superconducting circuits~\cite{wendin2017,krantz19} employ physical devices which are only approximately described by two-level systems~\cite{wu02}.
Among these, trapped ions may be described in the two-level approximation under conditions discussed in Ref.~\cite{leibfried03}.
Single-electron quantum dots can be described as two-level systems if the orbital wave function is neglected. However, two-electron~\cite{levy02,petta05} and three-electron systems~\cite{divincenzo00,gaudreau11} confined in quantum dots are again only approximately described by two-level systems, and leakage out of the computational space may need to be taken into account~\cite{divincenzo00,medford13,cerfontaine16}.

For superconducting circuits, it depends on the particular circuit design how well the two relevant energy levels, which define the qubit states, are separated from the higher energy levels. For instance, the phase qubit~\cite{martinis02,steffen06} and the transmon~\cite{koch2007}, an extension of the charge qubit~\cite{shnirman97,bouchiat1998,Nakamura1999}, have rather small anharmonicities. Thus, leakage to higher energy levels is a major issue when performing gate operations~\cite{motzoi09,lucero10,gambetta11,Wallman2016,willsch17,wood18} and is typically alleviated by the use of pulse-shaping techniques~\cite{motzoi09,chow10,lucero10,gambetta11,chen16,mckay17}.
Flux qubits, on the other hand, have a strong anharmonicity and are less prone to excitations to higher energy levels~\cite{Poletto2009,yoshihara14,billangeon15,chen16,krantz19} as long as the qubit is not driven too strongly~\cite{ferron10}.

In this study we focus on the superconducting quantum interference device (SQUID)-based flux qubit~\cite{chiarello00,Makhlin2001,Poletto2009}, which is used in the D-Wave quantum annealer~\cite{harris09_newjournal,harris10}.
Due to the large superconducting loop needed for the SQUID-based qubit, it is sensitive to flux noise which limits the coherence time~\cite{wendin2017}.

Other flux qubits, commonly used for gate-based quantum computing, are the three-junction qubit~\cite{Mooij99,orlando99,vanderWal00,grajcar04}, the C-shunt flux qubit~\cite{you07,steffen10,Yan2016}, and the fluxonium qubit~\cite{Manucharyan09,pop2014,nguyen18}.
For the three-junction qubit and the fluxonium qubit, the large inductance is realized by using two or more Josephson junctions.
In this way, the loop size and thus the sensitivity to flux noise can be reduced~\cite{wendin07}.
The C-shunt flux qubit is a capacitively shunted variant of the flux qubit with improved coherence when operated away from the degeneracy point~\cite{you07}. There is an ongoing discussion about the role of decoherence during quantum annealing~\cite{childs01,sarandy05,ashhab06,amin09_decoherence,boixo13,dickson13}.

In this paper we address three questions. First, we study the flux dynamics of the SQUIDs used in the D-Wave quantum annealer, addressing the issue of how well these dynamics are captured by a qubit model.

Second, we investigate whether the presence of the SQUID used as a tunable coupler in the D-Wave device affects the performance of the quantum annealing process. To answer the first two questions, we study the dynamics of the flux degrees of freedom of two SQUIDs functioning as qubits and their tunable coupler, a third SQUID, by solving the time-dependent Schr\"odinger equation (TDSE) for the model Hamiltonian in terms of flux degrees of freedom. So far, studies including higher energy levels have been limited to four-level qudits~\cite{johnson11,amin13}.

The approach that we adopt in this paper is to start from an idealized model of the SQUIDs which does not take into account fabrication variations of circuit elements or stray fluxes induced by the control lines. That is, in the idealized model, the two SQUIDs functioning as qubits are equal and the complete system can be regarded as a perfect device.

The third question we consider is to what extent the data produced by a D-Wave quantum annealer can be described by quantum annealing of the qubit model including environment effects. In order to do so, we study the dynamics of a two-qubit system interacting with an environment of two-level systems, representing, e.g., a heat bath~\cite{ZHAO16,RAED17b} or a collection of defects described by non-interacting two-level systems~\cite{shnirman05,mueller09,cole10}, by solving the corresponding TDSE.

The structure of the paper is as follows. In Sec.~\ref{sec:quantumannealing} we give a short introduction to quantum annealing and its relation to optimization problems. Section~\ref{sec:squids} contains a description of the SQUID-based model which we simulate. The mapping of the model onto the qubit model is given in Sec.~\ref{sec:mapping}. The flux dynamics of the SQUID model are simulated by solving the TDSE using the method described in Sec.~\ref{sec:method}. The results of the simulation are presented in Sec.~\ref{sec:results}. In Sec.~\ref{section2} we describe the two different models for the bath of two-level systems coupled to the qubit system and discuss the simulation and its results in comparison to data obtained from a D-Wave 2000Q quantum annealer. We conclude with a summary in Sec.~\ref{sec:summary}.

\section{Theoretical background}\label{sec:quantumannealing}
In general, the Hamiltonian describing a quantum annealing process can be written as
\begin{align}
   H(s) = A(s)H_\mathrm{init} + B(s)H_\mathrm{final},\label{eq:genH_qa}
\end{align}
where $H_\mathrm{init}$ is the initial Hamiltonian whose ground state defines the state in which the system is prepared initially, $H_\mathrm{final}$ denotes the Hamiltonian at the end of the annealing process and whose ground state is the one to be determined, $s=t/t_a\in[0,1]$ is the rescaled (dimensionless) time, and $t_a$ is the total annealing time. The functions $A(s)$ and $B(s)$ determine the energy scale (in our case GHz) and the annealing scheme. They satisfy $|A(0)|\gtrsim 1$ and $B(0)\approx 0$, and $A(1)\approx 0$ and $|B(1)|\gtrsim 1$ with respect to the corresponding energy scale.

From the adiabatic theorem~\cite{born28}, it follows that the system stays in the ground state of the instantaneous Hamiltonian $H(s)$ during the annealing process if $t_a\rightarrow \infty$ such that for $s=1$ the system is in the ground state of $H_\mathrm{final}$.
Let $\Delta E_j(s)$ denote the difference between the energy of the ground state $\ket{GS(s)}$ and the $j$-th excited state $\ket{ES_j(s)}$ at the rescaled time $s$.
A finite $t_a$ can be sufficient for the system to stay in the ground state if~\cite{amin09_adiabatictheorem,albash18}
\begin{align}
   \underset{s\in[0,1]}{\mathrm{max}}\frac{|\bra{ES_j(s)}\partial_s H(s)\ket{GS(s)}|}{\Delta E_j(s)^2} \ll t_a.
\end{align}

Quantum annealing can be used to solve optimization problems that can be mapped onto the Hamiltonian $H_\mathrm{final}$. The class of so-called quadratic unconstrained binary optimization (QUBO) problems can be mapped onto the Ising-spin model of the form
\begin{align}
  H_\mathrm{QUBO}= -\sum_{k=1}^N h_k S_k - \sum_{\mathclap{1\le j<k}}J_{jk} S_j S_k,\label{eq:Hqubo}
\end{align}
where $N$ is the number of binary variables $S_k\in \{-1, 1\}$, and $h_k$ and $J_{jk}$ are dimensionless real numbers defining the particular QUBO. The set of variables $\{S_k\}$ that minimizes Eq.~(\ref{eq:Hqubo}) gives the solution of the QUBO problem.

Quantum annealing can, at least in principle, find (one of) the ground state(s) of the Ising-spin Hamiltonian Eq.~(\ref{eq:Hqubo})~\cite{kadowaki98} or, equivalently, solve the corresponding QUBO problem.
For this purpose, the two-valued variables $S_k$ are replaced by the Pauli $Z$ matrices $\sigma_k^z$ with eigenvalues $\pm 1$ and eigenstates $\ket{\uparrow}$ and $\ket{\downarrow}$ such that Eq.~(\ref{eq:Hqubo}) transforms into
\begin{align}
   H_\mathrm{Ising} = -\sum_{k=1}^N h_k\sigma_k^z - \sum_{\mathclap{1\le j<k}}J_{jk}\sigma_j^z\sigma_k^z.\label{eq:Hising}
\end{align}
The product states of the $\sigma^z$ eigenstates define the so-called computational basis and are eigenstates of Eq.~(\ref{eq:Hising}).
The ground state of the Hamiltonian Eq.~(\ref{eq:Hising}) can then be found by quantum annealing with $H_\mathrm{final}$ replaced by $H_\mathrm{Ising}$ in Eq.~(\ref{eq:genH_qa}).
For quantum annealing, the simplest choice for $H_\mathrm{init}$ is the Hamiltonian of spins in a transverse field~\cite{kadowaki98}
\begin{align}
   H_\mathrm{trans} = -\sum_{k=1}^N \sigma_k^x,\label{eq:Htrans}
\end{align}
where $\sigma^x$ is the Pauli $X$ matrix. The ground state of this Hamiltonian is given by the product state $\ket{+\dots +}$, with $\ket{+} = (\ket{\uparrow} + \ket{\downarrow})/\sqrt{2}$.

Equation~(\ref{eq:Hising}) is used to formulate optimization problems for the quantum annealer manufactured by D-Wave Systems Inc.~\cite{harris09_newjournal}.
By design, the parameters of the final Hamiltonian are restricted to $h_k\in [-2,2]$ and $J_{jk}\in [-1,1]$, and the qubit connectivity is given by the Chimera graph such that, in the notation of Eq.~(\ref{eq:Hising}), some $J_{jk}$ have to be set to zero \cite{harris10_eightqubit}. In the following sections, we discuss the SQUID model and describe the mapping of the SQUID model onto the qubit model in terms of Eqs.~(\ref{eq:Hising}) and~(\ref{eq:Htrans}).

\section{SQUID Model}\label{sec:squids}

In this section we introduce the model Hamiltonian of the three-SQUID system that is used to simulate the flux dynamics during quantum annealing. Two of the three SQUIDs serve as qubits, each qubit subspace being defined by projection onto the two lowest energy states of the individual SQUIDs. The third SQUID acts as a tunable coupler between the two other SQUIDs. 

\begin{figure}[bt]
   \centering
   \includegraphics[width=0.3\textwidth]{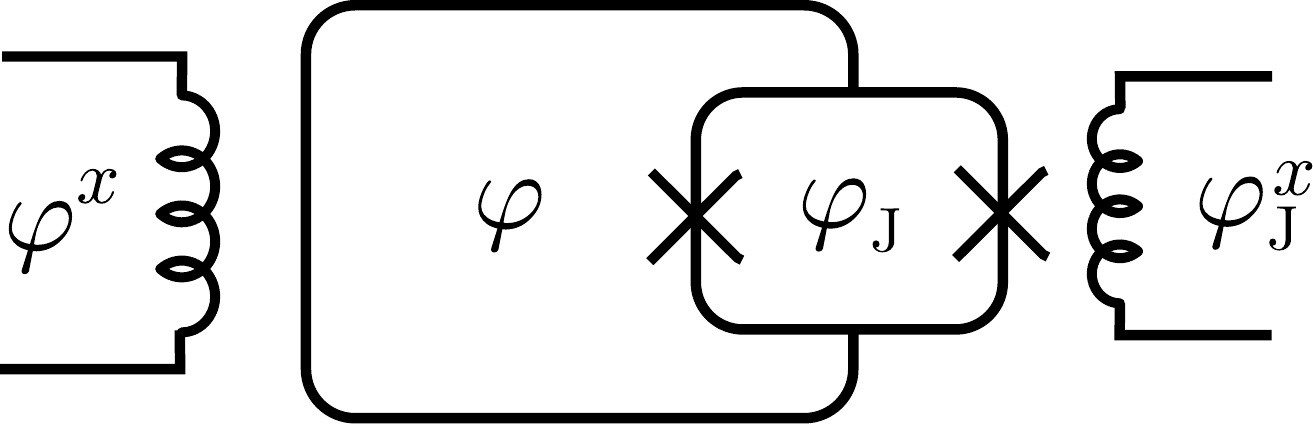}
   \caption{Sketch of a SQUID with a CJJ loop. The magnetic fluxes $\varphi$ and $\varphi_\mathrm{J}$ are the dynamical variables of the system. The external fluxes $\varphi^x$ and $\varphi^x_\mathrm{J}$ are used to control the operation of the device.}
   \label{fig:squid}
\end{figure}

Figure \ref{fig:squid} shows the circuit of a SQUID with a compound Josephson junction (CJJ) loop. It is used as a building block for the flux qubits and the effective coupling between them in the D-Wave quantum annealer. Including the CJJ loop effectively leads to a tunable Josephson junction~\cite{chiarello00}.
The two qubit states correspond to the left-circulating and right-circulating persistent current in the superconducting (main) loop and the tunable Josephson junction is used to control the annealing process.
For the coupler element, the tunable Josephson junction results in the tunable coupling strength \cite{harris09}.

The Hamiltonian of a SQUID with a CJJ loop is given by~\cite{han89,harris10,boixo16}
\begin{align}
    H_\mathrm{SQUID} = &-E_C\partial_\varphi^2 + E_L\frac{(\varphi-\varphi^x)^2}{2}\nonumber\\
    &- E_{C_{\mathrm{J}}}\partial_{\varphi_{\mathrm{J}}}^2 + E_{L_{\mathrm{J}}}\frac{(\varphi_{\mathrm{J}}
    -\varphi_{\mathrm{J}}^x)^2}{2}\nonumber\\ &- E_J\cos(\varphi)\cos\left( \frac{\varphi_{\mathrm{J}}}{2} \right) ,\label{eq:squid}
\end{align}
where $\varphi=2\pi\Phi/\Phi_0=2e\Phi$ (we use $\hbar =1$) is the dimensionless magnetic flux in the main loop and $\varphi_\mathrm{J}$ the dimensionless magnetic flux in the CJJ loop with
$\Phi_0$ denoting the magnetic flux quantum and $e$ the electron charge. In addition, $E_C$ and $E_{C_\mathrm{J}}$ are capacitive energies, $E_L$ and $E_{L_\mathrm{J}}$ are inductive energies, and $E_J$ is the Josephson energy.
For an uncoupled SQUID, the inductive energy $E_L$ is given by $E_L= 1/(4e^2L)$, where $L=L_\mathrm{main}+L_\mathrm{J}/4$ \cite{harris10} is the total inductance.
The potential of the flux variable $\varphi$, $V(\varphi)=E_L(\varphi-\varphi^x)^2/2-E_J\cos(\varphi_\mathrm{J}/2)\cos(\varphi)$, can be either a single potential well or a double-well potential depending on the value of $\varphi_\mathrm{J}$.
Thus, $\varphi_\mathrm{J}$ can be used to change the shape of the potential and also the barrier height between the double wells~\cite{han89,Chiarello2005}.
This property is used to control the annealing process via the external flux $\varphi_\mathrm{J}^x$~\cite{lanting14} and to set the coupling strength of the coupling SQUID~\cite{harris09}.
The external flux $\varphi^x$ can be used to tilt the potential, thereby lowering one of the wells and raising the other one~\cite{Chiarello2005}.
In terms of the qubit model (see Sec.~\ref{sec:mapping}), the external flux $\varphi^x$ can be used to set the parameter $h_k$ of the Ising Hamiltonian given in Eq.~(\ref{eq:Hising})~\cite{harris10_eightqubit}.

\subsection{Total Hamiltonian}

\begin{figure}[bt]
   \centering
   \includegraphics[width=0.48\textwidth]{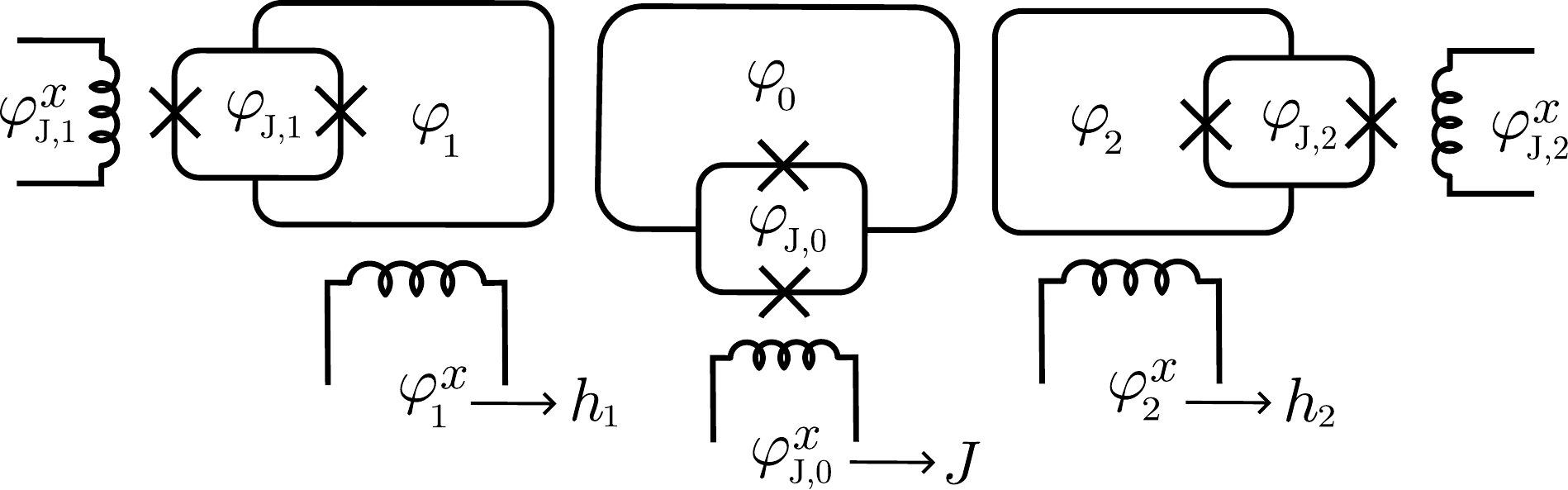}
   \caption{Sketch of three SQUIDs to realize a tunable coupling. The magnetic fluxes $\varphi_i$ and $\varphi_{\mathrm{J},i}$ are the dynamical variables of the system. The external fluxes $\varphi_i^x$ and $\varphi^x_{\mathrm{J},i}$ are used to control the operation of the device. The parameters $\varphi^x_1$, $\varphi^x_2$, and $\varphi_{\mathrm{J},0}^x$ determine the values of the parameters $h_1$, $h_2$, and $J$ of the qubit model Eq.~(\ref{eq:Hising}), respectively.}
   \label{fig:squids}
\end{figure}

So far, we have discussed the Hamiltonian of a single SQUID. In this section we introduce the Hamiltonian of the complete system consisting of the three SQUIDs.
A tunable coupling constant $J_{jk}$ (see Eq.~(\ref{eq:Hising})) is realized by inserting a SQUID as a coupler element between the other two SQUIDs \cite{harris07,harris09}. For the SQUID that functions as the coupler element, we denote the flux in the CJJ loop by $\varphi_{\mathrm{J},0}$ and the one in the main loop by $\varphi_0$. Accordingly, energies that correspond to the coupler main loop are labeled by an index ``$0$'', and those that correspond to the coupler CJJ loop by an index ``J,$0$''. The external control flux $\varphi_{\mathrm{J},0}^x$ can be used to tune the coupling strength between the SQUIDs. We label the fluxes of the SQUIDs corresponding to the qubits with indices ``$1$'' and ``$2$'', respectively. Since in the idealized model, the SQUIDs are equal and subject to the same annealing functions $A(s)$ and $B(s)$, their energies and their external fluxes $\varphi_{\mathrm{J}}^x$ are equal. Therefore, we drop the indices ``1'' and ``2'' in these cases. Although the external fluxes $\varphi_1^x$, $\varphi_2^x$, and $\varphi_\mathrm{J}^x$ depend on time, we do not write this explicitly for reasons of readability. A sketch of the complete system is shown in Fig.~\ref{fig:squids}.

By coupling SQUIDs, the inductive energies change such that they are given by $E_L(1+M^2/(LL_\mathrm{eff}))$ for the SQUIDs representing the qubits and by $E_{L_\mathrm{eff}}=E_{L_0}L_0/L_\mathrm{eff}$ for the coupler, where $L_\mathrm{eff}=L_0-2M^2/L$ is the effective inductance of the coupler element and $M$ is the mutual inductance between the coupler and the other SQUIDs' main loops. In addition to the modified Hamiltonians of the three SQUIDs, the interaction terms
\begin{align}
   H_\mathrm{int} &= \frac{M}{L_{\mathrm{eff}}}E_L\left( \varphi_1-\varphi_1^x \right)\left( \varphi_0-\varphi_0^x \right)\nonumber\\
   &+ \frac{M}{L_\mathrm{eff}}E_L\left( \varphi_2-\varphi_2^x \right)\left( \varphi_0-\varphi_0^x \right) \nonumber\\
   &+ \frac{M^2}{LL_\mathrm{eff}}E_L\left( \varphi_1-\varphi_1^x \right)\left( \varphi_2-\varphi_2^x \right),\label{eq:coupling}
\end{align}
have to be included \cite{vandenbrink05}. The tunable coupler can be operated without an external flux in the coupler main loop if the junction asymmetry is negligible, i.e., the difference between the critical currents (of the two junctions of the coupler SQUID) is much smaller than the sum of these critical currents~\cite{harris09}. Since we do not consider junction asymmetries in the idealized model, we set $\varphi_0^x=0$.
Collecting all terms, the total Hamiltonian is given by
\begin{widetext}
\begin{align}
   H_\mathrm{total} &= \sum\limits_{i=1}^2 \Bigg[  - E_{J}\cos(\varphi_i)\cos\left( \frac{\varphi_{\mathrm{J},i}}{2} \right)+ E_{L_{\mathrm{J}}}\frac{(\varphi_{\mathrm{J},i}-\varphi_{\mathrm{J}}^x)^2}{2}
    - E_{C_{\mathrm{J}}}\partial_{\varphi_{\mathrm{J},i}}^2+E_{L}\!\left(\! 1+\frac{M^2}{LL_\mathrm{eff}} \right)\frac{(\varphi_i-\varphi_i^x)^2}{2}-E_{C}\partial_{\varphi_i}^2  \Bigg]\nonumber\\
    &+E_{L_\mathrm{eff}}\frac{\varphi_0^2}{2}-E_{C_0}\partial_{\varphi_0}^2 + E_{L_{\mathrm{J},0}}\frac{(\varphi_{\mathrm{J},0}-\varphi_{\mathrm{J},0}^x)^2}{2}
    - E_{C_{\mathrm{J},0}}\partial_{\varphi_{\mathrm{J},0}}^2 - E_{J_0}\cos(\varphi_0)\cos\left( \frac{\varphi_{\mathrm{J},0}}{2} \right)\nonumber\\
    &+\frac{M}{L_{\mathrm{eff}}}E_L\left( \varphi_1-\varphi_1^x \right)\varphi_0 + \frac{M}{L_\mathrm{eff}}E_L\left( \varphi_2-\varphi_2^x \right) \varphi_0
   + \frac{M^2}{LL_\mathrm{eff}}E_L\left( \varphi_1-\varphi_1^x \right)\left( \varphi_2-\varphi_2^x \right).\label{eq:Htotal}
\end{align}
\end{widetext}
This is the final Hamiltonian for which we solve the TDSE without further simplification.

\subsection{Effective coupling}\label{sec:model_coupling}
The idea of inserting the coupler element is that it leads to a tunable effective coupling between the other two SQUIDs~\cite{vandenbrink05,harris07,harris09} such that the interaction Hamiltonian takes the form
\begin{align}
   H_\mathrm{int}^\mathrm{eff} = C(\varphi_{\mathrm{J},0}^x) \varphi_1 \varphi_2 ,\label{eq:Heff_int}
\end{align}
where $C(\varphi_{\mathrm{J},0}^x)$ is the effective coupling strength tunable by the external flux $\varphi_{\mathrm{J},0}^x$ of the coupler CJJ loop.

To derive an approximate effective Hamiltonian that exhibits such a coupling term, we first replace the flux of the coupler CJJ loop $\varphi_{\mathrm{J},0}$ by its approximate expectation value. To obtain this expectation value, we expand the Hamiltonian of the SQUID given in Eq.\ (\ref{eq:squid}) to second order in $\varphi_{\mathrm{J},0}-\varphi_{\mathrm{J},0}^x$ and set $\varphi_0=0$. The resulting Hamiltonian describes a shifted harmonic oscillator
\begin{align}
   H_\mathrm{co} =\frac{E_{L_{\mathrm{J},0}}'}{2} \Bigg(\varphi_{\mathrm{J},0} &-\left(\varphi_{\mathrm{J},0}^x -\frac{2E_{J_0}\sin\left(\varphi_{\mathrm{J},0}^x/2\right)}{4E_{L_{\mathrm{J},0}}'}\right)\Bigg)^2\nonumber\\
   &- E_{C_{\mathrm{J},0}}\partial_{\varphi_{\mathrm{J},0}}^2,
\end{align}
where $E_{L_{\mathrm{J},0}}' = E_{L_{\mathrm{J},0}}+E_{J_0}\cos(\varphi_{\mathrm{J},0}^x/2)/4$. The expectation value of $\varphi_{\mathrm{J},0}$ in its ground state can thus be identified as
\begin{align}
   \langle \varphi_{\mathrm{J},0} \rangle = \varphi_{\mathrm{J},0}^x -\frac{2E_{J_0}\sin\left(\varphi_{\mathrm{J},0}^x/2\right)}{4E_{L_{\mathrm{J},0}}+E_{J_0}\cos\big(\varphi_{\mathrm{J},0}^x/2\big)}.\label{eq:expv}
\end{align}
With $\varphi_{\mathrm{J},0}$ replaced by $\langle \varphi_{\mathrm{J},0}\rangle$, we can find a matrix $T$ such that the transformation
\begin{align}
   H^{\mathrm{eff}} &= e^{iT(t)}H_\mathrm{total}e^{-iT(t)} +i\left(\frac{\mathrm{d}}{\mathrm{d}t}e^{iT(t)}\right)e^{-iT(t)} \label{eq:trafo}
\end{align}
of the total Hamiltonian yields an effective Hamiltonian which contains an interaction term of the form Eq.\ (\ref{eq:Heff_int}). Choosing
\begin{align}
   T &= T(t) =i\frac{M}{L(1+\beta_\mathrm{eff})}\left( \varphi_1-\varphi_1^x +\varphi_2 -\varphi_2^x \right)\partial_{\varphi_0},
\end{align}
where the external fluxes $\varphi_1^x$ and $\varphi_2^x$ depend on time,
\begin{align}
  \beta_\mathrm{eff}=\frac{E_{J_0}}{E_{L_\mathrm{eff}}}\cos\left(\frac{\langle\varphi_{\mathrm{J},0}\rangle}{2}\right), \label{eq:beff}
\end{align}
and expanding
\begin{align}
   \cos\left( \varphi_0 -\frac{M}{L(1+\beta_\mathrm{eff})}\left( \varphi_1-\varphi_1^x +\varphi_2 -\varphi_2^x \right) \right)
\end{align}
to second order in (the products of) $\varphi_0$, $\varphi_1-\varphi_1^x$, and $\varphi_2-\varphi_2^x$, we obtain the effective Hamiltonian
\begin{widetext}
\begin{align}
   H^\mathrm{eff} =  \sum\limits_{i=1}^2\Bigg[ &E_L\left(1+\frac{M^2}{LL_\mathrm{eff}}\frac{\beta_\mathrm{eff}}{1+\beta_\mathrm{eff}}\right)\frac{\varphi_i^2}{2}
   - E_C\partial_{\varphi_i}^2 -E_J\cos(\varphi_i)\cos\left( \frac{\varphi_{\mathrm{J,}i}}{2} \right)
   - E_{C_{\mathrm{J}}}\partial_{\varphi_{\mathrm{J,}i}}^2 + E_{L_{\mathrm{J}}}\frac{(\varphi_{\mathrm{J},i}
    -\varphi_{\mathrm{J}}^x)^2}{2}\Bigg] \nonumber\\
   +\sum\limits_{i=1}^2\Bigg[&-E_L\left(1+\frac{M^2}{LL_\mathrm{eff}}\frac{\beta_\mathrm{eff}}{1+\beta_\mathrm{eff}}\right)\varphi_i^x\varphi_i \Bigg]
   +\sum\limits_{i=1}^2\Bigg[-E_L\frac{M^2}{LL_\mathrm{eff}}\frac{\beta_\mathrm{eff}}{1+\beta_\mathrm{eff}}\varphi_{j\neq i}^x\varphi_i\Bigg]
   +E_L\frac{M^2}{LL_\mathrm{eff}}\frac{\beta_\mathrm{eff}}{1+\beta_\mathrm{eff}} \varphi_1\varphi_2 \nonumber\\
   -\Big(E_{C_\mathrm{0}} &+\frac{2E_C M^2}{L^2(1+\beta_\mathrm{eff})^2} \Big)\partial_{\varphi_\mathrm{0}}^2+E_{L_\mathrm{eff}}(1+\beta_\mathrm{eff})\frac{\varphi_\mathrm{0}^2}{2}
   + \frac{M}{L(1+\beta_\mathrm{eff})}\left(i\frac{\mathrm{d}}{\mathrm{d}t}\left( \varphi_1^x+\varphi_2^x \right)-2E_C\left( \partial_{\varphi_1} +\partial_{\varphi_2}\right) \right)\partial_{\varphi_\mathrm{0}}.\label{H_eff}
\end{align}
\end{widetext}

In the basis defined by the transformation Eq.~(\ref{eq:trafo}), we obtain the term $C(\varphi_{\mathrm{J},0}^x)\varphi_1\varphi_2$ where the dependence on $\varphi_{\mathrm{J},0}^x$ is given via $\langle\varphi_{\mathrm{J},0}\rangle$ in $\beta_\mathrm{eff}$ (see Eqs.~(\ref{eq:expv}) and (\ref{eq:beff})). The only coupling term between the coupler element and the other two SQUIDs that remains is the last term in Eq.~(\ref{H_eff}) which is expected to be much smaller than the previous coupling terms since $E_C\ll E_L$.

Note that none of the approximations made to derive Eq.~(\ref{H_eff}) affect the simulation results, as these are obtained by solving the TDSE for the Hamiltonian Eq.~(\ref{eq:Htotal}). However, as discussed in the next section, the approximation Eq.~(\ref{H_eff}) is necessary to relate the external flux $\varphi_{\mathrm{J},0}^x$ to the coupling constant $J_{12}$ (denoted by $J$ for two qubits), which appears in the Ising Hamiltonian Eq.~(\ref{eq:Hising}).

\section{Mapping to the qubit model}\label{sec:mapping}
In this section we investigate the mapping of the flux model Eq.~(\ref{H_eff}) onto the qubit model Eq.~(\ref{eq:genH_qa}) with $H_\mathrm{final}$ and $H_\mathrm{init}$ given by Eqs.~(\ref{eq:Hising}) and (\ref{eq:Htrans}), respectively. The two-qubit Hamiltonian reads
\begin{align}
   H(s)=-A(s)( \sigma_1^x +\sigma_2^x) -B(s)( h_1\sigma_1^z +h_2\sigma_2^z + J\sigma_1^z\sigma_2^z).\label{eq:H_2level}
\end{align}
As we will see below, reducing Eq.~(\ref{H_eff}) to the generic form of Eq.~(\ref{eq:H_2level}) enforces a specific choice of the external fluxes $\varphi_i^x$ (see Eq.~(\ref{eq:phi_i})) and gives a relation between $J$ and $\varphi_{\mathrm{J},0}^x$ (see Eq.~(\ref{eq:J_of_phicox})).

Since we have assumed the two SQUIDs to be identical, the mapping to the qubit model is the same for both, and therefore we omit the SQUID indices in this section. The two lowest-energy states $\ket{g}$ and $\ket{e}$ of each SQUID for $\varphi^x=0$ define the computational subspace \cite{harris10}. We obtain them by diagonalizing the first part in square brackets of Eq.~(\ref{H_eff}) in $\varphi$- and $\varphi_\mathrm{J}$-space (see Sec.~\ref{sec:method} for the definition of the discretized basis).

Note that the first summand given in Eq.~(\ref{H_eff}) contains an effective change of the inductive energy depending on the value chosen for $\varphi_{\mathrm{J},0}^x$ (because $\beta_\mathrm{eff}$ depends on it). Therefore, the definition of the computational subspace changes with the coupling strength. This leads to slightly different annealing schemes, i.e., a dependence of $A(s)$ and $B(s)$ on $\varphi_{\mathrm{J},0}^x$, as observed experimentally \cite{harris09,harris10}.

The computational basis states $\ket{\uparrow}$ and $\ket{\downarrow}$ are defined as the eigenstates of the operator $\varphi$ (and thus of the second part in square brackets in Eq.~(\ref{H_eff})) inside the computational subspace $\mathrm{span}\{\ket{g},\ket{e}\}$. We obtain
\begin{align}
   \ket{\uparrow} &= a\ket{g} +b\ket{e} = \int\limits_{-\infty}^\infty \int\limits_{-\infty}^\infty \mathrm{d}\varphi\mathrm{d}\varphi_\mathrm{J} \,u(\varphi,\varphi_\mathrm{J}) \ket{\varphi\,\varphi_\mathrm{J}}, \label{eq:upstate}\\
   \ket{\downarrow} &= a\ket{g} -b\ket{e} =\int\limits_{-\infty}^\infty \int\limits_{-\infty}^\infty \mathrm{d}\varphi\mathrm{d}\varphi_\mathrm{J} \,d(\varphi,\varphi_\mathrm{J}) \ket{\varphi\,\varphi_\mathrm{J}}, \label{eq:downstate}
\end{align}
where $|a|=|b|=1/\sqrt{2}$~\cite{harris10} and $u(\varphi,\,\varphi_\mathrm{J})$ and $d(\varphi,\,\varphi_\mathrm{J})$ are the resulting amplitudes of the states $\ket{\uparrow}$ and $\ket{\downarrow}$ in $\varphi$- and $\varphi_\mathrm{J}$-space. Note that $\ket{g}$ and $\ket{e}$ depend on the time-dependent external flux $\varphi_\mathrm{J}^x$, implying that the definition of the computational states changes with time. The projection of the operator $\widetilde E_L\varphi$ with
\begin{align}\widetilde E_L =E_L\left(1+\frac{M^2}{LL_\mathrm{eff}} \frac{\beta_\mathrm{eff}}{1+\beta_\mathrm{eff}}\right)\label{eq:tilde_E_L}\end{align} has eigenstates $\ket{\uparrow}$ and $\ket{\downarrow}$
with eigenvalues $\pm I_p(s)/2e$, respectively. Thus, in this subspace, $\widetilde E_L\varphi$ is represented by $I_p(s)\sigma^z/2e$ and the first contributions in square brackets in Eq.~(\ref{H_eff}) are mapped to $-\Delta(s) \sigma^x/2$, where $\Delta(s) = E_1(s)-E_0(s)$ is the energy gap between the ground state $\ket{g}$ and the first excited state $\ket{e}$.

To derive the coupling terms, we write the SQUID indices $i$ again. For the terms in $\sigma_i^z$ and $\sigma_1^z\sigma_2^z$ to scale with the same annealing function $B(s)$~\cite{harris10_eightqubit}, $\varphi_i^x$ has to be set to
\begin{align}
  \varphi_i^x=h_i\gamma\frac{2eI_p(s)M^2}{L_\mathrm{eff}},\label{eq:phi_i}
\end{align}
where $\gamma = \mathrm{max}_{\varphi_{\mathrm{J},0}^x} \beta_\mathrm{eff}E_L^2 /(1+\beta_\mathrm{eff})\widetilde E_L^2$.
Disregarding the contribution of the last term in Eq.~(\ref{H_eff}), we find that the Hamiltonian for $\varphi_0$ effectively decouples from the qubit Hamiltonian and thus the effective qubit Hamiltonian can be written as
\begin{align}
   H^{\mathrm{eff},q}\approx &-\sum\limits_{i=1}^2 \left(\frac{\Delta(s)}{2}\sigma_i^x + h_i\gamma\frac{I_p^2(s)M^2}{L_\mathrm{eff}}\sigma_i^z \right)\nonumber\\ 
   &- \frac{E_L}{\widetilde E_L}\frac{I_p^2(s)M^4}{LL_\mathrm{eff}^2}\frac{\beta_\mathrm{eff}}{1+\beta_\mathrm{eff}}\gamma\left( h_1\sigma_2^z +h_2\sigma_1^z \right) \nonumber\\
   &+\frac{E_L^2}{\widetilde E_L^2}\frac{I_p^2(s)M^2}{L_\mathrm{eff}}\frac{\beta_\mathrm{eff}}{1+\beta_\mathrm{eff}}\sigma_1^z\sigma_2^z.\label{eq:Heff}
\end{align}

For all $J\in [-1,1]$, we have
\begin{align}
  -\gamma = -\underset{\varphi_{\mathrm{J},0}^x}{\mathrm{max}} \frac{\beta_\mathrm{eff}}{1+\beta_\mathrm{eff}}\frac{E_L^2}{\widetilde E_L^2} \le -J\gamma \le \underset{\varphi_{\mathrm{J},0}^x}{\mathrm{max}} \frac{\beta_\mathrm{eff}}{1+\beta_\mathrm{eff}}\frac{E_L^2}{\widetilde E_L^2} = \gamma.
\end{align}
Thus, and because $E_L^2\beta_\mathrm{eff}(\varphi_{\mathrm{J},0}^x)/(\widetilde E_L^2(\varphi_{\mathrm{J},0}^x)(1+\beta_\mathrm{eff}(\varphi_{\mathrm{J},0}^x)))$ is monotonic, it is possible to find $\varphi_{\mathrm{J},0}^x$ such that 
\begin{align}
\frac{\beta_\mathrm{eff}(\varphi_{\mathrm{J},0}^x)}{1+\beta_\mathrm{eff}(\varphi_{\mathrm{J},0}^x)}\frac{E_L^2}{\widetilde E_L^2(\varphi_{\mathrm{J},0}^x)}=-J\gamma \label{eq:J_of_phicox}
\end{align}
for all $J\in [-1,1]$, and Eq.~(\ref{eq:Heff}) becomes
\begin{align}
   H^{\mathrm{eff},q}\approx & -\sum\limits_{i=1}^2 \frac{\Delta(s)}{2}\sigma_i^x -\gamma\frac{I_p^2(s)M^2}{L_\mathrm{eff}}\Bigg( \sum\limits_{i=1}^2 h_i\sigma_i^z + J\sigma_1^z\sigma_2^z  \nonumber\\
   &- \frac{\widetilde E_L}{E_L}\frac{M^2}{LL_\mathrm{eff}} J\gamma\left( h_1\sigma_2^z +h_2\sigma_1^z \right)\Bigg), \label{eq:Heff_q}
\end{align}
which has the structure of an Ising model in a transverse field.
Comparing Eq.~(\ref{eq:Heff_q}) to Eqs.~(\ref{eq:Hising}) and~(\ref{eq:Htrans}), we can identify $A(s)=\Delta(s)/2$ and $B(s)=\gamma I_p^2(s)M^2/L_\mathrm{eff}$ and see that
\begin{align}
    H^{\mathrm{eff},q}\approx &-A(s)( \sigma_1^x +\sigma_2^x)  \nonumber\\
   &-B(s)\Bigg( h_1\sigma_1^z +h_2\sigma_2^z + J\sigma_1^z\sigma_2^z  \nonumber\\
   &- \frac{\widetilde E_L}{E_L}\frac{M^2}{LL_\mathrm{eff}} J\gamma\left( h_1\sigma_2^z +h_2\sigma_1^z \right)\Bigg),\label{eq:H_2l_model}
\end{align}
where the last term only adds a small contribution since $M^2\ll LL_\mathrm{eff}$.

\section{Simulation}\label{sec:method}
This section starts with a brief description of the numerical technique used to perform the simulation of the three-SQUID model. Then we discuss the choice of the model parameters that appear in Eq.~(\ref{eq:Htotal}) and explain the method by which we numerically extract the annealing scheme and the qubit-qubit coupling $J$, which appears in Eq.~(\ref{eq:H_2level}), from the dynamics of the fluxes.

For the simulation of the time evolution of the system defined by Eq.~(\ref{eq:Htotal}), we use the Suzuki-Trotter product-formula algorithm \cite{suzuki84,deraedt87} to numerically solve the TDSE
\begin{align}
   i\partial_t |\psi(t)\rangle = H(t)|\psi(t)\rangle.\label{eq:tdse}
\end{align}
The time-dependent Hamiltonian is discretized such that the state vector $|\psi(t)\rangle$ can be updated by a time step $\tau$ to $|\psi(t+\tau)\rangle$ using the time-evolution operator $U(t,t+\tau)=\exp( -i\tau H(t+\tau/2) )$. To implement the algorithm, we fix a basis for the description of $\ket{\psi(t)}$ and a decomposition of the Hamiltonian $H(t)=\sum_k A_{k}(t)$ such that
\begin{align}
   e^{-iH(t)\tau} \approx e^{-iA_{1}(t)\tau} e^{-iA_{2}(t)\tau} \cdots e^{-iA_{K}(t)\tau} = U_{t,1}(\tau),
\end{align}
is a good approximation for sufficiently small $\tau$ and the update of the state vector can be performed with two-component updates only. For a detailed description of how to choose the $A_k$, see Ref.~\cite{deraedt87}. In our simulation, we use the second-order approach given by
\begin{align}
   e^{-iH(t)\tau} \approx U_{t,1}\left(\tau/2\right)U^\dagger_{t,1}\left(-\tau/2\right).
\end{align} 
Note that there is no need to diagonalize the Hamiltonian or to store the full matrices representing the Hamiltonian or the time-evolution operator.

For the description of the state $\ket{\psi}$, the fluxes $\varphi_i$ through the main loops are discretized, i.e., the wave function is defined at $\lambda_i$ discrete points $\varphi_{i\mathrm{min}} + l_i\Delta \varphi_i$, $l_i=0,\dots ,\lambda_i-1$. By studying the convergence of the numerical results as a function of $\lambda_i$ and $\Delta\varphi_i$, we find that $\lambda_1=\lambda_2=47$ and $-2.0 \le \varphi_{1},\varphi_2\le 2.0$, and $\lambda_0=31$ and $-1.0\le \varphi_{0}\le 1.0$ provide a good compromise between accuracy and computational work to solve the TDSE. Since $E_{L_\mathrm{J}}\gg E_J$  and $E_{L_\mathrm{J,0}}\gg E_{J_0}$, the Hamiltonian for $\varphi_{\mathrm{J},i}$ resembles an oscillator with small anharmonicity. In the harmonic-oscillator basis, the evolution of $\varphi_{\mathrm{J},i}$ can be well described with the three lowest states and thus the fluxes $\varphi_{\mathrm{J},i}$ through the CJJ loops can be discretized in the harmonic-oscillator basis and labeled by $\ket{m_i}$, $m_i=0,1,2$. In summary, the state $\ket{\psi}$ is represented by
\begin{align}
|\psi\rangle = \sum\limits_{\mathclap{\substack{l_0,l_1,l_2,\\m_0,m_1,m_2}}} \phi_{l_0,m_0,l_1,m_1,l_2,m_2}\ket{l_0\, m_0\, l_1\, m_1\, l_2\, m_2},\label{eq:discr_psi}
\end{align}
where the amplitudes $\phi_{l_0,m_0,l_1,m_1,l_2,m_2}$ are stored as an array of $\lambda_1 \lambda_2 \lambda_0\times 3^3 \approx 1.85\times 10^6$ complex double-precision numbers. To store this array, approximately $30\,\mathrm{MB}$ of memory is needed. Parallelization of the state updates is implemented using OpenMP.
Testing with decreasing time steps $\tau$ and studying the convergence, we find that for $\tau=1.5\times 10^{-5}\,\mathrm{ns}$ the results are sufficiently accurate. Due to this small time step, one quantum annealing run of $5\,\mathrm{ns}$ takes about 16 hours on a 24-core node of the supercomputer JURECA~\cite{jureca}.

\subsection{Parameters}
\begin{figure}[bt]
   \centering
   \includegraphics[width=0.35\textwidth]{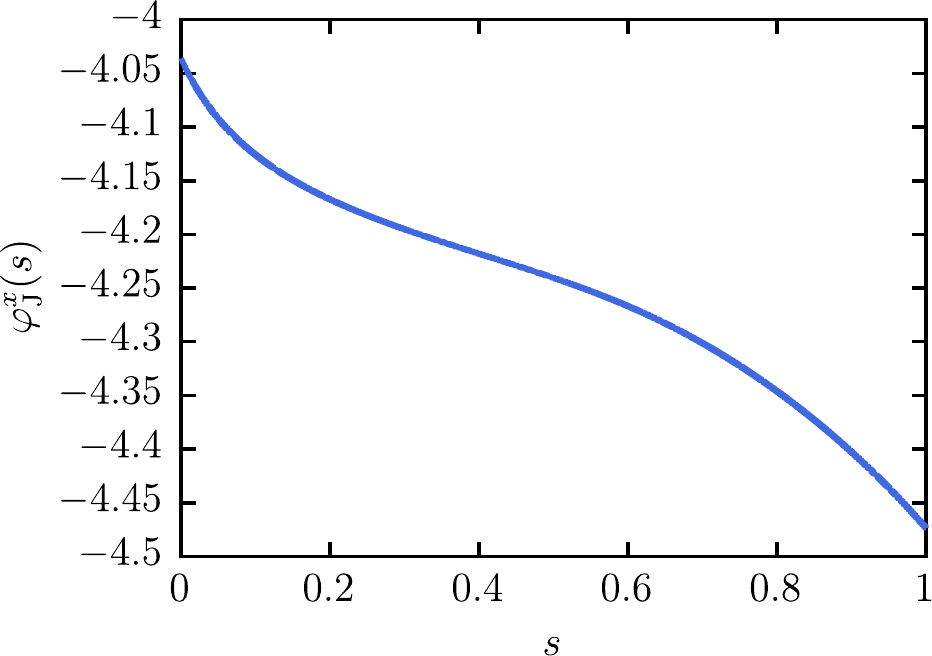}
   \caption{(color online) External flux $\varphi_\mathrm{J}^x=2\pi\Phi_\mathrm{J}^x/\Phi_0$ as a function of the normalized annealing time $s$, as provided to us by D-Wave Systems Inc.}
   \label{fig:phicjjx}
\end{figure}

The parameters of the Hamiltonian Eq.~(\ref{eq:Htotal}) and the values of the time-dependent flux $\varphi_\mathrm{J}^x(s)$, which determines the annealing scheme, were provided to us by D-Wave Systems Inc.\ and are typical values of the D-Wave 2000Q processor~\cite{dwave_private}. The device parameters used in the simulation are slightly modified and listed in Table~\ref{tab:parameters} and $\varphi_\mathrm{J}^x(s)$ is plotted in Fig.\ \ref{fig:phicjjx}. The external fluxes $\varphi_i^x$ and $\varphi_{\mathrm{J},0}^x$ are computed from Eq.~(\ref{eq:phi_i}) and by solving Eq.~(\ref{eq:J_of_phicox}) numerically for $\varphi_{\mathrm{J},0}^x(J)$.

\begin{figure}[bt]
   \centering
   \includegraphics[width=0.35\textwidth]{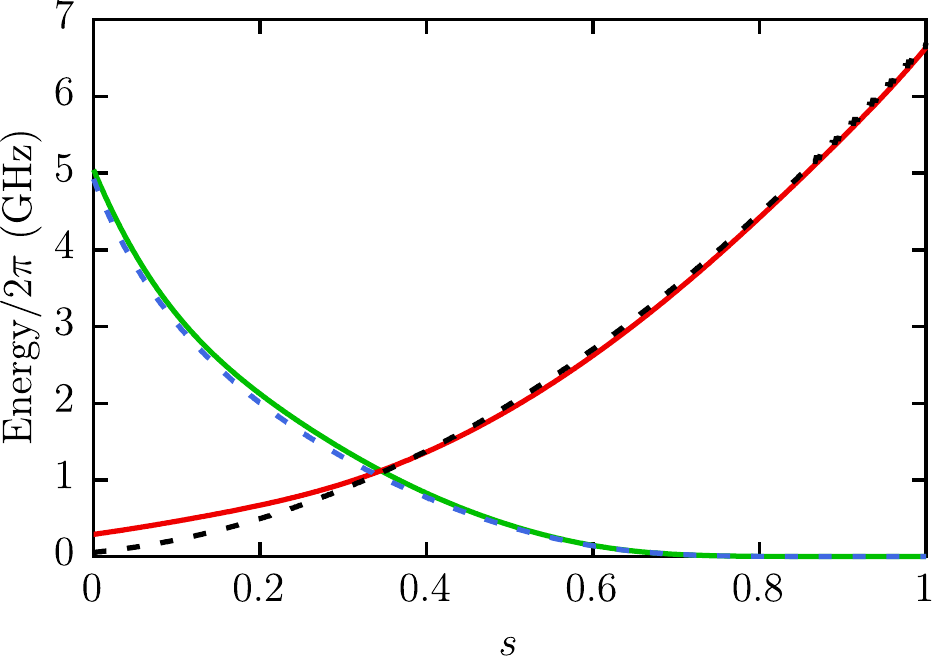}
   \caption{(color online) Functions $A(s)$ and $B(s)$ of the annealing scheme as provided to us by D-Wave Systems Inc.~ (dashed lines) in comparison to the annealing scheme obtained from the full Hamiltonian for an uncoupled SQUID (solid lines).}
   \label{fig:scheme}
\end{figure}

Using the provided parameters, we compute the annealing scheme of a single SQUID by exact diagonalization of Eq.~(\ref{eq:squid}) with $\varphi^x=0$ and computing $\Delta(s)$ and $I_p(s)$ as described in Sec.~\ref{sec:mapping}. However, the resulting annealing scheme (data not shown) does not match the data of the annealing scheme provided to us by D-Wave Systems Inc.~\cite{dwave_private} (see Fig.~\ref{fig:scheme} (dashed lines)). Better agreement between the two annealing schemes was found by using $E_C=4.68\,\mathrm{GHz}$ (which was computed from the provided capacitance directly) instead of $E_C=5.85\,\mathrm{GHz}$ (value provided by D-Wave Systems Inc.), $E_{L_\mathrm{J}}=54538\,\mathrm{GHz}$ instead of $E_{L_\mathrm{J}}=73388\,\mathrm{GHz}$, and $M=15.97\,\mathrm{pH}$ instead of $M=13.7\,\mathrm{pH}$, see Fig.~\ref{fig:scheme} (solid lines). The disagreement in $B(s)$ for small $s$ could not be removed by slight variation of the model parameters. Changing $\varphi_i^x$ would reduce the disagreement for the single-qubit terms, but at the same time Eq.~(\ref{eq:phi_i}) would be violated, effectively yielding different functions for the single-qubit and two-qubit $\sigma^z$ terms. Thus, we decided to keep $\varphi_i^x$ as given by Eq.~(\ref{eq:phi_i}).
\begin{table}[tb]
  \centering
  \caption{Values of the device parameters appearing in the Hamiltonian Eq.~(\ref{eq:Htotal}) and used in our numerical work.}
  \label{tab:parameters}
  \begin{minipage}[t]{0.23\textwidth}
   \begin{tabular}{l c}
    Parameter & Value \\
    \hline
    \hline
    $ E_C $ & $ 4.68\,\mathrm{GHz} $\\
    $ E_L $ & $ 3.48 \,\mathrm{THz} $\\
    $ E_{C_\mathrm{J}} $ & $ 133.02 \,\mathrm{GHz} $ \\
    $ E_{L_\mathrm{J}} $ & $ 54.54 \,\mathrm{THz} $ \\
    $ E_J $ & $ 7.80 \,\mathrm{THz} $ \\
    $ M $ & $ 15.97 \,\mathrm{pH} $  \\
    \end{tabular}
  \end{minipage}
\hfill
  \begin{minipage}[t]{0.23\textwidth}
     \begin{tabular}{l c}
    Parameter & Value \\
    \hline
    \hline
    $ E_{C_0} $ & $ 9.02\,\mathrm{GHz} $\\
    $ E_{L_0} $ & $ 15.67\,\mathrm{THz} $ \\
    $ E_{C_{\mathrm{J},0}} $ & $ 213.50\,\mathrm{GHz} $ \\
    $ E_{L_{\mathrm{J},0}} $ & $ 354.18\,\mathrm{THz} $ \\
    $ E_{J_0} $ & $ 18.72\,\mathrm{THz} $ \\
    \end{tabular}
  \end{minipage}
\end{table}
\subsection{Estimation of the coupling strength and the annealing scheme}\label{sec:method_scheme}
In order to map the full state Eq.~(\ref{eq:discr_psi}) to the computational space, we have to trace out the degrees of freedom of the coupler element and project the resulting reduced density matrix onto the computational subspace. To do so, we discretize Eqs.~(\ref{eq:upstate}) and (\ref{eq:downstate}) to obtain
\begin{align}
   \ket{\uparrow} &= \sum\limits_{l,m} u_{l,m} \ket{l\, m},\label{eq:def_up}\\
   \ket{\downarrow} &= \sum\limits_{l,m} d_{l,m} \ket{l\, m}\label{eq:def_down}
\end{align}
for a single qubit, and accordingly the product states for the two-qubit states, where $u_{lm}$ and $d_{lm}$ are the discretizations of $u(\varphi,\,\varphi_\mathrm{J})$ and $d(\varphi,\,\varphi_\mathrm{J})$, respectively. Since this projection is not a unitary transformation, the trace of the projected density matrix $\rho^\mathrm{comp}$ will be less than one if there is leakage to higher levels, i.e., if excitations to states outside the computational subspace occur. The deviation of the trace from one is a measure for the amount of leakage to higher levels.

To obtain the effective coupling strength and the annealing scheme we proceed as follows. We start with the ideal qubit Hamiltonian (for simplicity, with $h_1=h_2=0$)
\begin{align}
   H_2(s) = -\frac{\Delta(s)}{2}\left( \sigma_1^x + \sigma_2^x \right) + C(s)\sigma_1^z\sigma_2^z,
\end{align}
where $\Delta(s)$ and $C(s)$ are to be determined by comparison with the data obtained by simulating the model Eq.~(\ref{eq:Htotal}). For fixed $s\in [0,1]$, the evolution determined by $H_2(s)$ of the initial state $\ket{++} = \left(  \ket{\uparrow} + \ket{\downarrow}\right)\otimes\left(  \ket{\uparrow} + \ket{\downarrow}\right)/2$ and the expectation values $\langle \sigma_1^\alpha\sigma_2^\beta \rangle$ for $\sigma_i^\alpha,\sigma_i^\beta\in \{\mathbb{I}_i,\sigma_i^x,\sigma_i^y,\sigma_i^z\}$ in the evolved state can be calculated analytically. On the other hand, for any time $t$, we can compute these expectation values directly from the simulation of the time evolution of the initial state $\ket{{++}}$ expressed in flux degrees of freedom using Eqs.~(\ref{eq:def_up}) and (\ref{eq:def_down}) at a fixed value for $s$. The time evolution is governed by the full Hamiltonian Eq.~(\ref{eq:Htotal}) based on the flux degrees of freedom with fixed $s$. In this case, the expectation values are computed by $\mathrm{Tr}(\rho^\mathrm{comp}\sigma_1^\alpha\sigma_2^\beta)$. We can then estimate $\Delta (s)$ and $C(s)$, and thus the effective coupling strength and the annealing scheme, by fitting the analytical expressions to the simulation data.

\section{Results}\label{sec:results}
In this section, we present the results obtained from the simulation of the flux model described by the Hamiltonian Eq.~(\ref{eq:Htotal}). First, we show that the mapping between $J$ and $\varphi_\mathrm{J,0}^x$ given by Eq.~(\ref{eq:J_of_phicox}) leads to the desired effective coupling strength. Subsequently, we discuss the effective annealing scheme obtained by using the procedure described in Sec.~\ref{sec:method_scheme}. We check the results of the simulation based on the flux model and the results of the qubit model against each other by comparing the probabilities during and at the end of the annealing process. Finally, we briefly discuss the data obtained from the D-Wave quantum annealer in comparison to the simulation results.
\subsection{Effective coupling and annealing scheme}
\begin{figure}[bt]
   \centering
   \includegraphics[width=0.35\textwidth]{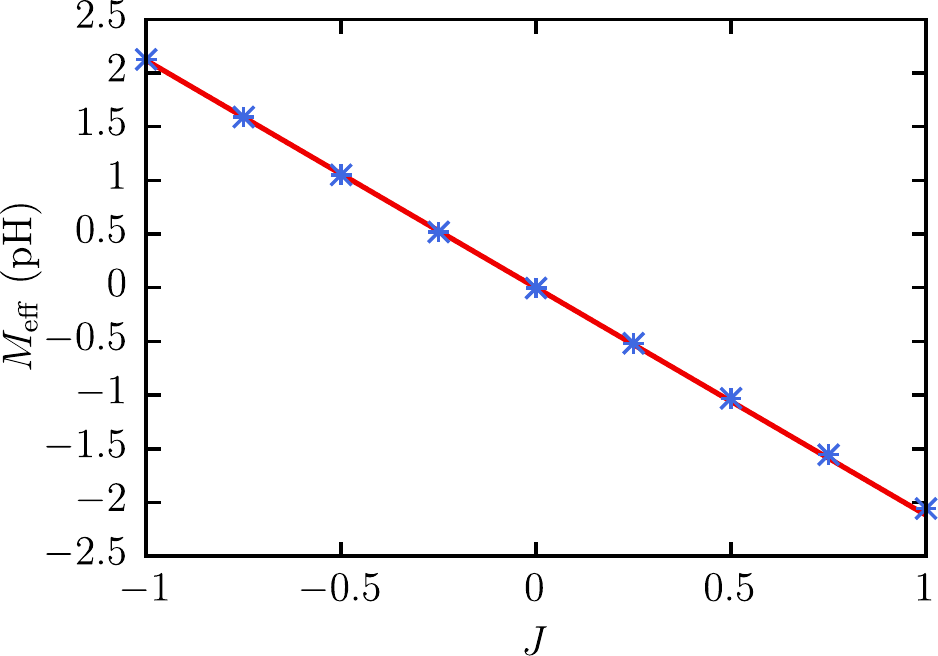}
   \caption{(color online) Effective mutual inductance between the qubits as a function of the qubit-qubit coupling $J$. The solid line shows the expected behavior based on the analytical calculation presented in Sec.~\ref{sec:mapping}. Asterisks show the simulation data.}
   \label{fig:J}
\end{figure}

In order to assess the mapping between $J$ and $\varphi_\mathrm{J,0}^x$ using Eq.~(\ref{eq:J_of_phicox}), we first study the effective mutual inductance $M_\mathrm{eff}$ as a function of $J$.
We utilize the method described in Sec.~\ref{sec:method_scheme} for various values of $J$ and $s=1$ (such that $\Delta(s)\approx 0$) to determine the coupling strength.
In this case, the analytical result for the expectation value $\langle \sigma_1^y\sigma_2^z \rangle=\sin(2C(1)t)$ can be used for fitting. The obtained value for $C(1)$ for each $J$ is then mapped onto the effective mutual inductance $M_\mathrm{eff}(J)=C(1)/I_p^2(1)=-J\gamma M^2/L_\mathrm{eff}$ and plotted against $J$.
The result for the effective inductance $M_\mathrm{eff}(J)$ between the qubits is presented in Fig.~\ref{fig:J} and shows good agreement between the theoretical linear curve from the approximation and the simulation result. For $J$ in the range $[-1,1]$, we can reach all values for $M_\mathrm{eff}$ in $[-M_\mathrm{eff,max},M_\mathrm{eff,max}]$ to good precision and have thus obtained a transformation $\varphi_\mathrm{J,0}^x \leftrightarrow J$ such that the mapping $J\leftrightarrow M_\mathrm{eff}$ is linear. Therefore, we can expect that the mapping onto the qubit model and the resultant mapping $J\leftrightarrow \varphi_\mathrm{J,0}^x$ work reasonably well.

\begin{figure}[bt]
   \centering
   \includegraphics[width=0.35\textwidth]{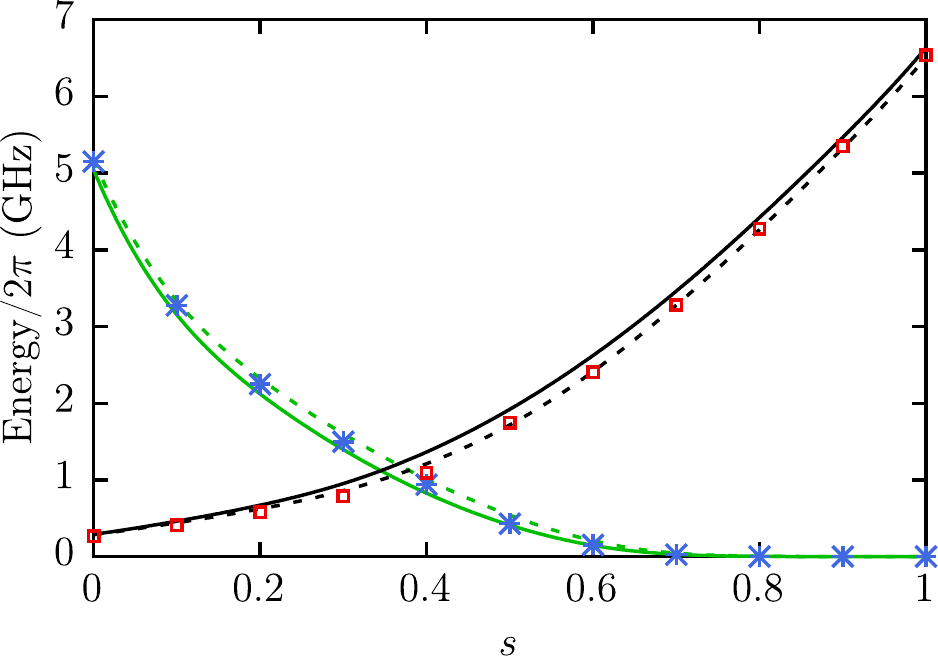}
   \caption{(color online) Annealing scheme for the uncoupled single SQUID model obtained by diagonalization of Eq.~(\ref{eq:squid}) with $\varphi^x=0$ (solid line) and with $E_L$ replaced by $\widetilde E_L$ (see Eq.~(\ref{eq:tilde_E_L})) where $\varphi_\mathrm{J,0}^x$ is set to correspond to $J=-1$ (dashed line) to show the effect of coupling on the annealing scheme. Blue asterisks and red squares are obtained from the simulation of the coupled model Eq.~(\ref{eq:Htotal}) as described in Sec.~\ref{sec:mapping}. Solid and dashed lines following the asterisks represent $A(s)$ and solid and dashed lines following the squares represent $B(s)$. The parameters are $J=-1$ and $h_1=h_2=0$.
   }
   \label{fig:annealing_v1-1}
\end{figure}

To assess the effective annealing scheme, we use the method described in Sec.~\ref{sec:method_scheme} for different values $s'\in[0,1]$, using the analytical expression of the expectation value
\begin{align}
  \langle \sigma^z_1\sigma_2^z\rangle = \frac{2\Delta(s')C(s')\sin^2\left(\sqrt{\Delta(s')^2+C(s')^2}t\right)}{\Delta(s')^2+C(s')^2}
\end{align}
for the fitting of $\Delta(s')$ and $C(s')$. Figure \ref{fig:annealing_v1-1} shows the effective annealing scheme (data points) obtained in this way.

We find that the data points in Fig.~\ref{fig:annealing_v1-1} deviate from the annealing scheme for an uncoupled qubit (solid lines, obtained by using Eq.~(\ref{eq:squid})), but they are in better agreement with the annealing scheme obtained by using $\widetilde E_L$ (see Eq.~(\ref{eq:tilde_E_L})) instead of $E_L$ (dashed lines).
Note that for the computation of the annealing scheme, the single-SQUID Hamiltonian Eq.~(\ref{eq:squid}) is mapped onto the effective Hamiltonian $H^{\mathrm{eff},q}\approx -A(s)\sigma_1^x-B(s)h_1\sigma_1^z$. In the qubit model (Eq.~(\ref{eq:H_2level})), this gives the same function for $B(s)$ as the term proportional to $\sigma_1^z\sigma_2^z$ in the case of two coupled qubits. Because of the choice for $\varphi_1^x$ (see Eq.~(\ref{eq:phi_i})), this should also be the case for the SQUID model if the mapping to the qubit model works well enough.

We find, in agreement with our analytical calculation, that the effective coupling between the SQUIDs induces a shift in the inductive energy, leading to shifts in the annealing scheme. Including this shift, the effective annealing scheme can be well described by the single-SQUID annealing scheme. The influence of the coupling on the inductive energy was also observed in experiments \cite{harris09}.
\subsection{Comparison to the qubit model}
The next step is to compare the overall performance and the final probabilities between the real-time simulation with the Hamiltonian given in Eq.~(\ref{eq:Htotal}) and the qubit Hamiltonian given in Eq.~(\ref{eq:H_2level}).

\begin{figure}[bt]
   \centering
   \includegraphics[width=0.35\textwidth]{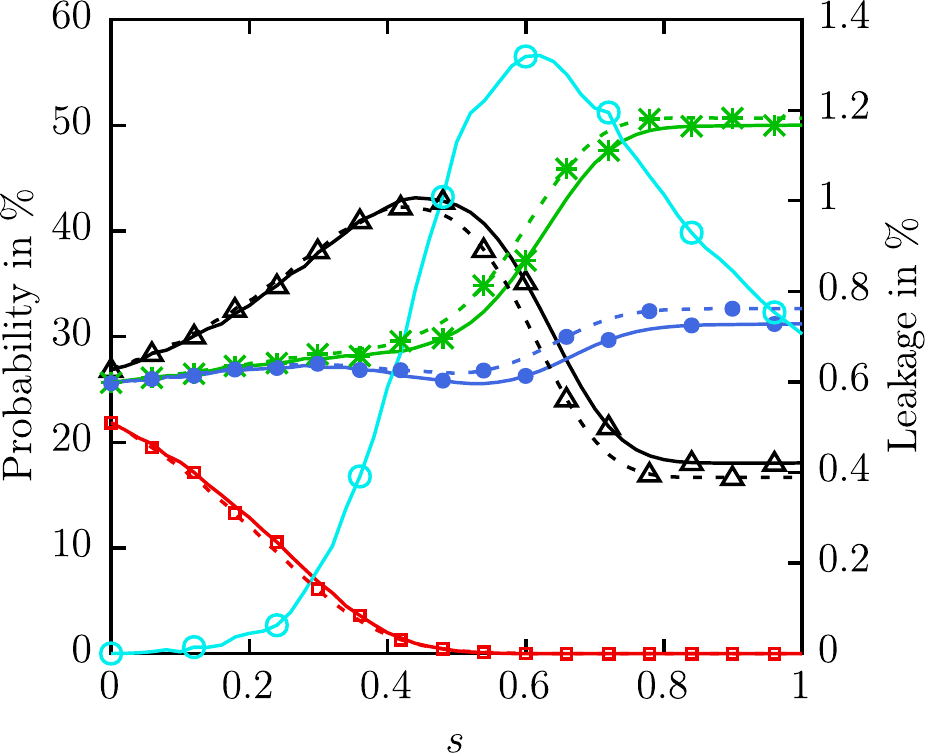}
   \caption{(color online) Probabilities of the four computational states $\ket{\uparrow \uparrow}$ (black triangles), $\ket{\uparrow \downarrow}$ (green asterisks), $\ket{\downarrow \uparrow}$ (blue solid circles), and $\ket{\downarrow \downarrow}$ (red squares) during the annealing process for the qubit model (dashed lines) and the full system (solid lines). Markers are used to better distinguish the lines of the different states. For the data from the simulation of the qubit model, every 120th data point is plotted with a marker and for the data from the simulation of the full system, every 6th point is plotted with a marker. For the full system, additionally the probability of leakage (cyan open circles) is shown using the right $y$ axis. The annealing time was set to $t_a=5\,\mathrm{ns}$. The parameters are $J=-1$, $h_1=0.96$, and $h_2=0.94$.}
   \label{fig:comp_states}
\end{figure}

As mentioned in Sec.~\ref{sec:method}, the amount of leakage to higher excited states can be computed by projecting the density matrix onto the computational subspace. The projected density matrix can also be used to obtain the probabilities of the computational basis states. As an illustration, in Fig.~\ref{fig:comp_states} we show the results for $J=-1$, $h_1=0.96$, and $h_2=0.94$.
For this choice of parameters, the ground state of Eq.~(\ref{eq:Hising}) is $\ket{{\uparrow\downarrow}}$.
The total annealing time was set to $t_a = 5\,\mathrm{ns}$ for the simulations of both the flux model and the qubit model.

This annealing time is much less than typically used on the D-Wave processors (order of microseconds), but for comparison of the results of the flux simulation with the qubit description only, this difference is unimportant. Note that in this section our aim is to scrutinize the validity of the qubit model as a description of the flux dynamics governed by the Hamiltonian Eq.~(\ref{eq:Htotal}), not to compare simulation results with experiments performed on the D-Wave quantum annealer (see Secs.~\ref{sec:compare_d_wave} and~\ref{section2} below).

As seen from Fig.~\ref{fig:comp_states}, there are small deviations from the probabilities obtained from the qubit representation. Some leakage which has its maximum at about $s=0.6$, where the change in the probabilities of the computational states is strongest, can also be observed. In general, the evolutions of both the full model and the qubit model show the same features.

In the following we refer to the probability of finding the system at the end of the annealing process in the ground state of Hamiltonian Eq.~(\ref{eq:Hising}) as success probability. For the example case shown in Fig.~\ref{fig:comp_states}, the success probability for the flux model and the qubit model differ only slightly.

\begin{figure}[bt]
   \centering
   \includegraphics[width=0.4\textwidth]{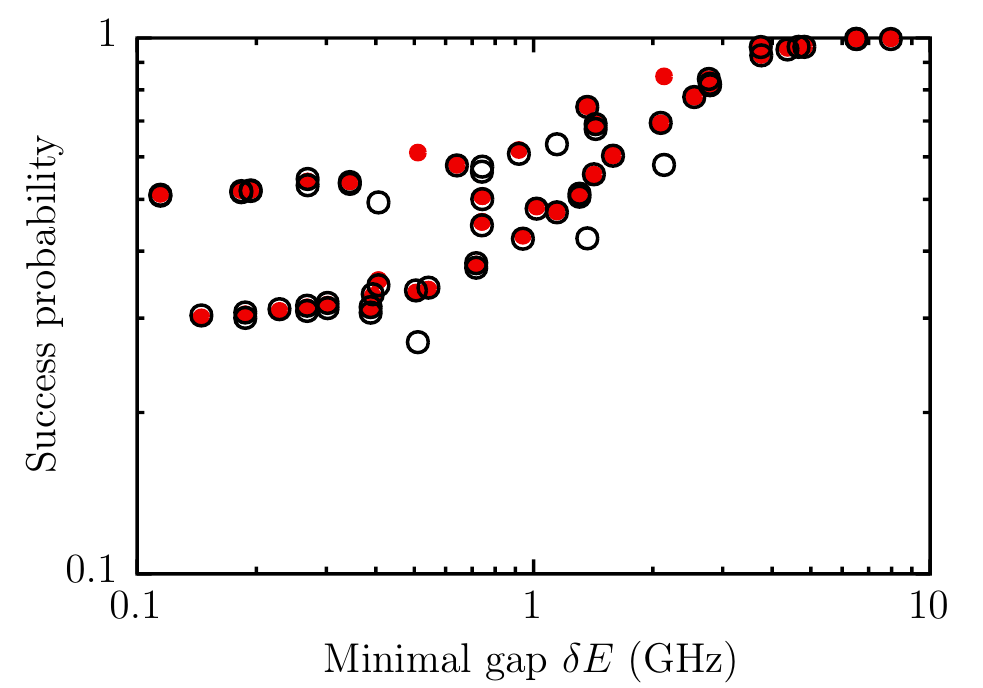}
   \caption{(color online) Success probability as a function of the minimal energy gap $\delta E = \min_s E_1(s)-E_0(s)$ (computed from the qubit model) during the annealing process. Each data point represents another problem, i.e., other values for the parameters $h_1$, $h_2$, and $J$. A list with all cases is given in Appendix~\ref{app:data}. Closed (red) circles show the results for the qubit model and open (black) circles originate from the simulation of the flux model.}
   \label{fig:gap_probs}
\end{figure}

In this example, the success probability is higher for the qubit model. However, Fig.~\ref{fig:gap_probs} shows that there are also cases in which the success probability is lower for the qubit model. Note that the annealing process does not start with equal probability for all states because we start the annealing in the ground state of the system instead of in the state $\ket{++}=\ket{+}_1\otimes\ket{+}_2$, since for $B(s=0)>0$, the ground state of the qubit model Eq.~(\ref{eq:Hising}) is not exactly the state $\ket{++}$ but a superposition of all basis states. A simulation of the qubit model comparing the annealing processes with the two different initial states shows deviations during the annealing process, but there is no significant difference in the success probability (data not shown).

In summary, we observed an influence of the coupling on the annealing scheme and some amount of leakage to higher levels. The important question, however, is whether these effects have consequences on the final success probability. Figure \ref{fig:gap_probs} shows the success probability for many different problems (defined in Appendix~\ref{app:data}) as a function of the minimal energy gap between the ground state and the first excited state during the annealing process, computed from the qubit model. As can be seen in Fig.~\ref{fig:gap_probs}, for most of the investigated cases, the effects on the success probability of using a subspace of a larger system as the qubit instead of an ideal qubit representation are rather small. The data points generated by the simulation based on Eq.~(\ref{eq:Htotal}) (black open circles) show, apart from a few exceptions, only small deviations from the data points of the ideal qubit model (red closed circles). Note that the success probability can be enhanced as well as reduced compared to the ideal qubit model.

\begin{figure*}[bt]
   \centering
   \includegraphics[width=0.9\textwidth]{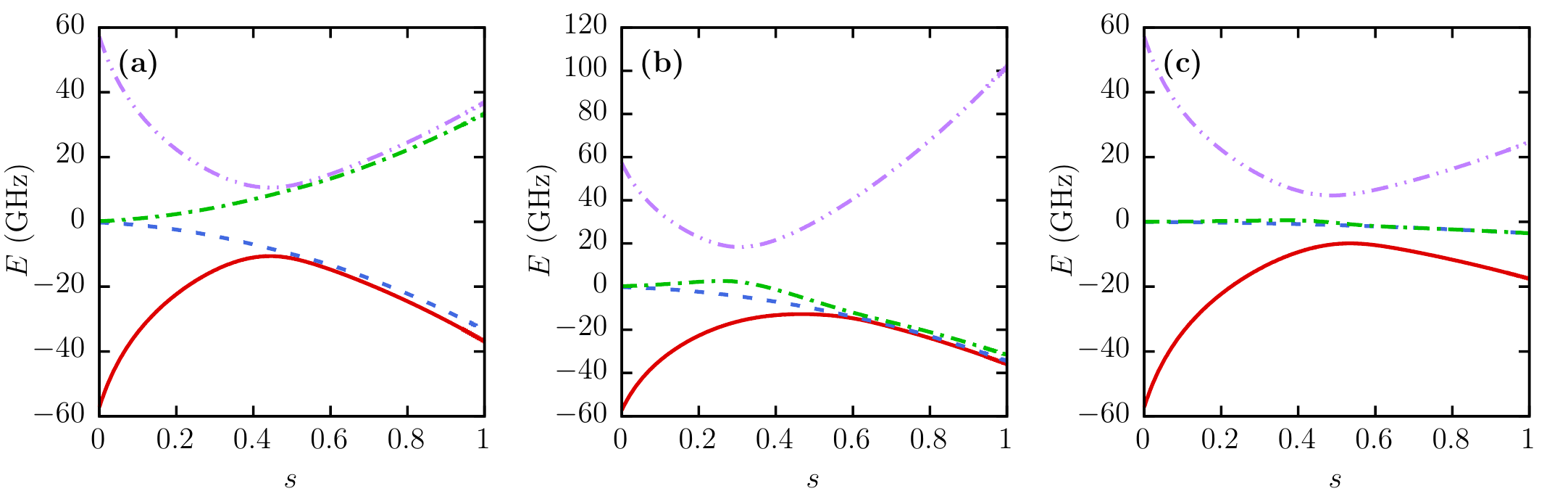}
   \caption{(color online) Energy spectra during the annealing process for the three cases listed in Table~\ref{tab:d_wave_data} with the parameters (a) parameters $J=-1$, $h_1=0$, and $h_2=0.05$; (b) $J=-1$, $h_1=0.96$, and $h_2=0.94$; and (c) $J=0.1$, $h_1=0.3$, and $h_2=-0.3$.}
   \label{fig:spec_all}
\end{figure*}

Another interesting observation can be made in Fig.~\ref{fig:gap_probs}. For large minimal energy gaps $\delta E = \min_s E_1(s)-E_0(s)$, the system shows Landau-Zener behavior~\cite{landau32,zener32}. For small minimal energy gaps $\delta E$, the success probabilities form two clusters, one cluster of probability approximately equal to $0.5$ and a second one of probability approximately equal to $ 0.3$. Considering the energy spectra of these instances, we can separate them into three classes. The first class would be the one where the energies of all three excited states come close to the ground-state energy and the success probability clusters at roughly $0.3$.
The second class of problem instances has a spectrum similar to the one shown in Fig.~\ref{fig:spec_all}(a). Only the energy of the first excited state comes close to the ground-state energy. The energies of the second and third excited states are much higher. For this class, the success probability clusters around $0.5$. The third class of problems has a spectrum similar to the one shown in Fig.~\ref{fig:spec_all}(b). The energies of the first and second excited states come close to the ground-state energy. For this class, the success probability depends on the particular problem instance. The reason is that for degenerate ground states, the probabilities to find these states are not necessarily equal~\cite{Matsuda09} and for problem instances which are close to these degenerate cases with unequal probabilities, this imbalance may have an influence when non-adiabatic transitions occur. All instances which show larger deviations between the success probabilities obtained from the qubit model and the flux model simulations (see Fig.~\ref{fig:gap_probs}) belong to the third class. A possible explanation for these deviations might be that, in some cases, due to the presence of the additional states in the flux model, these unequal probabilities in the computational subspace are different during the evolution than in the qubit model. We leave a more detailed study of this phenomenon for future work.

As noted in Ref.~\cite{albash15,dwave_manual} and confirmed by our analytical calculation in Sec.~\ref{sec:mapping}, there is some crosstalk between the qubits (last term in Eq.~(\ref{eq:H_2l_model})), leading to small offsets in the parameters $h_i$. Furthermore, a dependence of the annealing scheme on the parameter $\varphi_{\mathrm{J},0}^x$ was found, also leading to small discrepancies between the ideal qubit representation and the full system. Additionally, for the mapping of $J$ to $\varphi_{\mathrm{J},0}^x$, we had to draw on an approximate analytical calculation which may be another source for the small differences between the results obtained from the two models. Nevertheless, the results fit very well. Interestingly, the coupler element, which can be viewed as part of the environment and might be the source of additional noise, does not cause significant deviations in the results compared to the results of the qubit description.

\subsection{Comparison to D-Wave 2000Q data}\label{sec:compare_d_wave}
Because we find good agreement between the system described by the Hamiltonian Eq.~(\ref{eq:Htotal}) and the qubit model, we compare the success probability for both these systems with the percentage of successful runs on the D-Wave 2000Q quantum annealer.
In Appendix~\ref{app:data} (see Table~\ref{tab:cases}), we present the data obtained by at least ten repetitions of annealing simultaneously, 992 (976) copies of the two-qubit problems distributed over the D-Wave DW\_2000Q\_2 (DW\_2000Q\_2\_1) chip, for an annealing time $t_a=20\,\mu\mathrm{s}$. Postprocessing and autoscaling have been turned off for all experiments on the D-Wave 2000Q.

We find that although the annealing time on the D-Wave is much larger than for our simulations ($20\mu$s instead of $5$ns), a large fraction of the D-Wave data seems to agree (approximately) with the corresponding success probabilities obtained from the simulation of the SQUID model Eq.~(\ref{eq:Htotal}) and its two-level approximation.
This agreement is probably accidental. The annealing time of $5\,$ns was chosen to keep the real time to solve the TDSE of the SQUID model within acceptable limits as well as having some variation in the success probability at the end of the annealing process without having to use too-small values for, or differences between, the parameters $J$ and $h_i$.
In spite of the large difference in annealing times, the good agreement suggests that in the D-Wave device there are physical processes at work that affect the annealing, processes which are not incorporated in the SQUID model Eq.~(\ref{eq:Htotal}) or the corresponding qubit model Eq.~(\ref{eq:H_2l_model}).

Concrete evidence for the relevance of such processes is shown in Table~\ref{tab:d_wave_data}, where we present D-Wave data for three different cases whose energy spectra are shown in Figs.~\ref{fig:spec_all}(a)--(c). Because the spectra of these cases differ significantly, we assume that they are a representative subset of the cases studied previously. Data characterizing the problem instances such as the minimal energy gap $\delta E$ and the problem gap $\Delta_p$ of the final Hamiltonian are listed in Table~\ref{tab:d_wave_data} as well as the frequency of runs finding the ground states ($\ket{\downarrow\uparrow},\ket{\uparrow\downarrow},$ and $\ket{\uparrow\downarrow}$ for the three cases, respectively) on the D-Wave machine for four different annealing times. The results reported in Table~\ref{tab:d_wave_data} were obtained by putting 992 copies of the two-qubit problems on the Chimera graph and performing 1000 annealing runs.
\begin{table}[tb]
  \centering
  \caption{Percentage for finding the ground state (GS) on D-Wave's DW\_2000Q\_2 chip for three different problem instances and four different annealing times. The minimal and final gap (in GHz) are denoted by $\delta E$ and $\Delta_p$, respectively.}  \label{tab:d_wave_data}
    \begin{tabular}{c c c c c c c c c c}
     \multirow{2}{*}{Case} &\multirow{2}{*}{$J$} & \multirow{2}{*}{$h_1$} & \multirow{2}{*}{$h_2$} & \multirow{2}{*}{$\delta E$} & \multirow{2}{*}{$\Delta_p$} & \multicolumn{4}{c}{GS probability in \%}\\
      & & & & & & $1\,\mu$s & $20\,\mu$s & $100\,\mu$s & $1\,\mathrm{ms}$ \\
      \hline\hline
(a) & $-1$  & $0$    & $0.05$ & $1.206$ & $3.519$ & $63.0$ & $65.6$ & $67.1 $ & $69.7$\\
(b) & $-1$  & $0.96$ & $0.94$ & $0.627$ & $1.407$ & $51.4$ & $52.9$ & $53.6 $ & $55.6$ \\
(c) & $0.1$ & $0.3$  & $-0.3$ & $5.481$ & $14.07$ & $92.5$ & $96.2$ & $97.6 $ & $98.5$
      \end{tabular}
\end{table}

Table~\ref{tab:d_wave_data} shows some additional interesting facts.
First, recall that for the shortest annealing time considered, i.e., $1\,\mu\mathrm{s}$, simulation of the quantum annealing process in the qubit description yields the ground state with probability one for the three cases listed.
Clearly, as Table~\ref{tab:d_wave_data} shows, this is not the case for the D-Wave data, not even if we increase the annealing time to $1\,\mathrm{ms}$, as is most evident for case (c).
We emphasize that these differences in the observed frequencies for finding the ground state are not due to poor statistics nor can they be attributed to the differences in the minimal gaps $\delta E$.
Correlating these observations with the energy level spectra shown in Fig.~\ref{fig:spec_all} suggests that the observed reduction of the frequency for finding the ground state may be related to the distribution of energy levels for $s\rightarrow1$. However, the mechanism that causes the observed change of frequencies when going from case (a) to (c) cannot be found within the description of the quantum dynamics of the two-qubit system.
Explaining these experimental observations requires considering additional physical processes.

The first process that comes to mind is the interaction of the qubits with their environment at non-zero temperature.
In the next section, we address this issue by solving the TDSE of the two-qubit model Eq.~(\ref{eq:H_2l_model}) coupled to a bath of two-level systems, complementing previous work that investigated the effects of finite temperature on quantum annealing~\cite{amin08,johansson09,amin09,amin09_decoherence,amin09_nonmarkovian,dickson13,amin15}.

\section{System coupled to a bath}\label{section2}

The annealing process of the isolated two-qubit system may be understood in terms of the adiabatic theorem.
However, in the real world, the system modeling the two qubits is in contact with an environment at finite temperature.
In this section we scrutinize the extent to which the coupling to a heat bath, i.e., the presence of thermal
fluctuations, affects the annealing process.
This, we hope, may shed light on the annealing behavior observed on the D-Wave machine in the cases studied.

We do not assume the heat bath to be Markovian but instead we solve the TDSE of the system comprising the two-qubit system and the bath.
In order to be able to perform such simulations, it is necessary
to keep these models simple.
Therefore, it would be remarkable to find good quantitative agreement between the results
of the simulations and those obtained with the D-Wave machine.
Thus, the goal here is limited to find out if such models can reproduce, qualitatively, the trends
observed in the D-Wave data.

We model the heat bath as a collection of two-level systems~\cite{Phillips1972,anderson72}
which might represent e.g., defects in the material.
Such models have been used to discuss noise and dephasing in superconducting resonators and circuits~\cite{shnirman05,mueller09,cole10,burnett14,faoro15,Lisenfeld15,deGraaf18,mueller19}.
We assume that this heat bath is at thermal equilibrium, with a temperature
given by the operating temperature of the D-Wave machine. We denote the inverse of this operating temperature by $\beta^*=0.588\,\mathrm{ns}$ (in units of $\hbar=k_B=1$), corresponding to a temperature of $T\approx 13\,\mathrm{mK}$.

The Hamiltonian of the system (S) + bath (B) reads
\begin{eqnarray}
\mathcal{H}(t)&=& H_\mathrm{S}(s=t/t_a) + H_\mathrm{B} + \lambda H_\mathrm{SB}
,
\label{s40}
\end{eqnarray}
where $\lambda$ controls the overall strength of the system-bath interaction. The time evolution during the quantum annealing process of the closed quantum system defined by the Hamiltonian Eq.~(\ref{s40}) is obtained by solving the TDSE given in Eq.~(\ref{eq:tdse})
with the initial state
\begin{eqnarray}
|\Psi(t=0)\rangle=\ket{++}\otimes|\Phi(\beta)\rangle.
\end{eqnarray}
The method to prepare the thermal state $|\Phi(\beta)\rangle$ and other technical details are discussed in Appendix~\ref{app:technical_description}.

The system Hamiltonian is given by
\begin{align}
H_\mathrm{S}&=A(s) \left(-\sigma^x_{1}-\sigma^x_{2}\right) + B(s)\left( -J\sigma^z_{1}\sigma^z_{2} - h^z_1\sigma^z_{1} - h^z_2\sigma^z_{2}\right)
\label{s41}
\end{align}
and changes with time as a function of the annealing variable $s=t/t_a$.
We consider two extreme cases for $H_\mathrm{B}$ and $H_\mathrm{SB}$.

\subsection{Model I}
In the first case, the bath is modeled as a ring of two-level systems represented by the Pauli matrices
$\bm\mu_{n}=(\mu^x_{n},\mu^y_{n},\mu^z_{n})$, described by the Hamiltonian 
\begin{align}
H_\mathrm{B}&= -K\sum_{n=1}^{N_{\mathrm{B}}}\left(
r^x_n\mu^x_{n}\mu^x_{n+1}
+r^y_n\mu^y_{n}\mu^y_{n+1}
+r^z_n\mu^z_{n}\mu^z_{n+1}
\right)
.\label{eq:H_bath}
\end{align}
Here and in the following $N_{\mathrm{B}}$ denotes the number of bath particles.
The couplings $r^x_n$, $r^y_n$, and $r^z_n$ are taken to be uniform random numbers in the range $[-1,+1]$
and $K$ determines the spectral range of $H_\mathrm{B}$.
For random couplings it is unlikely that the model Eq.~(\ref{eq:H_bath}) is integrable (in the Bethe-ansatz sense)
or has any other special features such as conserved magnetization.
The bath Hamiltonian Eq.~(\ref{eq:H_bath}) with random couplings has the property that
the distribution of nearest-neighbor energy levels is Wigner-Dyson-like~\cite{ZHAO16}.
Extensive simulation work on spin baths with very different degrees
of connectivity~\cite{JIN13a}
suggests that as long as there is randomness in the system-bath coupling and
randomness in the intrabath coupling, the simple model Eq.~(\ref{eq:H_bath})
may be considered as a generic spin bath~\cite{ZHAO16}.
The Hamiltonian describing the interaction of the two-qubit system with the bath
is taken to be
\begin{align}
H_\mathrm{SB}&=
-r^x_{n,1}\mu^x_{n}\sigma^x_{1}
-r^y_{n,1}\mu^y_{n}\sigma^y_{1}
-r^z_{n,1}\mu^z_{n}\sigma^z_{1}\nonumber\\
&-r^x_{m,2}\mu^x_{m}\sigma^x_{2}
-r^y_{m,2}\mu^y_{m}\sigma^y_{2}
-r^z_{m,2}\mu^z_{m}\sigma^z_{2}
,
\label{s43b}
\end{align}
where $n$ and $m$ are chosen randomly from the set $\{1,\ldots,N_{\mathrm{B}}\}$ such that $n\not=m$.
The $r^\alpha_{n,1}$ and $r^\alpha_{m,2}$ are real-value random numbers in the range $[-1,+1]$.

\subsection{Model II}
In this case, the bath is modeled as a collection of non-interacting two-level systems given by the Hamiltonian
\begin{align}
   H_\mathrm{B} = -\Omega \sum\limits_{n=1}^{N_\mathrm{B}} r_n^x\mu_n^x +r_n^y\mu_n^y + r_n^z\mu_n^z,\label{eq:H_bathII}
\end{align}
where the parameter $\Omega$, together with the random numbers $r_n^x,\,r_n^y,\,r_n^z\in [-1,1]$, determines the level splitting of each two-level system.
The interaction between the qubits and the two-level systems of the bath is chosen such that each qubit interacts with a different half of the bath. This type of interaction is modeled by the Hamiltonian
\begin{align}
   H_\mathrm{SB} =\!\! \sum\limits_{\alpha=x,y,z}\left[ \sigma_1^\alpha\sum\limits_{n=1}^{N_{\mathrm{B}}/2}r_{n,1}^\alpha\mu_n^\alpha + \sigma_2^\alpha\,\sum\limits_{\mathclap{n=N_\mathrm{B}/2+1}}^{N_\mathrm{B}}\,r_{n,2}^\alpha\mu_n^\alpha \right].
\end{align}

\subsection{Parameters}
Obviously, to compare the simulation results with D-Wave results it is necessary to solve the TDSE for the same time interval as used on the D-Wave machine.
In practice, this requirement puts a severe constraint on the size of the problems for which one can solve the TDSE within a reasonable amount of real time.
Simulating the annealing process over $1\,\mu\mathrm{s}$ (the shortest annealing time possible on the D-Wave machine) for a system comprising $N_B=16$ on a BullSequana X1000 supercomputer (JUWELS~\cite{JUWELS}) takes about 4 hours using 40 compute cores.
Performing a simulation for $N_B=28$ two-level systems in the bath takes about one week (400000 time steps of $0.0025\,\mathrm{ns}$ using 5120 compute cores).
Earlier work which studied the TDSE dynamics of two spins coupled to a spin bath~\cite{RAED17b} shows that the results for $N_B=16$ and $N_B=28$ primarily differ in the statistical fluctuations on the data (see also Appendix~\ref{app:technical_description}).
Therefore, we have chosen to perform most simulations with $N_B=16$, repeating runs with different random numbers to collect statistics, and use a few runs with $N_B=28$ as an additional check on the data.

\begin{figure}[bt]
   \centering
   \includegraphics[width=0.35\textwidth]{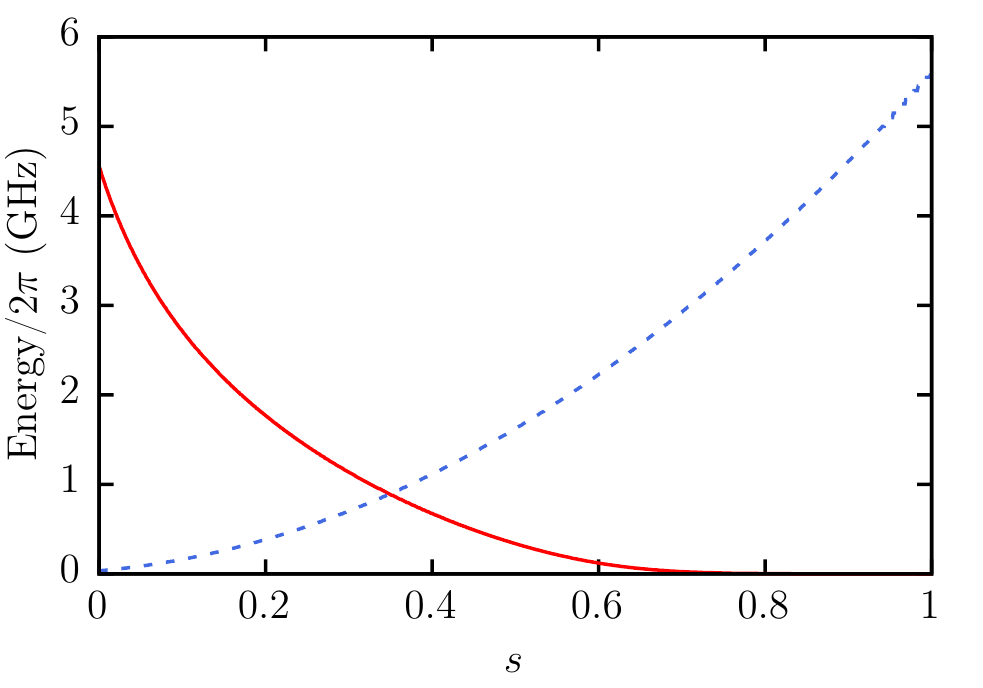}
   \caption{(color online) Annealing schedule of the DW\_2000Q\_2 chip which was used for the simulation with the environment. The red solid line corresponds to $A(s)$ and the blue dashed line corresponds to $B(s)$.}
   \label{fig:DW_schedule}
\end{figure}

We use the annealing schedule of the DW\_2000Q\_2 chip (see Fig.~\ref{fig:DW_schedule}) which is the machine that we used for our experiments.

In the case of model I, the initial state of the bath is prepared by projection with the operator $\exp(-\beta H_B)=\exp(-\beta K (H_B/K))$, as explained in Appendix~\ref{app:technical_description}. From Eqs.~(\ref{B2}) and (\ref{B3}) it is clear that baths with the same $\beta K$ (and the same $r_n^x$, $r_n^y$, and $r_n^z$; see Eq.~(\ref{eq:H_bath})) will have the same thermal equilibrium properties.
Therefore, we may use $K$ as an adjustable parameter to ``calibrate'' the temperature of the bath with respect to the operating temperature of the D-Wave machine on which we performed our experiments.
For instance, for a fixed choice of $r_n^x$, $r_n^y$ and $r_n^z$, baths with $( K=5\,\mathrm{GHz},\,\beta=0.2\,\mathrm{ns}\,(T\approx 38.2\,\mathrm{mK}))$ and $(K=5/3\,\mathrm{GHz},\,\beta=0.6\,\mathrm{ns}\,(T\approx 12.7\,\mathrm{mK}) )$ have the same thermal properties.
On the other hand, $K$ sets the time scale of the dynamics of the two-level systems of the bath. Simulations with $K$ in the range $[1\,\mathrm{GHz},5\,\mathrm{GHz}]$ (data not shown) reveal that the primary quantity of interest, the success probability of the two-qubit system at $s=1$, does not change significantly with $K$ (in the mentioned range and for the same value of $\lambda$). This leaves only the system-bath interaction $\lambda$ as a parameter to fit the simulation data to the D-Wave data.

In the case of model II, $\Omega$ plays the role of $K$ in model I, i.e., $\beta\Omega$ determines the thermal equilibrium properties of the bath. Note that for modest values of $N_B$, model II is too simple to act as a genuine heat bath, but as a model for a few defects interacting with the SQUIDs, it can be a realistic choice~\cite{shnirman05,mueller09,cole10}. Therefore, in this case, we set $\beta\approx\beta^*$ and use $\Omega$ and $\lambda$ as fitting parameters.

\subsection{Comparison to D-Wave 2000Q data}
\begin{figure*}[bt]
   \centering
   \includegraphics[width=0.95\textwidth]{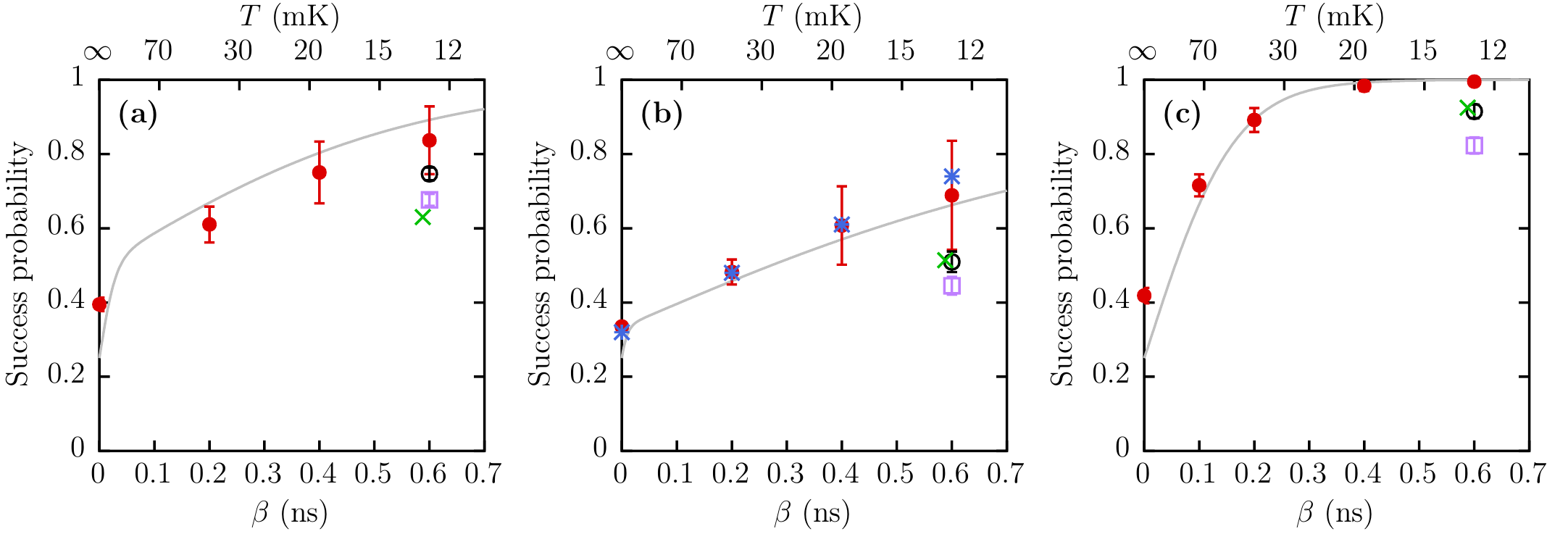}
   \caption{(color online) Success probability for three different problem instances (a) $J=-1$, $h_1=0$, and $h_2=0.05$; (b) $J=-1$, $h_1=0.96$, and $h_2=0.94$; and (c) $J=0.1$, $h_1=0.3$, and $h_2=-0.3$ as a function of the inverse temperature $\beta$ with annealing time $t_a=1\mu\mathrm{s}$. Data points are averages over ten simulation runs with different initializations of the heat bath described by model I (coupled two-level systems) with the parameters $K=5\,\mathrm{GHz}$, $\lambda=0.8\,\mathrm{GHz}$ and $N_B=16$ (red closed circles) and described by model II (independent two-level systems) for $\Omega=0.125\,\mathrm{GHz}$ and $\lambda=1\,\mathrm{GHz}$ (black open circle) and $\lambda=1.5\,\mathrm{GHz}$ (purple open square). Error bars indicate the standard deviation. Results for $N_B=28$ (blue asterisks) are for the same parameters as the red closed circles but for a single sample. The gray solid line shows the success probability in thermal equilibrium as a function of the inverse temperature $\beta$. The green cross represents the D-Wave result with annealing time $t_a=1\mu\mathrm{s}$ (see Table~\ref{tab:d_wave_data}).}
   \label{fig:bath_all}
\end{figure*}
Figures~\ref{fig:bath_all}(a)--(c) depict the results of the simulation with the heat bath (closed circles) averaged over ten different random initializations of the heat bath with $N_B=16$. Results for $N_B=28$ (asterisks) show that for each value of $\beta$, the averages of 10 samples of $N_\mathrm{B}=16$ data are in good agreement with the data obtained from one $N_\mathrm{B}=28$ sample.
The solid line indicates the probability $p_0$ of finding the ground state of the isolated qubit system in thermal equilibrium for $H_\mathrm{S}$ at $s=1$, i.e., $p_0=\exp(-\beta E_0)/Z$, where $E_0$ is the ground-state energy and $Z=\mathrm{Tr}(\exp(-\beta H_S))$ is the partition function. Qualitatively, the simulation data obtained using model I (closed circles) nicely match the equilibrium line. The deviations from the equilibrium line may be due to the freeze-out where thermal transitions stop~\cite{amin15} and/or too-short annealing times and/or the magnetic Foehn effect~\cite{saito01}. However, the simulation data do not match the data generated on the D-Wave machine (crosses). Assuming that the qubit system on the D-Wave machine is in thermal equilibrium, we would infer from Figs.~\ref{fig:bath_all}(a)--(c) that $\beta\approx 0.2\,\mathrm{ns}$ (corresponding to $T\approx 38.2\,$mK), which is about a factor of 3 smaller than the inverse operational temperature of about $\beta^*=0.588\,\mathrm{ns}$ (corresponding to $T\approx 13\,$mK).

For model II, we have searched the parameter space $1/8\,\mathrm{GHz}\le \Omega\le 2\,\mathrm{GHz}$ and $1/2\,\mathrm{GHz}\le\lambda\le 2\,\mathrm{GHz}$ for sets of ($\Omega,\lambda$) which would fit the D-Wave results best. These data are shown in Figs.~\ref{fig:bath_all}(a)--(c) as open circles ($\Omega=0.125\,\mathrm{GHz},\lambda=1\,\mathrm{GHz}$) and open squares ($\Omega=0.125\,\mathrm{GHz},\lambda=1.5\,\mathrm{GHz}$). For cases (b) and (c) (see Table~\ref{tab:d_wave_data}), the former fit remarkably well to the D-Wave data. However, we have not found a set ($\Omega,\lambda$) which fits all D-Wave data very well.

\section{Summary}\label{sec:summary}
We simulated the full system of three SQUIDs resembling two qubits and one tunable coupler element as used in the quantum annealer built by D-Wave Systems Inc.\ and found that the higher energy levels as well as the presence of the coupler element have observable effects on the annealing process which however do not have a strong influence on the final success probability compared to the ideal qubit model. In contrast to the investigation of the influence of the higher levels and resonators present in current systems for gate-based quantum computing~\cite{willsch17}, we found that, apart from a few exceptions, the effects are small for the cases of quantum annealing examined.

Furthermore, we investigated in detail the derivation of the qubit representation to obtain expressions for $\varphi_i^x$ and $J$ that led to satisfying results (see Figs.~\ref{fig:J}--\ref{fig:gap_probs}). The simulation results are in good agreement with effects encountered in this derivation such as the change in the annealing scheme depending on the choice of $\varphi_{\mathrm{J},0}^x$, which is also supported by findings in experiments~\cite{harris09}.

Simulation data for the SQUID model Eq.~(\ref{eq:Htotal}) and the corresponding two-level approximation Eq.~(\ref{eq:H_2l_model}) for an annealing time $t_a=5\,\mathrm{ns}$ show remarkably good agreement with the D-Wave data obtained with an annealing time $t_a=20\,\mu\mathrm{s}$, also in those cases for which the success probability is substantially less than one.
Although this agreement might be accidental, it suggests that the dynamics of the D-Wave quantum annealer are more complicated than what can be described by a closed-system model of the SQUIDs including higher levels and the tunable coupler.

Therefore, as a first step, we have studied quantum annealing in the presence of a heat bath. We solve the TDSE of the two-qubit system (Eq.~(\ref{eq:H_2l_model})) plus bath (Eq.~(\ref{eq:H_bath})) for three cases with qualitatively different energy spectra of the two-qubit system (see Fig.~\ref{fig:spec_all}).
Comparing D-Wave and simulation results for an annealing time of $1\,\mu\mathrm{s}$, we found that the simulation data for the success probabilities of the two-qubit systems quite nicely agree with the corresponding thermal equilibrium values but also that these probabilities are significantly larger than those obtained with the D-Wave annealer.
We have not found a common set of parameters $(\beta,K,\lambda)$ of the two-qubit-bath model that reproduces the D-Wave results for the three different cases considered.

Modeling the environment as a collection of non-interacting two-level defects (see Eq.~(\ref{eq:H_bathII})) was found to yield a much more appropriate description of the D-Wave data.
Although we could readily find values of the bath parameters $\Omega$ and $\lambda$ for which the solution of the TDSE yields results that are in excellent agreement with D-Wave data for two of the three different cases considered, we could not find a similar level of agreement with these data for all three cases if we impose the elementary requirement that the bath parameters $\Omega$ and $\lambda$ do not depend on the $J$ and $h$ that define the problem Hamiltonian.

Unlike in the case of a time-independent problem, the exchange of energy between the two qubits in the time-dependent (annealing) field and the bath of two-level systems seems to be an intricate process which, according to our simulation data, depends on the model parameters in a complicated manner. We leave a detailed study of this challenging problem for future research.

\begin{acknowledgments}
We would like to thank Mauricio Reis of D-Wave Systems Inc.\ for providing us with technical information and Seiji Miyashita for helpful discussions.

Access and compute time on the D-Wave machine located at the headquarters of D-Wave Systems Inc.~in Burnaby (Canada) were provided by D-Wave Systems Inc.

The authors gratefully acknowledge the Gauss Centre for Supercomputing e.V.\ for funding this project by providing computing time on the GCS Supercomputer JUWELS at J\"ulich Supercomputing Centre (JSC) and the computing time granted through JARA on the supercomputer JURECA at Forschungszentrum J\"ulich.

D.W.~was supported by the Initiative and Networking Fund of the Helmholtz Association through the Strategic Future Field of Research project ``Scalable solid state quantum computing'' (ZT-0013).
\end{acknowledgments}

\appendix

\begin{table}[tbp]
  \centering
  \scriptsize
  \caption{\footnotesize Parameter values of the problems shown in Fig.~\ref{fig:gap_probs}. (a) Success probability for the qubit model ($t_a=5$ns). (b) Success probability for the full model ($t_a=5$ns). (c) Percentage of successful runs on D-Wave's DW\_2000Q\_2 and DW\_2000Q\_2\_1 chips ($t_a=20\mu$s).}  \label{tab:cases}
  \resizebox{0.88\columnwidth}{!}{
    \begin{tabular}{l l l c c c c}
      \multirow{2}{*}{$h_1$} & \multirow{2}{*}{$h_2$} & \multirow{2}{*}{$J$} &Minimal gap &  \multicolumn{3}{c}{Success probability in \%}\\
      & & & $\delta E$ in GHz & (a) & (b) & (c) \\
      \hline
      \hline
$0.2$ & $0.2$ & $0.2$        & $7.958914$ & $99.9$ & $99.5$ & $99.7 $ \\
$0.2$ & $-0.2$ & $0$         & $6.524809$ & $99.8$ & $99.6$ & $95.8 $ \\
$0.3$ & $-0.3$ & $0.1$       & $6.509859$ & $99.9$ & $99.6$ & $96.2 $ \\
$0.1$ & $-0.1$ & $-0.1$      & $4.817172$ & $96.5$ & $96.3$ & $94.0 $ \\
$0.9$ & $0.7$ & $-1$         & $4.660374$ & $96.3$ & $96.2$ & $93.9 $ \\
$0.99$ & $-0.8$ & $1$        & $4.367788$ & \phantom{0}$95.8$\phantom{0} & \phantom{0}$95.2$\phantom{0} & \phantom{0}$94.1$\phantom{0}\\
$0.1$ & $0.1$ & $0$          & $3.750846$ & $93.0$ & $92.8$ & $80.5 $ \\
$0.3$ & $0.3$ & $-0.2$       & $3.740396$ & $96.2$ & $96.2$ & $81.4 $ \\
$0.07$ & $-0.07$ & $0$       & $2.786031$ & $82.1$ & $81.6$ & $66.8 $ \\
$0.07$ & $0.07$ & $0$        & $2.786031$ & $82.1$ & $82.1$ & $66.9 $ \\
$0.9$ & $-0.8$ & $1$         & $2.766581$ & $83.8$ & $83.9$ & $80.0$\\
$0.02$ & $0.08$ & $0.05$     & $2.542547$ & $77.7$ & $77.6$ & $71.2 $ \\
$0.99$ & $-1$ & $0.94$       & $2.134413$ & $84.8$ & $57.9$ & $48.3 $ \\
$0.05$ & $0.05$ & $0$        & $2.092326$ & $69.3$ & $69.4$ & $56.8 $ \\
\hline
$0$ & $-0.05$ & $0.05$       & $1.585987$ & $60.3$ & $60.3$ & $55.6 $ \\
$0$ & $0.05$ & $1$           & $1.433807$ & $69.0$ & $67.6$ & $64.4 $ \\
$0$ & $0.05$ & $-1$          & $1.433807$ & $69.0$ & $69.1$ & $65.3 $ \\
$0.01$ & $0.04$ & $0.025$    & $1.419405$ & $55.6$ & $55.7$ & $50.4$ \\
$0.99$ & $-1$ & $0.96$       & $1.366784$ & $74.7$ & $42.3$ & $38.1 $ \\
$0.99$ & $1$ & $-0.96$       & $1.366784$ & $74.7$ & $74.4$ & $46.9 $ \\
$0.02$ & $-0.02$ & $-0.02$   & $1.305954$ & $51.1$ & $50.6$ & $43.8 $ \\
$0.02$ & $0.02$ & $0.02$     & $1.305954$ & $51.1$ & $51.2$ & $43.9 $ \\
$0.95$ & $-0.99$ & $0.98$    & $1.145772$ & $47.5$ & $63.4$ & $55.7 $ \\
$0.95$ & $0.99$ & $-0.98$    & $1.145772$ & $47.5$ & $47.3$ & $52.3 $ \\
$0.99$ & $0.96$ & $-1$       & $1.018001$ & $48.4$ & $48.0$ & $54.8 $ \\
$0.02$ & $-0.02$ & $0$       & $0.939407$ & $42.7$ & $42.2$ & $36.2 $ \\
$0$ & $0.03$ & $1$           & $0.917871$ & $61.8$ & $60.8$ & $59.0 $ \\
$0.96$ & $-0.94$ & $1$       & $0.742309$ & $50.7$ & $57.5$ & $55.1 $ \\
$0.96$ & $0.94$ & $-1$       & $0.742309$ & $50.7$ & $50.0$ & $54.0 $ \\
$0.98$ & $-0.96$ & $1$       & $0.740480$ & $45.4$ & $56.3$ & $53.4 $ \\
$0.98$ & $0.96$ & $-1$       & $0.740480$ & $45.4$ & $44.7$ & $51.0 $ \\
$0.01$ & $-0.01$ & $-0.01$   & $0.716800$ & $37.8$ & $37.3$ & $33.2 $ \\
$0.01$ & $0.01$ & $0.01$     & $0.716800$ & $37.8$ & $38.0$ & $33.9 $ \\
$0$ & $0.02$ & $-1$          & $0.640932$ & $57.9$ & $57.8$ & $57.0 $ \\
$0.01$ & $0.009$ & $0.002$   & $0.543047$ & $33.9$ & $34.2$ &  $31.4$\\
$0.99$ & $-1$ & $0.98$       & $0.510495$ & $61.1$ & $27.0$ & $27.5 $ \\
$0.01$ & $0.01$ & $0$        & $0.505201$ & $33.5$ & $33.8$ & $31.7 $ \\
$0.99$ & $-0.98$ & $1$       & $0.406202$ & $35.3$ & $49.3$ & $47.2 $ \\
$0.99$ & $0.98$ & $-1$       & $0.406202$ & $35.3$ & $34.6$ & $45.6 $ \\
$0$ & $-0.01$ & $0.01$       & $0.392828$ & $33.0$ & $33.2$ & $31.6 $ \\
$0.005$ & $0.005$ & $0.005$  & $0.388173$ & $31.2$ & $31.5$ &  $29.8$\\
$0.005$ & $-0.005$ & $-0.005$& $0.388173$ & $31.2$ & $30.7$ &  $29.1$\\
$0$ & $0.01$ & $1$           & $0.343987$ & $54.0$ & $53.4$ & $52.7 $ \\
$0$ & $0.01$ & $-1$          & $0.343987$ & $54.0$ & $53.8$ & $53.1 $ \\
$0.007$ & $0$ & $-0.01$      & $0.302380$ & $31.8$ & $31.3$ &  $30.4$\\
$0.007$ & $0$ & $0.01$       & $0.302380$ & $31.8$ & $32.0$ &  $30.5$\\
$1$ & $0$ & $-0.005$         & $0.269309$ & $54.0$ & $53.1$ & $52.8$\\
$1$ & $0$ & $0.005$          & $0.269309$ & $54.0$ & $54.6$ &  $52.8$\\
$0.005$ & $0.001$ & $0.01$   & $0.268263$ & $31.4$ & $31.6$ &  $30.2$\\
$0.005$ & $-0.001$ & $-0.01$ & $0.268263$ & $31.4$ & $30.9$ &  $30.0$\\
$0.005$ & $0$ & $0.01$       & $0.228548$ & $30.9$ & $31.2$ &  $29.6$\\
$0$ & $0.005$ & $0.5$        & $0.193353$ & $52.0$ & $51.8$ &  $51.3$\\
$0$ & $0.005$ & $-0.5$       & $0.193353$ & $52.0$ & $51.9$ &  $52.3$\\
$0.005$ & $-0.001$ & $0.01$  & $0.187420$ & $30.5$ & $30.7$ &  $29.3$\\
$0.005$ & $0.001$ & $-0.01$  & $0.187420$ & $30.5$ & $30.0$ &  $28.4$\\
$0$ & $0.005$ & $1$          & $0.183074$ & $52.0$ & $51.5$ &  $50.7$\\
$0.005$ & $0$ & $-1$         & $0.183074$ & $52.0$ & $51.8$ &  $50.7$\\
$0.003$ & $0$ & $0.01$       & $0.145233$ & $30.1$ & $30.3$ &  $29.2$\\
$0$ & $0.003$ & $-1$         & $0.114519$ & $51.2$ & $51.0$ &  $51.0$\\
$0$ & $0.003$ & $1$          & $0.114519$ & $51.2$ & $50.8$ &  $50.1$
      \end{tabular}
    }
\end{table}

\section{Data}\label{app:data}
Table \ref{tab:cases} contains a list with the parameters $h_i$ and $J$, the minimal energy gap $\delta E$, and the success probabilities for the problems used to generate Fig.\ \ref{fig:gap_probs}.
\section{Numerical Solution of the TDSE}\label{app:technical_description}
The numerical solution of the TDSE for a pure state of $N_\mathrm{B}+2$ two-level systems requires
computational resources (memory and CPU time) proportional to $2^{N_\mathrm{B}+2}$.
For a brute force calculation of thermal expectation values
$\mathbf{Tr}( e^{-\beta \mathcal{H}}{\cal A}(t))/\mathbf{Tr} (e^{-\beta \mathcal{H}})$
this number changes to $2^{N_\mathrm{B}+2}\times 2^{N_\mathrm{B}+2}$.
Fortunately, this increase in cost can be avoided by making use of random-state technology,
reducing the cost to that of solving the TDSE for one pure state~\cite{HAMS00}.
If $|\Phi\rangle$ is a pure state, picked uniformly from the $D=2^{{N_\mathrm{B}+2}}$-dimensional unit hypersphere,
one can show that for Hermitian matrices $X$~\cite{HAMS00},
\begin{equation}
\mathbf{Tr} (X) \approx D\langle\Phi|X|\Phi\rangle
,
\label{s46b}
\end{equation}
and that the statistical errors resulting from approximating
$\mathbf{Tr} (X) $ by $D\langle\Phi| X |\Phi\rangle$ are small if $D$ is large~\cite{HAMS00}.
For large baths, this property of the random pure state renders the problem amenable to numerical simulation.

In the case at hand, we proceed as follows.
First, we generate a thermal random state of the bath system, meaning that
we compute the pure state
\begin{eqnarray}
|\Phi(\beta)\rangle&=& \frac{e^{-\beta H_\mathrm{B}/2}|\Phi\rangle}{ \langle\Phi|e^{-\beta H_\mathrm{B}}|\Phi\rangle^{1/2} }
,
\label{B2}
\end{eqnarray}
where $\beta$ denotes the inverse temperature.
For any bath observable ${\cal A}(t)$ we have~\cite{HAMS00}
\begin{equation}
\langle {\cal A}(t)\rangle=\frac{\mathbf{Tr} (e^{-\beta H_\mathrm{B}}{\cal A}(t))}{\mathbf{Tr} (e^{-\beta H_\mathrm{B}})}
\approx \langle\Phi(\beta)| {\cal A}(t)|\Phi(\beta)\rangle
.
\label{B3}
\end{equation}
The initial state of the whole system is then a product state of the ground state of the two qubits at
$s=0$ and the thermal pure state $|\Phi(\beta)\rangle$, i.e.,
\begin{eqnarray}
|\Psi(t=0)\rangle=\ket{++}\otimes|\Phi(\beta)\rangle
.
\label{s4k1}
\end{eqnarray}

Since the Hamiltonian Eq.~(\ref{s40}) depends explicitly on time,
we can only solve Eq.~(\ref{eq:tdse}) numerically by time stepping.
For this purpose, we use a Suzuki-Trotter product-formula-based algorithm~\cite{RAED06}.
This algorithm employs a decomposition in terms of unitary matrices and is unconditionally stable
by construction.
All our simulations for the two qubits coupled to a heat bath were carried out
with the massively parallel quantum-spin dynamics simulator (in-house software), which is
based on the same computational kernel as the massively parallel quantum computer simulator~\cite{deraedt18}.

\bibliographystyle{apsrev4-1}
\bibliography{../bibliography.bib}

\begin{thebibliography}{104}%
\makeatletter
\providecommand \@ifxundefined [1]{%
 \@ifx{#1\undefined}
}%
\providecommand \@ifnum [1]{%
 \ifnum #1\expandafter \@firstoftwo
 \else \expandafter \@secondoftwo
 \fi
}%
\providecommand \@ifx [1]{%
 \ifx #1\expandafter \@firstoftwo
 \else \expandafter \@secondoftwo
 \fi
}%
\providecommand \natexlab [1]{#1}%
\providecommand \enquote  [1]{``#1''}%
\providecommand \bibnamefont  [1]{#1}%
\providecommand \bibfnamefont [1]{#1}%
\providecommand \citenamefont [1]{#1}%
\providecommand \href@noop [0]{\@secondoftwo}%
\providecommand \href [0]{\begingroup \@sanitize@url \@href}%
\providecommand \@href[1]{\@@startlink{#1}\@@href}%
\providecommand \@@href[1]{\endgroup#1\@@endlink}%
\providecommand \@sanitize@url [0]{\catcode `\\12\catcode `\$12\catcode
  `\&12\catcode `\#12\catcode `\^12\catcode `\_12\catcode `\%12\relax}%
\providecommand \@@startlink[1]{}%
\providecommand \@@endlink[0]{}%
\providecommand \url  [0]{\begingroup\@sanitize@url \@url }%
\providecommand \@url [1]{\endgroup\@href {#1}{\urlprefix }}%
\providecommand \urlprefix  [0]{URL }%
\providecommand \Eprint [0]{\href }%
\providecommand \doibase [0]{http://dx.doi.org/}%
\providecommand \selectlanguage [0]{\@gobble}%
\providecommand \bibinfo  [0]{\@secondoftwo}%
\providecommand \bibfield  [0]{\@secondoftwo}%
\providecommand \translation [1]{[#1]}%
\providecommand \BibitemOpen [0]{}%
\providecommand \bibitemStop [0]{}%
\providecommand \bibitemNoStop [0]{.\EOS\space}%
\providecommand \EOS [0]{\spacefactor3000\relax}%
\providecommand \BibitemShut  [1]{\csname bibitem#1\endcsname}%
\let\auto@bib@innerbib\@empty
\bibitem [{\citenamefont {Nielsen}\ and\ \citenamefont
  {Chuang}(2010)}]{nielsen_chuang}%
  \BibitemOpen
  \bibfield  {author} {\bibinfo {author} {\bibfnamefont {M.~A.}\ \bibnamefont
  {Nielsen}}\ and\ \bibinfo {author} {\bibfnamefont {I.~L.}\ \bibnamefont
  {Chuang}},\ }\href {\doibase 10.1017/CB09780511976667} {\emph {\bibinfo
  {title} {Quantum Computation and Quantum Information}}}\ (\bibinfo
  {publisher} {Cambridge University Press},\ \bibinfo {year}
  {2010})\BibitemShut {NoStop}%
\bibitem [{\citenamefont {Cirac}\ and\ \citenamefont {Zoller}(1995)}]{cirac95}%
  \BibitemOpen
  \bibfield  {author} {\bibinfo {author} {\bibfnamefont {J.~I.}\ \bibnamefont
  {Cirac}}\ and\ \bibinfo {author} {\bibfnamefont {P.}~\bibnamefont {Zoller}},\
  }\href {\doibase 10.1103/PhysRevLett.74.4091} {\bibfield  {journal} {\bibinfo
   {journal} {Phys. Rev. Lett.}\ }\textbf {\bibinfo {volume} {74}},\ \bibinfo
  {pages} {4091} (\bibinfo {year} {1995})}\BibitemShut {NoStop}%
\bibitem [{\citenamefont {Monroe}\ \emph {et~al.}(1995)\citenamefont {Monroe},
  \citenamefont {Meekhof}, \citenamefont {King}, \citenamefont {Itano},\ and\
  \citenamefont {Wineland}}]{monroe95}%
  \BibitemOpen
  \bibfield  {author} {\bibinfo {author} {\bibfnamefont {C.}~\bibnamefont
  {Monroe}}, \bibinfo {author} {\bibfnamefont {D.~M.}\ \bibnamefont {Meekhof}},
  \bibinfo {author} {\bibfnamefont {B.~E.}\ \bibnamefont {King}}, \bibinfo
  {author} {\bibfnamefont {W.~M.}\ \bibnamefont {Itano}}, \ and\ \bibinfo
  {author} {\bibfnamefont {D.~J.}\ \bibnamefont {Wineland}},\ }\href {\doibase
  10.1103/PhysRevLett.75.4714} {\bibfield  {journal} {\bibinfo  {journal}
  {Phys. Rev. Lett.}\ }\textbf {\bibinfo {volume} {75}},\ \bibinfo {pages}
  {4714} (\bibinfo {year} {1995})}\BibitemShut {NoStop}%
\bibitem [{\citenamefont {H\"affner}\ \emph {et~al.}(2003)\citenamefont
  {H\"affner}, \citenamefont {Gulde}, \citenamefont {Riebe}, \citenamefont
  {Lancaster}, \citenamefont {Becher}, \citenamefont {Eschner}, \citenamefont
  {Schmidt-Kaler},\ and\ \citenamefont {Blatt}}]{haeffner03}%
  \BibitemOpen
  \bibfield  {author} {\bibinfo {author} {\bibfnamefont {H.}~\bibnamefont
  {H\"affner}}, \bibinfo {author} {\bibfnamefont {S.}~\bibnamefont {Gulde}},
  \bibinfo {author} {\bibfnamefont {M.}~\bibnamefont {Riebe}}, \bibinfo
  {author} {\bibfnamefont {G.}~\bibnamefont {Lancaster}}, \bibinfo {author}
  {\bibfnamefont {C.}~\bibnamefont {Becher}}, \bibinfo {author} {\bibfnamefont
  {J.}~\bibnamefont {Eschner}}, \bibinfo {author} {\bibfnamefont
  {F.}~\bibnamefont {Schmidt-Kaler}}, \ and\ \bibinfo {author} {\bibfnamefont
  {R.}~\bibnamefont {Blatt}},\ }\href {\doibase 10.1103/PhysRevLett.90.143602}
  {\bibfield  {journal} {\bibinfo  {journal} {Phys. Rev. Lett.}\ }\textbf
  {\bibinfo {volume} {90}},\ \bibinfo {pages} {143602} (\bibinfo {year}
  {2003})}\BibitemShut {NoStop}%
\bibitem [{\citenamefont {Hanneke}\ \emph {et~al.}(2009)\citenamefont
  {Hanneke}, \citenamefont {Home}, \citenamefont {Jost}, \citenamefont {Amini},
  \citenamefont {Leibfried},\ and\ \citenamefont {Wineland}}]{Hanneke2009}%
  \BibitemOpen
  \bibfield  {author} {\bibinfo {author} {\bibfnamefont {D.}~\bibnamefont
  {Hanneke}}, \bibinfo {author} {\bibfnamefont {J.~P.}\ \bibnamefont {Home}},
  \bibinfo {author} {\bibfnamefont {J.~D.}\ \bibnamefont {Jost}}, \bibinfo
  {author} {\bibfnamefont {J.~M.}\ \bibnamefont {Amini}}, \bibinfo {author}
  {\bibfnamefont {D.}~\bibnamefont {Leibfried}}, \ and\ \bibinfo {author}
  {\bibfnamefont {D.~J.}\ \bibnamefont {Wineland}},\ }\href {\doibase
  10.1038/nphys1453} {\bibfield  {journal} {\bibinfo  {journal} {Nat. Phys.}\
  }\textbf {\bibinfo {volume} {6}},\ \bibinfo {pages} {13} (\bibinfo {year}
  {2009})}\BibitemShut {NoStop}%
\bibitem [{\citenamefont {Schindler}\ \emph {et~al.}(2013)\citenamefont
  {Schindler}, \citenamefont {Nigg}, \citenamefont {Monz}, \citenamefont
  {Barreiro}, \citenamefont {Martinez}, \citenamefont {Wang}, \citenamefont
  {Quint}, \citenamefont {Brandl}, \citenamefont {Nebendahl}, \citenamefont
  {Roos}, \citenamefont {Chwalla}, \citenamefont {Hennrich},\ and\
  \citenamefont {Blatt}}]{Schindler2013}%
  \BibitemOpen
  \bibfield  {author} {\bibinfo {author} {\bibfnamefont {P.}~\bibnamefont
  {Schindler}}, \bibinfo {author} {\bibfnamefont {D.}~\bibnamefont {Nigg}},
  \bibinfo {author} {\bibfnamefont {T.}~\bibnamefont {Monz}}, \bibinfo {author}
  {\bibfnamefont {J.~T.}\ \bibnamefont {Barreiro}}, \bibinfo {author}
  {\bibfnamefont {E.}~\bibnamefont {Martinez}}, \bibinfo {author}
  {\bibfnamefont {S.~X.}\ \bibnamefont {Wang}}, \bibinfo {author}
  {\bibfnamefont {S.}~\bibnamefont {Quint}}, \bibinfo {author} {\bibfnamefont
  {M.~F.}\ \bibnamefont {Brandl}}, \bibinfo {author} {\bibfnamefont
  {V.}~\bibnamefont {Nebendahl}}, \bibinfo {author} {\bibfnamefont {C.~F.}\
  \bibnamefont {Roos}}, \bibinfo {author} {\bibfnamefont {M.}~\bibnamefont
  {Chwalla}}, \bibinfo {author} {\bibfnamefont {M.}~\bibnamefont {Hennrich}}, \
  and\ \bibinfo {author} {\bibfnamefont {R.}~\bibnamefont {Blatt}},\ }\href
  {\doibase 10.1088/1367-2630/15/12/123012} {\bibfield  {journal} {\bibinfo
  {journal} {New J. Phys.}\ }\textbf {\bibinfo {volume} {15}},\ \bibinfo
  {pages} {123012} (\bibinfo {year} {2013})}\BibitemShut {NoStop}%
\bibitem [{\citenamefont {Ballance}\ \emph {et~al.}(2016)\citenamefont
  {Ballance}, \citenamefont {Harty}, \citenamefont {Linke}, \citenamefont
  {Sepiol},\ and\ \citenamefont {Lucas}}]{ballance16}%
  \BibitemOpen
  \bibfield  {author} {\bibinfo {author} {\bibfnamefont {C.~J.}\ \bibnamefont
  {Ballance}}, \bibinfo {author} {\bibfnamefont {T.~P.}\ \bibnamefont {Harty}},
  \bibinfo {author} {\bibfnamefont {N.~M.}\ \bibnamefont {Linke}}, \bibinfo
  {author} {\bibfnamefont {M.~A.}\ \bibnamefont {Sepiol}}, \ and\ \bibinfo
  {author} {\bibfnamefont {D.~M.}\ \bibnamefont {Lucas}},\ }\href {\doibase
  10.1103/PhysRevLett.117.060504} {\bibfield  {journal} {\bibinfo  {journal}
  {Phys. Rev. Lett.}\ }\textbf {\bibinfo {volume} {117}},\ \bibinfo {pages}
  {060504} (\bibinfo {year} {2016})}\BibitemShut {NoStop}%
\bibitem [{\citenamefont {Loss}\ and\ \citenamefont
  {DiVincenzo}(1998)}]{loss98}%
  \BibitemOpen
  \bibfield  {author} {\bibinfo {author} {\bibfnamefont {D.}~\bibnamefont
  {Loss}}\ and\ \bibinfo {author} {\bibfnamefont {D.~P.}\ \bibnamefont
  {DiVincenzo}},\ }\href {\doibase 10.1103/PhysRevA.57.120} {\bibfield
  {journal} {\bibinfo  {journal} {Phys. Rev. A}\ }\textbf {\bibinfo {volume}
  {57}},\ \bibinfo {pages} {120} (\bibinfo {year} {1998})}\BibitemShut
  {NoStop}%
\bibitem [{\citenamefont {Levy}(2002)}]{levy02}%
  \BibitemOpen
  \bibfield  {author} {\bibinfo {author} {\bibfnamefont {J.}~\bibnamefont
  {Levy}},\ }\href {\doibase 10.1103/PhysRevLett.89.147902} {\bibfield
  {journal} {\bibinfo  {journal} {Phys. Rev. Lett.}\ }\textbf {\bibinfo
  {volume} {89}},\ \bibinfo {pages} {147902} (\bibinfo {year}
  {2002})}\BibitemShut {NoStop}%
\bibitem [{\citenamefont {Wendin}(2017)}]{wendin2017}%
  \BibitemOpen
  \bibfield  {author} {\bibinfo {author} {\bibfnamefont {G.}~\bibnamefont
  {Wendin}},\ }\href {\doibase 10.1088/1361-6633/aa7e1a} {\bibfield  {journal}
  {\bibinfo  {journal} {Rep. Prog. Phys.}\ }\textbf {\bibinfo {volume} {80}},\
  \bibinfo {pages} {106001} (\bibinfo {year} {2017})}\BibitemShut {NoStop}%
\bibitem [{\citenamefont {Krantz}\ \emph {et~al.}(2019)\citenamefont {Krantz},
  \citenamefont {Kjaergaard}, \citenamefont {Yan}, \citenamefont {Orlando},
  \citenamefont {Gustavsson},\ and\ \citenamefont {Oliver}}]{krantz19}%
  \BibitemOpen
  \bibfield  {author} {\bibinfo {author} {\bibfnamefont {P.}~\bibnamefont
  {Krantz}}, \bibinfo {author} {\bibfnamefont {M.}~\bibnamefont {Kjaergaard}},
  \bibinfo {author} {\bibfnamefont {F.}~\bibnamefont {Yan}}, \bibinfo {author}
  {\bibfnamefont {T.~P.}\ \bibnamefont {Orlando}}, \bibinfo {author}
  {\bibfnamefont {S.}~\bibnamefont {Gustavsson}}, \ and\ \bibinfo {author}
  {\bibfnamefont {W.~D.}\ \bibnamefont {Oliver}},\ }\href {\doibase
  10.1063/1.5089550} {\bibfield  {journal} {\bibinfo  {journal} {Appl. Phys.
  Rev.}\ }\textbf {\bibinfo {volume} {6}},\ \bibinfo {pages} {021318} (\bibinfo
  {year} {2019})}\BibitemShut {NoStop}%
\bibitem [{\citenamefont {Wu}\ and\ \citenamefont {Lidar}(2002)}]{wu02}%
  \BibitemOpen
  \bibfield  {author} {\bibinfo {author} {\bibfnamefont {L.-A.}\ \bibnamefont
  {Wu}}\ and\ \bibinfo {author} {\bibfnamefont {D.~A.}\ \bibnamefont {Lidar}},\
  }\href {\doibase 10.1063/1.1499208} {\bibfield  {journal} {\bibinfo
  {journal} {J. Math. Phys.}\ }\textbf {\bibinfo {volume} {43}},\ \bibinfo
  {pages} {4506} (\bibinfo {year} {2002})}\BibitemShut {NoStop}%
\bibitem [{\citenamefont {Leibfried}\ \emph {et~al.}(2003)\citenamefont
  {Leibfried}, \citenamefont {Blatt}, \citenamefont {Monroe},\ and\
  \citenamefont {Wineland}}]{leibfried03}%
  \BibitemOpen
  \bibfield  {author} {\bibinfo {author} {\bibfnamefont {D.}~\bibnamefont
  {Leibfried}}, \bibinfo {author} {\bibfnamefont {R.}~\bibnamefont {Blatt}},
  \bibinfo {author} {\bibfnamefont {C.}~\bibnamefont {Monroe}}, \ and\ \bibinfo
  {author} {\bibfnamefont {D.}~\bibnamefont {Wineland}},\ }\href {\doibase
  10.1103/RevModPhys.75.281} {\bibfield  {journal} {\bibinfo  {journal} {Rev.
  Mod. Phys.}\ }\textbf {\bibinfo {volume} {75}},\ \bibinfo {pages} {281}
  (\bibinfo {year} {2003})}\BibitemShut {NoStop}%
\bibitem [{\citenamefont {Petta}\ \emph {et~al.}(2005)\citenamefont {Petta},
  \citenamefont {Johnson}, \citenamefont {Taylor}, \citenamefont {Laird},
  \citenamefont {Yacoby}, \citenamefont {Lukin}, \citenamefont {Marcus},
  \citenamefont {Hanson},\ and\ \citenamefont {Gossard}}]{petta05}%
  \BibitemOpen
  \bibfield  {author} {\bibinfo {author} {\bibfnamefont {J.~R.}\ \bibnamefont
  {Petta}}, \bibinfo {author} {\bibfnamefont {A.~C.}\ \bibnamefont {Johnson}},
  \bibinfo {author} {\bibfnamefont {J.~M.}\ \bibnamefont {Taylor}}, \bibinfo
  {author} {\bibfnamefont {E.~A.}\ \bibnamefont {Laird}}, \bibinfo {author}
  {\bibfnamefont {A.}~\bibnamefont {Yacoby}}, \bibinfo {author} {\bibfnamefont
  {M.~D.}\ \bibnamefont {Lukin}}, \bibinfo {author} {\bibfnamefont {C.~M.}\
  \bibnamefont {Marcus}}, \bibinfo {author} {\bibfnamefont {M.~P.}\
  \bibnamefont {Hanson}}, \ and\ \bibinfo {author} {\bibfnamefont {A.~C.}\
  \bibnamefont {Gossard}},\ }\href {\doibase 10.1126/science.1116955}
  {\bibfield  {journal} {\bibinfo  {journal} {Science}\ }\textbf {\bibinfo
  {volume} {309}},\ \bibinfo {pages} {2180} (\bibinfo {year}
  {2005})}\BibitemShut {NoStop}%
\bibitem [{\citenamefont {DiVincenzo}\ \emph {et~al.}(2000)\citenamefont
  {DiVincenzo}, \citenamefont {Bacon}, \citenamefont {Kempe}, \citenamefont
  {Burkard},\ and\ \citenamefont {Whaley}}]{divincenzo00}%
  \BibitemOpen
  \bibfield  {author} {\bibinfo {author} {\bibfnamefont {D.~P.}\ \bibnamefont
  {DiVincenzo}}, \bibinfo {author} {\bibfnamefont {D.}~\bibnamefont {Bacon}},
  \bibinfo {author} {\bibfnamefont {J.}~\bibnamefont {Kempe}}, \bibinfo
  {author} {\bibfnamefont {G.}~\bibnamefont {Burkard}}, \ and\ \bibinfo
  {author} {\bibfnamefont {K.~B.}\ \bibnamefont {Whaley}},\ }\href {\doibase
  10.1038/35042541} {\bibfield  {journal} {\bibinfo  {journal} {Nature}\
  }\textbf {\bibinfo {volume} {408}},\ \bibinfo {pages} {339} (\bibinfo {year}
  {2000})}\BibitemShut {NoStop}%
\bibitem [{\citenamefont {Gaudreau}\ \emph {et~al.}(2011)\citenamefont
  {Gaudreau}, \citenamefont {Granger}, \citenamefont {Kam}, \citenamefont
  {Aers}, \citenamefont {Studenikin}, \citenamefont {Zawadzki}, \citenamefont
  {Pioro-Ladri\`ere}, \citenamefont {Wasilewski},\ and\ \citenamefont
  {Sachrajda}}]{gaudreau11}%
  \BibitemOpen
  \bibfield  {author} {\bibinfo {author} {\bibfnamefont {L.}~\bibnamefont
  {Gaudreau}}, \bibinfo {author} {\bibfnamefont {G.}~\bibnamefont {Granger}},
  \bibinfo {author} {\bibfnamefont {A.}~\bibnamefont {Kam}}, \bibinfo {author}
  {\bibfnamefont {G.~C.}\ \bibnamefont {Aers}}, \bibinfo {author}
  {\bibfnamefont {S.~A.}\ \bibnamefont {Studenikin}}, \bibinfo {author}
  {\bibfnamefont {P.}~\bibnamefont {Zawadzki}}, \bibinfo {author}
  {\bibfnamefont {M.}~\bibnamefont {Pioro-Ladri\`ere}}, \bibinfo {author}
  {\bibfnamefont {Z.~R.}\ \bibnamefont {Wasilewski}}, \ and\ \bibinfo {author}
  {\bibfnamefont {A.~S.}\ \bibnamefont {Sachrajda}},\ }\href {\doibase
  10.1038/nphys2149} {\bibfield  {journal} {\bibinfo  {journal} {Nat. Phys.}\
  }\textbf {\bibinfo {volume} {8}},\ \bibinfo {pages} {54} (\bibinfo {year}
  {2011})}\BibitemShut {NoStop}%
\bibitem [{\citenamefont {Medford}\ \emph {et~al.}(2013)\citenamefont
  {Medford}, \citenamefont {Beil}, \citenamefont {Taylor}, \citenamefont
  {Bartlett}, \citenamefont {Doherty}, \citenamefont {Rashba}, \citenamefont
  {DiVincenzo}, \citenamefont {Lu}, \citenamefont {Gossard},\ and\
  \citenamefont {Marcus}}]{medford13}%
  \BibitemOpen
  \bibfield  {author} {\bibinfo {author} {\bibfnamefont {J.}~\bibnamefont
  {Medford}}, \bibinfo {author} {\bibfnamefont {J.}~\bibnamefont {Beil}},
  \bibinfo {author} {\bibfnamefont {J.~M.}\ \bibnamefont {Taylor}}, \bibinfo
  {author} {\bibfnamefont {S.~D.}\ \bibnamefont {Bartlett}}, \bibinfo {author}
  {\bibfnamefont {A.~C.}\ \bibnamefont {Doherty}}, \bibinfo {author}
  {\bibfnamefont {E.~I.}\ \bibnamefont {Rashba}}, \bibinfo {author}
  {\bibfnamefont {D.~P.}\ \bibnamefont {DiVincenzo}}, \bibinfo {author}
  {\bibfnamefont {H.}~\bibnamefont {Lu}}, \bibinfo {author} {\bibfnamefont
  {A.~C.}\ \bibnamefont {Gossard}}, \ and\ \bibinfo {author} {\bibfnamefont
  {C.~M.}\ \bibnamefont {Marcus}},\ }\href {\doibase nnano.2013.168} {\bibfield
   {journal} {\bibinfo  {journal} {Nat. Nanotechnol.}\ }\textbf {\bibinfo
  {volume} {8}},\ \bibinfo {pages} {654} (\bibinfo {year} {2013})}\BibitemShut
  {NoStop}%
\bibitem [{\citenamefont {Cerfontaine}\ \emph {et~al.}(2016)\citenamefont
  {Cerfontaine}, \citenamefont {Botzem}, \citenamefont {Humpohl}, \citenamefont
  {Schuh}, \citenamefont {Bougeard},\ and\ \citenamefont
  {Bluhm}}]{cerfontaine16}%
  \BibitemOpen
  \bibfield  {author} {\bibinfo {author} {\bibfnamefont {P.}~\bibnamefont
  {Cerfontaine}}, \bibinfo {author} {\bibfnamefont {T.}~\bibnamefont {Botzem}},
  \bibinfo {author} {\bibfnamefont {S.~S.}\ \bibnamefont {Humpohl}}, \bibinfo
  {author} {\bibfnamefont {D.}~\bibnamefont {Schuh}}, \bibinfo {author}
  {\bibfnamefont {D.}~\bibnamefont {Bougeard}}, \ and\ \bibinfo {author}
  {\bibfnamefont {H.}~\bibnamefont {Bluhm}},\ }\href@noop {} {\enquote
  {\bibinfo {title} {Feedback-tuned noise-resilient gates for encoded spin
  qubits},}\ } (\bibinfo {year} {2016}),\ \Eprint
  {http://arxiv.org/abs/arXiv:1606.01897} {arXiv:1606.01897} \BibitemShut
  {NoStop}%
\bibitem [{\citenamefont {Martinis}\ \emph {et~al.}(2002)\citenamefont
  {Martinis}, \citenamefont {Nam}, \citenamefont {Aumentado},\ and\
  \citenamefont {Urbina}}]{martinis02}%
  \BibitemOpen
  \bibfield  {author} {\bibinfo {author} {\bibfnamefont {J.~M.}\ \bibnamefont
  {Martinis}}, \bibinfo {author} {\bibfnamefont {S.}~\bibnamefont {Nam}},
  \bibinfo {author} {\bibfnamefont {J.}~\bibnamefont {Aumentado}}, \ and\
  \bibinfo {author} {\bibfnamefont {C.}~\bibnamefont {Urbina}},\ }\href
  {\doibase 10.1103/PhysRevLett.89.117901} {\bibfield  {journal} {\bibinfo
  {journal} {Phys. Rev. Lett.}\ }\textbf {\bibinfo {volume} {89}},\ \bibinfo
  {pages} {117901} (\bibinfo {year} {2002})}\BibitemShut {NoStop}%
\bibitem [{\citenamefont {Steffen}\ \emph {et~al.}(2006)\citenamefont
  {Steffen}, \citenamefont {Ansmann}, \citenamefont {McDermott}, \citenamefont
  {Katz}, \citenamefont {Bialczak}, \citenamefont {Lucero}, \citenamefont
  {Neeley}, \citenamefont {Weig}, \citenamefont {Cleland},\ and\ \citenamefont
  {Martinis}}]{steffen06}%
  \BibitemOpen
  \bibfield  {author} {\bibinfo {author} {\bibfnamefont {M.}~\bibnamefont
  {Steffen}}, \bibinfo {author} {\bibfnamefont {M.}~\bibnamefont {Ansmann}},
  \bibinfo {author} {\bibfnamefont {R.}~\bibnamefont {McDermott}}, \bibinfo
  {author} {\bibfnamefont {N.}~\bibnamefont {Katz}}, \bibinfo {author}
  {\bibfnamefont {R.~C.}\ \bibnamefont {Bialczak}}, \bibinfo {author}
  {\bibfnamefont {E.}~\bibnamefont {Lucero}}, \bibinfo {author} {\bibfnamefont
  {M.}~\bibnamefont {Neeley}}, \bibinfo {author} {\bibfnamefont {E.~M.}\
  \bibnamefont {Weig}}, \bibinfo {author} {\bibfnamefont {A.~N.}\ \bibnamefont
  {Cleland}}, \ and\ \bibinfo {author} {\bibfnamefont {J.~M.}\ \bibnamefont
  {Martinis}},\ }\href {\doibase 10.1103/PhysRevLett.97.050502} {\bibfield
  {journal} {\bibinfo  {journal} {Phys. Rev. Lett.}\ }\textbf {\bibinfo
  {volume} {97}},\ \bibinfo {pages} {050502} (\bibinfo {year}
  {2006})}\BibitemShut {NoStop}%
\bibitem [{\citenamefont {Koch}\ \emph {et~al.}(2007)\citenamefont {Koch},
  \citenamefont {Yu}, \citenamefont {Gambetta}, \citenamefont {Houck},
  \citenamefont {Schuster}, \citenamefont {Majer}, \citenamefont {Blais},
  \citenamefont {Devoret}, \citenamefont {Girvin},\ and\ \citenamefont
  {Schoelkopf}}]{koch2007}%
  \BibitemOpen
  \bibfield  {author} {\bibinfo {author} {\bibfnamefont {J.}~\bibnamefont
  {Koch}}, \bibinfo {author} {\bibfnamefont {T.~M.}\ \bibnamefont {Yu}},
  \bibinfo {author} {\bibfnamefont {J.}~\bibnamefont {Gambetta}}, \bibinfo
  {author} {\bibfnamefont {A.~A.}\ \bibnamefont {Houck}}, \bibinfo {author}
  {\bibfnamefont {D.~I.}\ \bibnamefont {Schuster}}, \bibinfo {author}
  {\bibfnamefont {J.}~\bibnamefont {Majer}}, \bibinfo {author} {\bibfnamefont
  {A.}~\bibnamefont {Blais}}, \bibinfo {author} {\bibfnamefont {M.~H.}\
  \bibnamefont {Devoret}}, \bibinfo {author} {\bibfnamefont {S.~M.}\
  \bibnamefont {Girvin}}, \ and\ \bibinfo {author} {\bibfnamefont {R.~J.}\
  \bibnamefont {Schoelkopf}},\ }\href {\doibase 10.1103/PhysRevA.76.042319}
  {\bibfield  {journal} {\bibinfo  {journal} {Phys. Rev. A}\ }\textbf {\bibinfo
  {volume} {76}},\ \bibinfo {pages} {042319} (\bibinfo {year}
  {2007})}\BibitemShut {NoStop}%
\bibitem [{\citenamefont {Shnirman}\ \emph {et~al.}(1997)\citenamefont
  {Shnirman}, \citenamefont {Sch\"on},\ and\ \citenamefont
  {Hermon}}]{shnirman97}%
  \BibitemOpen
  \bibfield  {author} {\bibinfo {author} {\bibfnamefont {A.}~\bibnamefont
  {Shnirman}}, \bibinfo {author} {\bibfnamefont {G.}~\bibnamefont {Sch\"on}}, \
  and\ \bibinfo {author} {\bibfnamefont {Z.}~\bibnamefont {Hermon}},\ }\href
  {\doibase 10.1103/PhysRevLett.79.2371} {\bibfield  {journal} {\bibinfo
  {journal} {Phys. Rev. Lett.}\ }\textbf {\bibinfo {volume} {79}},\ \bibinfo
  {pages} {2371} (\bibinfo {year} {1997})}\BibitemShut {NoStop}%
\bibitem [{\citenamefont {Bouchiat}\ \emph {et~al.}(1998)\citenamefont
  {Bouchiat}, \citenamefont {Vion}, \citenamefont {Joyez}, \citenamefont
  {Esteve},\ and\ \citenamefont {Devoret}}]{bouchiat1998}%
  \BibitemOpen
  \bibfield  {author} {\bibinfo {author} {\bibfnamefont {V.}~\bibnamefont
  {Bouchiat}}, \bibinfo {author} {\bibfnamefont {D.}~\bibnamefont {Vion}},
  \bibinfo {author} {\bibfnamefont {P.}~\bibnamefont {Joyez}}, \bibinfo
  {author} {\bibfnamefont {D.}~\bibnamefont {Esteve}}, \ and\ \bibinfo {author}
  {\bibfnamefont {M.~H.}\ \bibnamefont {Devoret}},\ }\href
  {https://iopscience.iop.org/article/10.1238/Physica.Topical.076a00165/meta}
  {\bibfield  {journal} {\bibinfo  {journal} {Phys. Scr.}\ }\textbf {\bibinfo
  {volume} {T76}},\ \bibinfo {pages} {165} (\bibinfo {year}
  {1998})}\BibitemShut {NoStop}%
\bibitem [{\citenamefont {Nakamura}\ \emph {et~al.}(1999)\citenamefont
  {Nakamura}, \citenamefont {Pashkin},\ and\ \citenamefont
  {Tsai}}]{Nakamura1999}%
  \BibitemOpen
  \bibfield  {author} {\bibinfo {author} {\bibfnamefont {Y.}~\bibnamefont
  {Nakamura}}, \bibinfo {author} {\bibfnamefont {Y.~A.}\ \bibnamefont
  {Pashkin}}, \ and\ \bibinfo {author} {\bibfnamefont {J.~S.}\ \bibnamefont
  {Tsai}},\ }\href {\doibase 10.1038/19718} {\bibfield  {journal} {\bibinfo
  {journal} {Nature}\ }\textbf {\bibinfo {volume} {398}},\ \bibinfo {pages}
  {786} (\bibinfo {year} {1999})}\BibitemShut {NoStop}%
\bibitem [{\citenamefont {Motzoi}\ \emph {et~al.}(2009)\citenamefont {Motzoi},
  \citenamefont {Gambetta}, \citenamefont {Rebentrost},\ and\ \citenamefont
  {Wilhelm}}]{motzoi09}%
  \BibitemOpen
  \bibfield  {author} {\bibinfo {author} {\bibfnamefont {F.}~\bibnamefont
  {Motzoi}}, \bibinfo {author} {\bibfnamefont {J.~M.}\ \bibnamefont
  {Gambetta}}, \bibinfo {author} {\bibfnamefont {P.}~\bibnamefont
  {Rebentrost}}, \ and\ \bibinfo {author} {\bibfnamefont {F.~K.}\ \bibnamefont
  {Wilhelm}},\ }\href {\doibase 10.1103/PhysRevLett.103.110501} {\bibfield
  {journal} {\bibinfo  {journal} {Phys. Rev. Lett.}\ }\textbf {\bibinfo
  {volume} {103}},\ \bibinfo {pages} {110501} (\bibinfo {year}
  {2009})}\BibitemShut {NoStop}%
\bibitem [{\citenamefont {Lucero}\ \emph {et~al.}(2010)\citenamefont {Lucero},
  \citenamefont {Kelly}, \citenamefont {Bialczak}, \citenamefont {Lenander},
  \citenamefont {Mariantoni}, \citenamefont {Neeley}, \citenamefont
  {O'Connell}, \citenamefont {Sank}, \citenamefont {Wang}, \citenamefont
  {Weides}, \citenamefont {Wenner}, \citenamefont {Yamamoto}, \citenamefont
  {Cleland},\ and\ \citenamefont {Martinis}}]{lucero10}%
  \BibitemOpen
  \bibfield  {author} {\bibinfo {author} {\bibfnamefont {E.}~\bibnamefont
  {Lucero}}, \bibinfo {author} {\bibfnamefont {J.}~\bibnamefont {Kelly}},
  \bibinfo {author} {\bibfnamefont {R.~C.}\ \bibnamefont {Bialczak}}, \bibinfo
  {author} {\bibfnamefont {M.}~\bibnamefont {Lenander}}, \bibinfo {author}
  {\bibfnamefont {M.}~\bibnamefont {Mariantoni}}, \bibinfo {author}
  {\bibfnamefont {M.}~\bibnamefont {Neeley}}, \bibinfo {author} {\bibfnamefont
  {A.~D.}\ \bibnamefont {O'Connell}}, \bibinfo {author} {\bibfnamefont
  {D.}~\bibnamefont {Sank}}, \bibinfo {author} {\bibfnamefont {H.}~\bibnamefont
  {Wang}}, \bibinfo {author} {\bibfnamefont {M.}~\bibnamefont {Weides}},
  \bibinfo {author} {\bibfnamefont {J.}~\bibnamefont {Wenner}}, \bibinfo
  {author} {\bibfnamefont {T.}~\bibnamefont {Yamamoto}}, \bibinfo {author}
  {\bibfnamefont {A.~N.}\ \bibnamefont {Cleland}}, \ and\ \bibinfo {author}
  {\bibfnamefont {J.~M.}\ \bibnamefont {Martinis}},\ }\href {\doibase
  10.1103/PhysRevA.82.042339} {\bibfield  {journal} {\bibinfo  {journal} {Phys.
  Rev. A}\ }\textbf {\bibinfo {volume} {82}},\ \bibinfo {pages} {042339}
  (\bibinfo {year} {2010})}\BibitemShut {NoStop}%
\bibitem [{\citenamefont {Gambetta}\ \emph {et~al.}(2011)\citenamefont
  {Gambetta}, \citenamefont {Motzoi}, \citenamefont {Merkel},\ and\
  \citenamefont {Wilhelm}}]{gambetta11}%
  \BibitemOpen
  \bibfield  {author} {\bibinfo {author} {\bibfnamefont {J.~M.}\ \bibnamefont
  {Gambetta}}, \bibinfo {author} {\bibfnamefont {F.}~\bibnamefont {Motzoi}},
  \bibinfo {author} {\bibfnamefont {S.~T.}\ \bibnamefont {Merkel}}, \ and\
  \bibinfo {author} {\bibfnamefont {F.~K.}\ \bibnamefont {Wilhelm}},\ }\href
  {\doibase 10.1103/PhysRevA.83.012308} {\bibfield  {journal} {\bibinfo
  {journal} {Phys. Rev. A}\ }\textbf {\bibinfo {volume} {83}},\ \bibinfo
  {pages} {012308} (\bibinfo {year} {2011})}\BibitemShut {NoStop}%
\bibitem [{\citenamefont {Wallman}\ \emph {et~al.}(2016)\citenamefont
  {Wallman}, \citenamefont {Barnhill},\ and\ \citenamefont
  {Emerson}}]{Wallman2016}%
  \BibitemOpen
  \bibfield  {author} {\bibinfo {author} {\bibfnamefont {J.~J.}\ \bibnamefont
  {Wallman}}, \bibinfo {author} {\bibfnamefont {M.}~\bibnamefont {Barnhill}}, \
  and\ \bibinfo {author} {\bibfnamefont {J.}~\bibnamefont {Emerson}},\ }\href
  {\doibase 10.1088/1367-2630/18/4/043021} {\bibfield  {journal} {\bibinfo
  {journal} {New J. Phys.}\ }\textbf {\bibinfo {volume} {18}},\ \bibinfo
  {pages} {043021} (\bibinfo {year} {2016})}\BibitemShut {NoStop}%
\bibitem [{\citenamefont {Willsch}\ \emph {et~al.}(2017)\citenamefont
  {Willsch}, \citenamefont {Nocon}, \citenamefont {Jin}, \citenamefont {{De
  Raedt}},\ and\ \citenamefont {Michielsen}}]{willsch17}%
  \BibitemOpen
  \bibfield  {author} {\bibinfo {author} {\bibfnamefont {D.}~\bibnamefont
  {Willsch}}, \bibinfo {author} {\bibfnamefont {M.}~\bibnamefont {Nocon}},
  \bibinfo {author} {\bibfnamefont {F.}~\bibnamefont {Jin}}, \bibinfo {author}
  {\bibfnamefont {H.}~\bibnamefont {{De Raedt}}}, \ and\ \bibinfo {author}
  {\bibfnamefont {K.}~\bibnamefont {Michielsen}},\ }\href {\doibase
  10.1103/PhysRevA.96.062302} {\bibfield  {journal} {\bibinfo  {journal} {Phys.
  Rev. A}\ }\textbf {\bibinfo {volume} {96}},\ \bibinfo {pages} {062302}
  (\bibinfo {year} {2017})}\BibitemShut {NoStop}%
\bibitem [{\citenamefont {Wood}\ and\ \citenamefont {Gambetta}(2018)}]{wood18}%
  \BibitemOpen
  \bibfield  {author} {\bibinfo {author} {\bibfnamefont {C.~J.}\ \bibnamefont
  {Wood}}\ and\ \bibinfo {author} {\bibfnamefont {J.~M.}\ \bibnamefont
  {Gambetta}},\ }\href {\doibase 10.1103/PhysRevA.97.032306} {\bibfield
  {journal} {\bibinfo  {journal} {Phys. Rev. A}\ }\textbf {\bibinfo {volume}
  {97}},\ \bibinfo {pages} {032306} (\bibinfo {year} {2018})}\BibitemShut
  {NoStop}%
\bibitem [{\citenamefont {Chow}\ \emph {et~al.}(2010)\citenamefont {Chow},
  \citenamefont {DiCarlo}, \citenamefont {Gambetta}, \citenamefont {Motzoi},
  \citenamefont {Frunzio}, \citenamefont {Girvin},\ and\ \citenamefont
  {Schoelkopf}}]{chow10}%
  \BibitemOpen
  \bibfield  {author} {\bibinfo {author} {\bibfnamefont {J.~M.}\ \bibnamefont
  {Chow}}, \bibinfo {author} {\bibfnamefont {L.}~\bibnamefont {DiCarlo}},
  \bibinfo {author} {\bibfnamefont {J.~M.}\ \bibnamefont {Gambetta}}, \bibinfo
  {author} {\bibfnamefont {F.}~\bibnamefont {Motzoi}}, \bibinfo {author}
  {\bibfnamefont {L.}~\bibnamefont {Frunzio}}, \bibinfo {author} {\bibfnamefont
  {S.~M.}\ \bibnamefont {Girvin}}, \ and\ \bibinfo {author} {\bibfnamefont
  {R.~J.}\ \bibnamefont {Schoelkopf}},\ }\href {\doibase
  10.1103/PhysRevA.82.040305} {\bibfield  {journal} {\bibinfo  {journal} {Phys.
  Rev. A}\ }\textbf {\bibinfo {volume} {82}},\ \bibinfo {pages} {040305(R)}
  (\bibinfo {year} {2010})}\BibitemShut {NoStop}%
\bibitem [{\citenamefont {Chen}\ \emph {et~al.}(2016)\citenamefont {Chen},
  \citenamefont {Kelly}, \citenamefont {Quintana}, \citenamefont {Barends},
  \citenamefont {Campbell}, \citenamefont {Chen}, \citenamefont {Chiaro},
  \citenamefont {Dunsworth}, \citenamefont {Fowler}, \citenamefont {Lucero},
  \citenamefont {Jeffrey}, \citenamefont {Megrant}, \citenamefont {Mutus},
  \citenamefont {Neeley}, \citenamefont {Neill}, \citenamefont {O'Malley},
  \citenamefont {Roushan}, \citenamefont {Sank}, \citenamefont {Vainsencher},
  \citenamefont {Wenner}, \citenamefont {White}, \citenamefont {Korotkov},\
  and\ \citenamefont {Martinis}}]{chen16}%
  \BibitemOpen
  \bibfield  {author} {\bibinfo {author} {\bibfnamefont {Z.}~\bibnamefont
  {Chen}}, \bibinfo {author} {\bibfnamefont {J.}~\bibnamefont {Kelly}},
  \bibinfo {author} {\bibfnamefont {C.}~\bibnamefont {Quintana}}, \bibinfo
  {author} {\bibfnamefont {R.}~\bibnamefont {Barends}}, \bibinfo {author}
  {\bibfnamefont {B.}~\bibnamefont {Campbell}}, \bibinfo {author}
  {\bibfnamefont {Y.}~\bibnamefont {Chen}}, \bibinfo {author} {\bibfnamefont
  {B.}~\bibnamefont {Chiaro}}, \bibinfo {author} {\bibfnamefont
  {A.}~\bibnamefont {Dunsworth}}, \bibinfo {author} {\bibfnamefont {A.~G.}\
  \bibnamefont {Fowler}}, \bibinfo {author} {\bibfnamefont {E.}~\bibnamefont
  {Lucero}}, \bibinfo {author} {\bibfnamefont {E.}~\bibnamefont {Jeffrey}},
  \bibinfo {author} {\bibfnamefont {A.}~\bibnamefont {Megrant}}, \bibinfo
  {author} {\bibfnamefont {J.}~\bibnamefont {Mutus}}, \bibinfo {author}
  {\bibfnamefont {M.}~\bibnamefont {Neeley}}, \bibinfo {author} {\bibfnamefont
  {C.}~\bibnamefont {Neill}}, \bibinfo {author} {\bibfnamefont {P.~J.~J.}\
  \bibnamefont {O'Malley}}, \bibinfo {author} {\bibfnamefont {P.}~\bibnamefont
  {Roushan}}, \bibinfo {author} {\bibfnamefont {D.}~\bibnamefont {Sank}},
  \bibinfo {author} {\bibfnamefont {A.}~\bibnamefont {Vainsencher}}, \bibinfo
  {author} {\bibfnamefont {J.}~\bibnamefont {Wenner}}, \bibinfo {author}
  {\bibfnamefont {T.~C.}\ \bibnamefont {White}}, \bibinfo {author}
  {\bibfnamefont {A.~N.}\ \bibnamefont {Korotkov}}, \ and\ \bibinfo {author}
  {\bibfnamefont {J.~M.}\ \bibnamefont {Martinis}},\ }\href {\doibase
  10.1103/PhysRevLett.116.020501} {\bibfield  {journal} {\bibinfo  {journal}
  {Phys. Rev. Lett.}\ }\textbf {\bibinfo {volume} {116}},\ \bibinfo {pages}
  {020501} (\bibinfo {year} {2016})}\BibitemShut {NoStop}%
\bibitem [{\citenamefont {McKay}\ \emph {et~al.}(2017)\citenamefont {McKay},
  \citenamefont {Wood}, \citenamefont {Sheldon}, \citenamefont {Chow},\ and\
  \citenamefont {Gambetta}}]{mckay17}%
  \BibitemOpen
  \bibfield  {author} {\bibinfo {author} {\bibfnamefont {D.~C.}\ \bibnamefont
  {McKay}}, \bibinfo {author} {\bibfnamefont {C.~J.}\ \bibnamefont {Wood}},
  \bibinfo {author} {\bibfnamefont {S.}~\bibnamefont {Sheldon}}, \bibinfo
  {author} {\bibfnamefont {J.~M.}\ \bibnamefont {Chow}}, \ and\ \bibinfo
  {author} {\bibfnamefont {J.~M.}\ \bibnamefont {Gambetta}},\ }\href {\doibase
  10.1103/PhysRevA.96.022330} {\bibfield  {journal} {\bibinfo  {journal} {Phys.
  Rev. A}\ }\textbf {\bibinfo {volume} {96}},\ \bibinfo {pages} {022330}
  (\bibinfo {year} {2017})}\BibitemShut {NoStop}%
\bibitem [{\citenamefont {Poletto}\ \emph {et~al.}(2009)\citenamefont
  {Poletto}, \citenamefont {Chiarello}, \citenamefont {Castellano},
  \citenamefont {Lisenfeld}, \citenamefont {Lukashenko}, \citenamefont
  {Carelli},\ and\ \citenamefont {Ustinov}}]{Poletto2009}%
  \BibitemOpen
  \bibfield  {author} {\bibinfo {author} {\bibfnamefont {S.}~\bibnamefont
  {Poletto}}, \bibinfo {author} {\bibfnamefont {F.}~\bibnamefont {Chiarello}},
  \bibinfo {author} {\bibfnamefont {M.~G.}\ \bibnamefont {Castellano}},
  \bibinfo {author} {\bibfnamefont {J.}~\bibnamefont {Lisenfeld}}, \bibinfo
  {author} {\bibfnamefont {A.}~\bibnamefont {Lukashenko}}, \bibinfo {author}
  {\bibfnamefont {P.}~\bibnamefont {Carelli}}, \ and\ \bibinfo {author}
  {\bibfnamefont {A.~V.}\ \bibnamefont {Ustinov}},\ }\href {\doibase
  10.1088/0031-8949/2009/t137/014011} {\bibfield  {journal} {\bibinfo
  {journal} {Phys. Scr.}\ }\textbf {\bibinfo {volume} {T137}},\ \bibinfo
  {pages} {014011} (\bibinfo {year} {2009})}\BibitemShut {NoStop}%
\bibitem [{\citenamefont {Yoshihara}\ \emph {et~al.}(2014)\citenamefont
  {Yoshihara}, \citenamefont {Nakamura}, \citenamefont {Yan}, \citenamefont
  {Gustavsson}, \citenamefont {Bylander}, \citenamefont {Oliver},\ and\
  \citenamefont {Tsai}}]{yoshihara14}%
  \BibitemOpen
  \bibfield  {author} {\bibinfo {author} {\bibfnamefont {F.}~\bibnamefont
  {Yoshihara}}, \bibinfo {author} {\bibfnamefont {Y.}~\bibnamefont {Nakamura}},
  \bibinfo {author} {\bibfnamefont {F.}~\bibnamefont {Yan}}, \bibinfo {author}
  {\bibfnamefont {S.}~\bibnamefont {Gustavsson}}, \bibinfo {author}
  {\bibfnamefont {J.}~\bibnamefont {Bylander}}, \bibinfo {author}
  {\bibfnamefont {W.~D.}\ \bibnamefont {Oliver}}, \ and\ \bibinfo {author}
  {\bibfnamefont {J.-S.}\ \bibnamefont {Tsai}},\ }\href {\doibase
  10.1103/PhysRevB.89.020503} {\bibfield  {journal} {\bibinfo  {journal} {Phys.
  Rev. B}\ }\textbf {\bibinfo {volume} {89}},\ \bibinfo {pages} {020503(R)}
  (\bibinfo {year} {2014})}\BibitemShut {NoStop}%
\bibitem [{\citenamefont {Billangeon}\ \emph {et~al.}(2015)\citenamefont
  {Billangeon}, \citenamefont {Tsai},\ and\ \citenamefont
  {Nakamura}}]{billangeon15}%
  \BibitemOpen
  \bibfield  {author} {\bibinfo {author} {\bibfnamefont {P.-M.}\ \bibnamefont
  {Billangeon}}, \bibinfo {author} {\bibfnamefont {J.~S.}\ \bibnamefont
  {Tsai}}, \ and\ \bibinfo {author} {\bibfnamefont {Y.}~\bibnamefont
  {Nakamura}},\ }\href {\doibase 10.1103/PhysRevB.91.094517} {\bibfield
  {journal} {\bibinfo  {journal} {Phys. Rev. B}\ }\textbf {\bibinfo {volume}
  {91}},\ \bibinfo {pages} {094517} (\bibinfo {year} {2015})}\BibitemShut
  {NoStop}%
\bibitem [{\citenamefont {Ferr\'on}\ and\ \citenamefont
  {Dom\'{\i}nguez}(2010)}]{ferron10}%
  \BibitemOpen
  \bibfield  {author} {\bibinfo {author} {\bibfnamefont {A.}~\bibnamefont
  {Ferr\'on}}\ and\ \bibinfo {author} {\bibfnamefont {D.}~\bibnamefont
  {Dom\'{\i}nguez}},\ }\href {\doibase 10.1103/PhysRevB.81.104505} {\bibfield
  {journal} {\bibinfo  {journal} {Phys. Rev. B}\ }\textbf {\bibinfo {volume}
  {81}},\ \bibinfo {pages} {104505} (\bibinfo {year} {2010})}\BibitemShut
  {NoStop}%
\bibitem [{\citenamefont {Chiarello}(2000)}]{chiarello00}%
  \BibitemOpen
  \bibfield  {author} {\bibinfo {author} {\bibfnamefont {F.}~\bibnamefont
  {Chiarello}},\ }\href {\doibase 10.1016/S0375-9601(00)00714-3} {\bibfield
  {journal} {\bibinfo  {journal} {Phys. Lett. A}\ }\textbf {\bibinfo {volume}
  {277}},\ \bibinfo {pages} {189} (\bibinfo {year} {2000})}\BibitemShut
  {NoStop}%
\bibitem [{\citenamefont {Makhlin}\ \emph {et~al.}(2001)\citenamefont
  {Makhlin}, \citenamefont {Sch\"on},\ and\ \citenamefont
  {Shnirman}}]{Makhlin2001}%
  \BibitemOpen
  \bibfield  {author} {\bibinfo {author} {\bibfnamefont {Y.}~\bibnamefont
  {Makhlin}}, \bibinfo {author} {\bibfnamefont {G.}~\bibnamefont {Sch\"on}}, \
  and\ \bibinfo {author} {\bibfnamefont {A.}~\bibnamefont {Shnirman}},\ }\href
  {\doibase 10.1103/RevModPhys.73.357} {\bibfield  {journal} {\bibinfo
  {journal} {Rev. Mod. Phys.}\ }\textbf {\bibinfo {volume} {73}},\ \bibinfo
  {pages} {357} (\bibinfo {year} {2001})}\BibitemShut {NoStop}%
\bibitem [{\citenamefont {Harris}\ \emph
  {et~al.}(2009{\natexlab{a}})\citenamefont {Harris}, \citenamefont {Brito},
  \citenamefont {Berkley}, \citenamefont {Johansson}, \citenamefont {Johnson},
  \citenamefont {Lanting}, \citenamefont {Bunyk}, \citenamefont {Ladizinsky},
  \citenamefont {Bumble}, \citenamefont {Fung}, \citenamefont {Kaul},
  \citenamefont {Kleinsasser},\ and\ \citenamefont
  {Han}}]{harris09_newjournal}%
  \BibitemOpen
  \bibfield  {author} {\bibinfo {author} {\bibfnamefont {R.}~\bibnamefont
  {Harris}}, \bibinfo {author} {\bibfnamefont {F.}~\bibnamefont {Brito}},
  \bibinfo {author} {\bibfnamefont {A.~J.}\ \bibnamefont {Berkley}}, \bibinfo
  {author} {\bibfnamefont {J.}~\bibnamefont {Johansson}}, \bibinfo {author}
  {\bibfnamefont {M.~W.}\ \bibnamefont {Johnson}}, \bibinfo {author}
  {\bibfnamefont {T.}~\bibnamefont {Lanting}}, \bibinfo {author} {\bibfnamefont
  {P.}~\bibnamefont {Bunyk}}, \bibinfo {author} {\bibfnamefont
  {E.}~\bibnamefont {Ladizinsky}}, \bibinfo {author} {\bibfnamefont
  {B.}~\bibnamefont {Bumble}}, \bibinfo {author} {\bibfnamefont
  {A.}~\bibnamefont {Fung}}, \bibinfo {author} {\bibfnamefont {A.}~\bibnamefont
  {Kaul}}, \bibinfo {author} {\bibfnamefont {A.}~\bibnamefont {Kleinsasser}}, \
  and\ \bibinfo {author} {\bibfnamefont {S.}~\bibnamefont {Han}},\ }\href
  {http://stacks.iop.org/1367-2630/11/i=12/a=123022} {\bibfield  {journal}
  {\bibinfo  {journal} {New J. Phys.}\ }\textbf {\bibinfo {volume} {11}},\
  \bibinfo {pages} {123022} (\bibinfo {year} {2009}{\natexlab{a}})}\BibitemShut
  {NoStop}%
\bibitem [{\citenamefont {Harris}\ \emph
  {et~al.}(2010{\natexlab{a}})\citenamefont {Harris}, \citenamefont
  {Johansson}, \citenamefont {Berkley}, \citenamefont {Johnson}, \citenamefont
  {Lanting}, \citenamefont {Han}, \citenamefont {Bunyk}, \citenamefont
  {Ladizinsky}, \citenamefont {Oh}, \citenamefont {Perminov}, \citenamefont
  {Tolkacheva}, \citenamefont {Uchaikin}, \citenamefont {Chapple},
  \citenamefont {Enderud}, \citenamefont {Rich}, \citenamefont {Thom},
  \citenamefont {Wang}, \citenamefont {Wilson},\ and\ \citenamefont
  {Rose}}]{harris10}%
  \BibitemOpen
  \bibfield  {author} {\bibinfo {author} {\bibfnamefont {R.}~\bibnamefont
  {Harris}}, \bibinfo {author} {\bibfnamefont {J.}~\bibnamefont {Johansson}},
  \bibinfo {author} {\bibfnamefont {A.~J.}\ \bibnamefont {Berkley}}, \bibinfo
  {author} {\bibfnamefont {M.~W.}\ \bibnamefont {Johnson}}, \bibinfo {author}
  {\bibfnamefont {T.}~\bibnamefont {Lanting}}, \bibinfo {author} {\bibfnamefont
  {S.}~\bibnamefont {Han}}, \bibinfo {author} {\bibfnamefont {P.}~\bibnamefont
  {Bunyk}}, \bibinfo {author} {\bibfnamefont {E.}~\bibnamefont {Ladizinsky}},
  \bibinfo {author} {\bibfnamefont {T.}~\bibnamefont {Oh}}, \bibinfo {author}
  {\bibfnamefont {I.}~\bibnamefont {Perminov}}, \bibinfo {author}
  {\bibfnamefont {E.}~\bibnamefont {Tolkacheva}}, \bibinfo {author}
  {\bibfnamefont {S.}~\bibnamefont {Uchaikin}}, \bibinfo {author}
  {\bibfnamefont {E.~M.}\ \bibnamefont {Chapple}}, \bibinfo {author}
  {\bibfnamefont {C.}~\bibnamefont {Enderud}}, \bibinfo {author} {\bibfnamefont
  {C.}~\bibnamefont {Rich}}, \bibinfo {author} {\bibfnamefont {M.}~\bibnamefont
  {Thom}}, \bibinfo {author} {\bibfnamefont {J.}~\bibnamefont {Wang}}, \bibinfo
  {author} {\bibfnamefont {B.}~\bibnamefont {Wilson}}, \ and\ \bibinfo {author}
  {\bibfnamefont {G.}~\bibnamefont {Rose}},\ }\href {\doibase
  10.1103/PhysRevB.81.134510} {\bibfield  {journal} {\bibinfo  {journal} {Phys.
  Rev. B}\ }\textbf {\bibinfo {volume} {81}},\ \bibinfo {pages} {134510}
  (\bibinfo {year} {2010}{\natexlab{a}})}\BibitemShut {NoStop}%
\bibitem [{\citenamefont {Mooij}\ \emph {et~al.}(1999)\citenamefont {Mooij},
  \citenamefont {Orlando}, \citenamefont {Levitov}, \citenamefont {Tian},
  \citenamefont {van~der Wal},\ and\ \citenamefont {Lloyd}}]{Mooij99}%
  \BibitemOpen
  \bibfield  {author} {\bibinfo {author} {\bibfnamefont {J.~E.}\ \bibnamefont
  {Mooij}}, \bibinfo {author} {\bibfnamefont {T.~P.}\ \bibnamefont {Orlando}},
  \bibinfo {author} {\bibfnamefont {L.}~\bibnamefont {Levitov}}, \bibinfo
  {author} {\bibfnamefont {L.}~\bibnamefont {Tian}}, \bibinfo {author}
  {\bibfnamefont {C.~H.}\ \bibnamefont {van~der Wal}}, \ and\ \bibinfo {author}
  {\bibfnamefont {S.}~\bibnamefont {Lloyd}},\ }\href {\doibase
  10.1126/science.285.5430.1036} {\bibfield  {journal} {\bibinfo  {journal}
  {Science}\ }\textbf {\bibinfo {volume} {285}},\ \bibinfo {pages} {1036}
  (\bibinfo {year} {1999})}\BibitemShut {NoStop}%
\bibitem [{\citenamefont {Orlando}\ \emph {et~al.}(1999)\citenamefont
  {Orlando}, \citenamefont {Mooij}, \citenamefont {Tian}, \citenamefont
  {van~der Wal}, \citenamefont {Levitov}, \citenamefont {Lloyd},\ and\
  \citenamefont {Mazo}}]{orlando99}%
  \BibitemOpen
  \bibfield  {author} {\bibinfo {author} {\bibfnamefont {T.~P.}\ \bibnamefont
  {Orlando}}, \bibinfo {author} {\bibfnamefont {J.~E.}\ \bibnamefont {Mooij}},
  \bibinfo {author} {\bibfnamefont {L.}~\bibnamefont {Tian}}, \bibinfo {author}
  {\bibfnamefont {C.~H.}\ \bibnamefont {van~der Wal}}, \bibinfo {author}
  {\bibfnamefont {L.~S.}\ \bibnamefont {Levitov}}, \bibinfo {author}
  {\bibfnamefont {S.}~\bibnamefont {Lloyd}}, \ and\ \bibinfo {author}
  {\bibfnamefont {J.~J.}\ \bibnamefont {Mazo}},\ }\href {\doibase
  10.1103/PhysRevB.60.15398} {\bibfield  {journal} {\bibinfo  {journal} {Phys.
  Rev. B}\ }\textbf {\bibinfo {volume} {60}},\ \bibinfo {pages} {15398}
  (\bibinfo {year} {1999})}\BibitemShut {NoStop}%
\bibitem [{\citenamefont {van~der Wal}\ \emph {et~al.}(2000)\citenamefont
  {van~der Wal}, \citenamefont {ter Haar}, \citenamefont {Wilhelm},
  \citenamefont {Schouten}, \citenamefont {Harmans}, \citenamefont {Orlando},
  \citenamefont {Lloyd},\ and\ \citenamefont {Mooij}}]{vanderWal00}%
  \BibitemOpen
  \bibfield  {author} {\bibinfo {author} {\bibfnamefont {C.~H.}\ \bibnamefont
  {van~der Wal}}, \bibinfo {author} {\bibfnamefont {A.~C.~J.}\ \bibnamefont
  {ter Haar}}, \bibinfo {author} {\bibfnamefont {F.~K.}\ \bibnamefont
  {Wilhelm}}, \bibinfo {author} {\bibfnamefont {R.~N.}\ \bibnamefont
  {Schouten}}, \bibinfo {author} {\bibfnamefont {C.~J. P.~M.}\ \bibnamefont
  {Harmans}}, \bibinfo {author} {\bibfnamefont {T.~P.}\ \bibnamefont
  {Orlando}}, \bibinfo {author} {\bibfnamefont {S.}~\bibnamefont {Lloyd}}, \
  and\ \bibinfo {author} {\bibfnamefont {J.~E.}\ \bibnamefont {Mooij}},\ }\href
  {\doibase 10.1126/science.290.5492.773} {\bibfield  {journal} {\bibinfo
  {journal} {Science}\ }\textbf {\bibinfo {volume} {290}},\ \bibinfo {pages}
  {773} (\bibinfo {year} {2000})}\BibitemShut {NoStop}%
\bibitem [{\citenamefont {Grajcar}\ \emph {et~al.}(2004)\citenamefont
  {Grajcar}, \citenamefont {Izmalkov}, \citenamefont {Il'ichev}, \citenamefont
  {Wagner}, \citenamefont {Oukhanski}, \citenamefont {H\"ubner}, \citenamefont
  {May}, \citenamefont {Zhilyaev}, \citenamefont {Hoenig}, \citenamefont
  {Greenberg}, \citenamefont {Shnyrkov}, \citenamefont {Born}, \citenamefont
  {Krech}, \citenamefont {Meyer}, \citenamefont {{Maassen van den Brink}},\
  and\ \citenamefont {Amin}}]{grajcar04}%
  \BibitemOpen
  \bibfield  {author} {\bibinfo {author} {\bibfnamefont {M.}~\bibnamefont
  {Grajcar}}, \bibinfo {author} {\bibfnamefont {A.}~\bibnamefont {Izmalkov}},
  \bibinfo {author} {\bibfnamefont {E.}~\bibnamefont {Il'ichev}}, \bibinfo
  {author} {\bibfnamefont {T.}~\bibnamefont {Wagner}}, \bibinfo {author}
  {\bibfnamefont {N.}~\bibnamefont {Oukhanski}}, \bibinfo {author}
  {\bibfnamefont {U.}~\bibnamefont {H\"ubner}}, \bibinfo {author}
  {\bibfnamefont {T.}~\bibnamefont {May}}, \bibinfo {author} {\bibfnamefont
  {I.}~\bibnamefont {Zhilyaev}}, \bibinfo {author} {\bibfnamefont {H.~E.}\
  \bibnamefont {Hoenig}}, \bibinfo {author} {\bibfnamefont {Y.~S.}\
  \bibnamefont {Greenberg}}, \bibinfo {author} {\bibfnamefont {V.~I.}\
  \bibnamefont {Shnyrkov}}, \bibinfo {author} {\bibfnamefont {D.}~\bibnamefont
  {Born}}, \bibinfo {author} {\bibfnamefont {W.}~\bibnamefont {Krech}},
  \bibinfo {author} {\bibfnamefont {H.-G.}\ \bibnamefont {Meyer}}, \bibinfo
  {author} {\bibfnamefont {A.}~\bibnamefont {{Maassen van den Brink}}}, \ and\
  \bibinfo {author} {\bibfnamefont {M.~H.~S.}\ \bibnamefont {Amin}},\ }\href
  {\doibase 10.1103/PhysRevB.69.060501} {\bibfield  {journal} {\bibinfo
  {journal} {Phys. Rev. B}\ }\textbf {\bibinfo {volume} {69}},\ \bibinfo
  {pages} {060501(R)} (\bibinfo {year} {2004})}\BibitemShut {NoStop}%
\bibitem [{\citenamefont {You}\ \emph {et~al.}(2007)\citenamefont {You},
  \citenamefont {Hu}, \citenamefont {Ashhab},\ and\ \citenamefont
  {Nori}}]{you07}%
  \BibitemOpen
  \bibfield  {author} {\bibinfo {author} {\bibfnamefont {J.~Q.}\ \bibnamefont
  {You}}, \bibinfo {author} {\bibfnamefont {X.}~\bibnamefont {Hu}}, \bibinfo
  {author} {\bibfnamefont {S.}~\bibnamefont {Ashhab}}, \ and\ \bibinfo {author}
  {\bibfnamefont {F.}~\bibnamefont {Nori}},\ }\href {\doibase
  10.1103/PhysRevB.75.140515} {\bibfield  {journal} {\bibinfo  {journal} {Phys.
  Rev. B}\ }\textbf {\bibinfo {volume} {75}},\ \bibinfo {pages} {140515(R)}
  (\bibinfo {year} {2007})}\BibitemShut {NoStop}%
\bibitem [{\citenamefont {Steffen}\ \emph {et~al.}(2010)\citenamefont
  {Steffen}, \citenamefont {Kumar}, \citenamefont {DiVincenzo}, \citenamefont
  {Rozen}, \citenamefont {Keefe}, \citenamefont {Rothwell},\ and\ \citenamefont
  {Ketchen}}]{steffen10}%
  \BibitemOpen
  \bibfield  {author} {\bibinfo {author} {\bibfnamefont {M.}~\bibnamefont
  {Steffen}}, \bibinfo {author} {\bibfnamefont {S.}~\bibnamefont {Kumar}},
  \bibinfo {author} {\bibfnamefont {D.~P.}\ \bibnamefont {DiVincenzo}},
  \bibinfo {author} {\bibfnamefont {J.~R.}\ \bibnamefont {Rozen}}, \bibinfo
  {author} {\bibfnamefont {G.~A.}\ \bibnamefont {Keefe}}, \bibinfo {author}
  {\bibfnamefont {M.~B.}\ \bibnamefont {Rothwell}}, \ and\ \bibinfo {author}
  {\bibfnamefont {M.~B.}\ \bibnamefont {Ketchen}},\ }\href {\doibase
  10.1103/PhysRevLett.105.100502} {\bibfield  {journal} {\bibinfo  {journal}
  {Phys. Rev. Lett.}\ }\textbf {\bibinfo {volume} {105}},\ \bibinfo {pages}
  {100502} (\bibinfo {year} {2010})}\BibitemShut {NoStop}%
\bibitem [{\citenamefont {Yan}\ \emph {et~al.}(2016)\citenamefont {Yan},
  \citenamefont {Gustavsson}, \citenamefont {Kamal}, \citenamefont {Birenbaum},
  \citenamefont {Sears}, \citenamefont {Hover}, \citenamefont {Gudmundsen},
  \citenamefont {Rosenberg}, \citenamefont {Samach}, \citenamefont {Weber},
  \citenamefont {Yoder}, \citenamefont {Orlando}, \citenamefont {Clarke},
  \citenamefont {Kerman},\ and\ \citenamefont {Oliver}}]{Yan2016}%
  \BibitemOpen
  \bibfield  {author} {\bibinfo {author} {\bibfnamefont {F.}~\bibnamefont
  {Yan}}, \bibinfo {author} {\bibfnamefont {S.}~\bibnamefont {Gustavsson}},
  \bibinfo {author} {\bibfnamefont {A.}~\bibnamefont {Kamal}}, \bibinfo
  {author} {\bibfnamefont {J.}~\bibnamefont {Birenbaum}}, \bibinfo {author}
  {\bibfnamefont {A.~P.}\ \bibnamefont {Sears}}, \bibinfo {author}
  {\bibfnamefont {D.}~\bibnamefont {Hover}}, \bibinfo {author} {\bibfnamefont
  {T.~J.}\ \bibnamefont {Gudmundsen}}, \bibinfo {author} {\bibfnamefont
  {D.}~\bibnamefont {Rosenberg}}, \bibinfo {author} {\bibfnamefont
  {G.}~\bibnamefont {Samach}}, \bibinfo {author} {\bibfnamefont
  {S.}~\bibnamefont {Weber}}, \bibinfo {author} {\bibfnamefont {J.~L.}\
  \bibnamefont {Yoder}}, \bibinfo {author} {\bibfnamefont {T.~P.}\ \bibnamefont
  {Orlando}}, \bibinfo {author} {\bibfnamefont {J.}~\bibnamefont {Clarke}},
  \bibinfo {author} {\bibfnamefont {A.~J.}\ \bibnamefont {Kerman}}, \ and\
  \bibinfo {author} {\bibfnamefont {W.~D.}\ \bibnamefont {Oliver}},\ }\href
  {https://doi.org/10.1038/ncomms12964} {\bibfield  {journal} {\bibinfo
  {journal} {Nat. Commun.}\ }\textbf {\bibinfo {volume} {7}},\ \bibinfo {pages}
  {12964} (\bibinfo {year} {2016})}\BibitemShut {NoStop}%
\bibitem [{\citenamefont {Manucharyan}\ \emph {et~al.}(2009)\citenamefont
  {Manucharyan}, \citenamefont {Koch}, \citenamefont {Glazman},\ and\
  \citenamefont {Devoret}}]{Manucharyan09}%
  \BibitemOpen
  \bibfield  {author} {\bibinfo {author} {\bibfnamefont {V.~E.}\ \bibnamefont
  {Manucharyan}}, \bibinfo {author} {\bibfnamefont {J.}~\bibnamefont {Koch}},
  \bibinfo {author} {\bibfnamefont {L.~I.}\ \bibnamefont {Glazman}}, \ and\
  \bibinfo {author} {\bibfnamefont {M.~H.}\ \bibnamefont {Devoret}},\ }\href
  {\doibase 10.1126/science.1175552} {\bibfield  {journal} {\bibinfo  {journal}
  {Science}\ }\textbf {\bibinfo {volume} {326}},\ \bibinfo {pages} {113}
  (\bibinfo {year} {2009})}\BibitemShut {NoStop}%
\bibitem [{\citenamefont {Pop}\ \emph {et~al.}(2014)\citenamefont {Pop},
  \citenamefont {Geerlings}, \citenamefont {Catelani}, \citenamefont
  {Schoelkopf}, \citenamefont {Glazman},\ and\ \citenamefont
  {Devoret}}]{pop2014}%
  \BibitemOpen
  \bibfield  {author} {\bibinfo {author} {\bibfnamefont {I.~M.}\ \bibnamefont
  {Pop}}, \bibinfo {author} {\bibfnamefont {K.}~\bibnamefont {Geerlings}},
  \bibinfo {author} {\bibfnamefont {G.}~\bibnamefont {Catelani}}, \bibinfo
  {author} {\bibfnamefont {R.~J.}\ \bibnamefont {Schoelkopf}}, \bibinfo
  {author} {\bibfnamefont {L.~I.}\ \bibnamefont {Glazman}}, \ and\ \bibinfo
  {author} {\bibfnamefont {M.~H.}\ \bibnamefont {Devoret}},\ }\href {\doibase
  10.1038/nature13017} {\bibfield  {journal} {\bibinfo  {journal} {Nature}\
  }\textbf {\bibinfo {volume} {508}},\ \bibinfo {pages} {369} (\bibinfo {year}
  {2014})}\BibitemShut {NoStop}%
\bibitem [{\citenamefont {Nguyen}\ \emph {et~al.}(2019)\citenamefont {Nguyen},
  \citenamefont {Lin}, \citenamefont {Somoroff}, \citenamefont {Mencia},
  \citenamefont {Grabon},\ and\ \citenamefont {Manucharyan}}]{nguyen18}%
  \BibitemOpen
  \bibfield  {author} {\bibinfo {author} {\bibfnamefont {L.~B.}\ \bibnamefont
  {Nguyen}}, \bibinfo {author} {\bibfnamefont {Y.-H.}\ \bibnamefont {Lin}},
  \bibinfo {author} {\bibfnamefont {A.}~\bibnamefont {Somoroff}}, \bibinfo
  {author} {\bibfnamefont {R.}~\bibnamefont {Mencia}}, \bibinfo {author}
  {\bibfnamefont {N.}~\bibnamefont {Grabon}}, \ and\ \bibinfo {author}
  {\bibfnamefont {V.~E.}\ \bibnamefont {Manucharyan}},\ }\href {\doibase
  10.1103/PhysRevX.9.041041} {\bibfield  {journal} {\bibinfo  {journal} {Phys.
  Rev. X}\ }\textbf {\bibinfo {volume} {9}},\ \bibinfo {pages} {041041}
  (\bibinfo {year} {2019})}\BibitemShut {NoStop}%
\bibitem [{\citenamefont {Wendin}\ and\ \citenamefont
  {Shumeiko}(2007)}]{wendin07}%
  \BibitemOpen
  \bibfield  {author} {\bibinfo {author} {\bibfnamefont {G.}~\bibnamefont
  {Wendin}}\ and\ \bibinfo {author} {\bibfnamefont {V.~S.}\ \bibnamefont
  {Shumeiko}},\ }\href {\doibase 10.1063/1.2780165} {\bibfield  {journal}
  {\bibinfo  {journal} {Low Temp. Phys.}\ }\textbf {\bibinfo {volume} {33}},\
  \bibinfo {pages} {724} (\bibinfo {year} {2007})}\BibitemShut {NoStop}%
\bibitem [{\citenamefont {Childs}\ \emph {et~al.}(2001)\citenamefont {Childs},
  \citenamefont {Farhi},\ and\ \citenamefont {Preskill}}]{childs01}%
  \BibitemOpen
  \bibfield  {author} {\bibinfo {author} {\bibfnamefont {A.~M.}\ \bibnamefont
  {Childs}}, \bibinfo {author} {\bibfnamefont {E.}~\bibnamefont {Farhi}}, \
  and\ \bibinfo {author} {\bibfnamefont {J.}~\bibnamefont {Preskill}},\ }\href
  {\doibase 10.1103/PhysRevA.65.012322} {\bibfield  {journal} {\bibinfo
  {journal} {Phys. Rev. A}\ }\textbf {\bibinfo {volume} {65}},\ \bibinfo
  {pages} {012322} (\bibinfo {year} {2001})}\BibitemShut {NoStop}%
\bibitem [{\citenamefont {Sarandy}\ and\ \citenamefont
  {Lidar}(2005)}]{sarandy05}%
  \BibitemOpen
  \bibfield  {author} {\bibinfo {author} {\bibfnamefont {M.~S.}\ \bibnamefont
  {Sarandy}}\ and\ \bibinfo {author} {\bibfnamefont {D.~A.}\ \bibnamefont
  {Lidar}},\ }\href {\doibase 10.1103/PhysRevLett.95.250503} {\bibfield
  {journal} {\bibinfo  {journal} {Phys. Rev. Lett.}\ }\textbf {\bibinfo
  {volume} {95}},\ \bibinfo {pages} {250503} (\bibinfo {year}
  {2005})}\BibitemShut {NoStop}%
\bibitem [{\citenamefont {Ashhab}\ \emph {et~al.}(2006)\citenamefont {Ashhab},
  \citenamefont {Johansson},\ and\ \citenamefont {Nori}}]{ashhab06}%
  \BibitemOpen
  \bibfield  {author} {\bibinfo {author} {\bibfnamefont {S.}~\bibnamefont
  {Ashhab}}, \bibinfo {author} {\bibfnamefont {J.~R.}\ \bibnamefont
  {Johansson}}, \ and\ \bibinfo {author} {\bibfnamefont {F.}~\bibnamefont
  {Nori}},\ }\href {\doibase 10.1103/PhysRevA.74.052330} {\bibfield  {journal}
  {\bibinfo  {journal} {Phys. Rev. A}\ }\textbf {\bibinfo {volume} {74}},\
  \bibinfo {pages} {052330} (\bibinfo {year} {2006})}\BibitemShut {NoStop}%
\bibitem [{\citenamefont {Amin}\ \emph
  {et~al.}(2009{\natexlab{a}})\citenamefont {Amin}, \citenamefont {Truncik},\
  and\ \citenamefont {Averin}}]{amin09_decoherence}%
  \BibitemOpen
  \bibfield  {author} {\bibinfo {author} {\bibfnamefont {M.~H.~S.}\
  \bibnamefont {Amin}}, \bibinfo {author} {\bibfnamefont {C.~J.~S.}\
  \bibnamefont {Truncik}}, \ and\ \bibinfo {author} {\bibfnamefont {D.~V.}\
  \bibnamefont {Averin}},\ }\href {\doibase 10.1103/PhysRevA.80.022303}
  {\bibfield  {journal} {\bibinfo  {journal} {Phys. Rev. A}\ }\textbf {\bibinfo
  {volume} {80}},\ \bibinfo {pages} {022303} (\bibinfo {year}
  {2009}{\natexlab{a}})}\BibitemShut {NoStop}%
\bibitem [{\citenamefont {Boixo}\ \emph {et~al.}(2013)\citenamefont {Boixo},
  \citenamefont {Albash}, \citenamefont {Spedalieri}, \citenamefont
  {Chancellor},\ and\ \citenamefont {Lidar}}]{boixo13}%
  \BibitemOpen
  \bibfield  {author} {\bibinfo {author} {\bibfnamefont {S.}~\bibnamefont
  {Boixo}}, \bibinfo {author} {\bibfnamefont {T.}~\bibnamefont {Albash}},
  \bibinfo {author} {\bibfnamefont {F.~M.}\ \bibnamefont {Spedalieri}},
  \bibinfo {author} {\bibfnamefont {N.}~\bibnamefont {Chancellor}}, \ and\
  \bibinfo {author} {\bibfnamefont {D.~A.}\ \bibnamefont {Lidar}},\ }\href
  {\doibase 10.1038/ncomms3067} {\bibfield  {journal} {\bibinfo  {journal}
  {Nat. Commun.}\ }\textbf {\bibinfo {volume} {4}},\ \bibinfo {pages} {2067}
  (\bibinfo {year} {2013})}\BibitemShut {NoStop}%
\bibitem [{\citenamefont {Dickson}\ \emph {et~al.}(2013)\citenamefont
  {Dickson}, \citenamefont {Johnson}, \citenamefont {Amin}, \citenamefont
  {Harris}, \citenamefont {Altomare}, \citenamefont {Berkley}, \citenamefont
  {Bunyk}, \citenamefont {Cai}, \citenamefont {Chapple}, \citenamefont
  {Chavez}, \citenamefont {Cioata}, \citenamefont {Cirip}, \citenamefont
  {deBuen}, \citenamefont {Drew-Brook}, \citenamefont {Enderud}, \citenamefont
  {Gildert}, \citenamefont {Hamze}, \citenamefont {Hilton}, \citenamefont
  {Hoskinson}, \citenamefont {Karimi}, \citenamefont {Ladizinsky},
  \citenamefont {Ladizinsky}, \citenamefont {Lanting}, \citenamefont {Mahon},
  \citenamefont {Neufeld}, \citenamefont {Oh}, \citenamefont {Perminov},
  \citenamefont {Petroff}, \citenamefont {Przybysz}, \citenamefont {Rich},
  \citenamefont {Spear}, \citenamefont {Tcaciuc}, \citenamefont {Thom},
  \citenamefont {Tolkacheva}, \citenamefont {Uchaikin}, \citenamefont {Wang},
  \citenamefont {Wilson}, \citenamefont {Merali},\ and\ \citenamefont
  {Rose}}]{dickson13}%
  \BibitemOpen
  \bibfield  {author} {\bibinfo {author} {\bibfnamefont {N.~G.}\ \bibnamefont
  {Dickson}}, \bibinfo {author} {\bibfnamefont {M.~W.}\ \bibnamefont
  {Johnson}}, \bibinfo {author} {\bibfnamefont {M.~H.}\ \bibnamefont {Amin}},
  \bibinfo {author} {\bibfnamefont {R.}~\bibnamefont {Harris}}, \bibinfo
  {author} {\bibfnamefont {F.}~\bibnamefont {Altomare}}, \bibinfo {author}
  {\bibfnamefont {A.~J.}\ \bibnamefont {Berkley}}, \bibinfo {author}
  {\bibfnamefont {P.}~\bibnamefont {Bunyk}}, \bibinfo {author} {\bibfnamefont
  {J.}~\bibnamefont {Cai}}, \bibinfo {author} {\bibfnamefont {E.~M.}\
  \bibnamefont {Chapple}}, \bibinfo {author} {\bibfnamefont {P.}~\bibnamefont
  {Chavez}}, \bibinfo {author} {\bibfnamefont {F.}~\bibnamefont {Cioata}},
  \bibinfo {author} {\bibfnamefont {T.}~\bibnamefont {Cirip}}, \bibinfo
  {author} {\bibfnamefont {P.}~\bibnamefont {deBuen}}, \bibinfo {author}
  {\bibfnamefont {M.}~\bibnamefont {Drew-Brook}}, \bibinfo {author}
  {\bibfnamefont {C.}~\bibnamefont {Enderud}}, \bibinfo {author} {\bibfnamefont
  {S.}~\bibnamefont {Gildert}}, \bibinfo {author} {\bibfnamefont
  {F.}~\bibnamefont {Hamze}}, \bibinfo {author} {\bibfnamefont {J.~P.}\
  \bibnamefont {Hilton}}, \bibinfo {author} {\bibfnamefont {E.}~\bibnamefont
  {Hoskinson}}, \bibinfo {author} {\bibfnamefont {K.}~\bibnamefont {Karimi}},
  \bibinfo {author} {\bibfnamefont {E.}~\bibnamefont {Ladizinsky}}, \bibinfo
  {author} {\bibfnamefont {N.}~\bibnamefont {Ladizinsky}}, \bibinfo {author}
  {\bibfnamefont {T.}~\bibnamefont {Lanting}}, \bibinfo {author} {\bibfnamefont
  {T.}~\bibnamefont {Mahon}}, \bibinfo {author} {\bibfnamefont
  {R.}~\bibnamefont {Neufeld}}, \bibinfo {author} {\bibfnamefont
  {T.}~\bibnamefont {Oh}}, \bibinfo {author} {\bibfnamefont {I.}~\bibnamefont
  {Perminov}}, \bibinfo {author} {\bibfnamefont {C.}~\bibnamefont {Petroff}},
  \bibinfo {author} {\bibfnamefont {A.}~\bibnamefont {Przybysz}}, \bibinfo
  {author} {\bibfnamefont {C.}~\bibnamefont {Rich}}, \bibinfo {author}
  {\bibfnamefont {P.}~\bibnamefont {Spear}}, \bibinfo {author} {\bibfnamefont
  {A.}~\bibnamefont {Tcaciuc}}, \bibinfo {author} {\bibfnamefont {M.~C.}\
  \bibnamefont {Thom}}, \bibinfo {author} {\bibfnamefont {E.}~\bibnamefont
  {Tolkacheva}}, \bibinfo {author} {\bibfnamefont {S.}~\bibnamefont
  {Uchaikin}}, \bibinfo {author} {\bibfnamefont {J.}~\bibnamefont {Wang}},
  \bibinfo {author} {\bibfnamefont {A.~B.}\ \bibnamefont {Wilson}}, \bibinfo
  {author} {\bibfnamefont {Z.}~\bibnamefont {Merali}}, \ and\ \bibinfo {author}
  {\bibfnamefont {G.}~\bibnamefont {Rose}},\ }\href {\doibase
  10.1038/ncomms2920} {\bibfield  {journal} {\bibinfo  {journal} {Nat.
  Commun.}\ }\textbf {\bibinfo {volume} {4}},\ \bibinfo {pages} {1903}
  (\bibinfo {year} {2013})}\BibitemShut {NoStop}%
\bibitem [{\citenamefont {Johnson}\ \emph {et~al.}(2011)\citenamefont
  {Johnson}, \citenamefont {Amin}, \citenamefont {Gildert}, \citenamefont
  {Lanting}, \citenamefont {Hamze}, \citenamefont {Dickson}, \citenamefont
  {Harris}, \citenamefont {Berkley}, \citenamefont {Johansson}, \citenamefont
  {Bunyk}, \citenamefont {Chapple}, \citenamefont {Enderud}, \citenamefont
  {Hilton}, \citenamefont {Karimi}, \citenamefont {Ladizinsky}, \citenamefont
  {Ladizinsky}, \citenamefont {Oh}, \citenamefont {Perminov}, \citenamefont
  {Rich}, \citenamefont {Thom}, \citenamefont {Tolkacheva}, \citenamefont
  {Truncik}, \citenamefont {Uchaikin}, \citenamefont {Wang}, \citenamefont
  {Wilson},\ and\ \citenamefont {Rose}}]{johnson11}%
  \BibitemOpen
  \bibfield  {author} {\bibinfo {author} {\bibfnamefont {M.~W.}\ \bibnamefont
  {Johnson}}, \bibinfo {author} {\bibfnamefont {M.~H.~S.}\ \bibnamefont
  {Amin}}, \bibinfo {author} {\bibfnamefont {S.}~\bibnamefont {Gildert}},
  \bibinfo {author} {\bibfnamefont {T.}~\bibnamefont {Lanting}}, \bibinfo
  {author} {\bibfnamefont {F.}~\bibnamefont {Hamze}}, \bibinfo {author}
  {\bibfnamefont {N.}~\bibnamefont {Dickson}}, \bibinfo {author} {\bibfnamefont
  {R.}~\bibnamefont {Harris}}, \bibinfo {author} {\bibfnamefont {A.~J.}\
  \bibnamefont {Berkley}}, \bibinfo {author} {\bibfnamefont {J.}~\bibnamefont
  {Johansson}}, \bibinfo {author} {\bibfnamefont {P.}~\bibnamefont {Bunyk}},
  \bibinfo {author} {\bibfnamefont {E.~M.}\ \bibnamefont {Chapple}}, \bibinfo
  {author} {\bibfnamefont {C.}~\bibnamefont {Enderud}}, \bibinfo {author}
  {\bibfnamefont {J.~P.}\ \bibnamefont {Hilton}}, \bibinfo {author}
  {\bibfnamefont {K.}~\bibnamefont {Karimi}}, \bibinfo {author} {\bibfnamefont
  {E.}~\bibnamefont {Ladizinsky}}, \bibinfo {author} {\bibfnamefont
  {N.}~\bibnamefont {Ladizinsky}}, \bibinfo {author} {\bibfnamefont
  {T.}~\bibnamefont {Oh}}, \bibinfo {author} {\bibfnamefont {I.}~\bibnamefont
  {Perminov}}, \bibinfo {author} {\bibfnamefont {C.}~\bibnamefont {Rich}},
  \bibinfo {author} {\bibfnamefont {M.~C.}\ \bibnamefont {Thom}}, \bibinfo
  {author} {\bibfnamefont {E.}~\bibnamefont {Tolkacheva}}, \bibinfo {author}
  {\bibfnamefont {C.~J.~S.}\ \bibnamefont {Truncik}}, \bibinfo {author}
  {\bibfnamefont {S.}~\bibnamefont {Uchaikin}}, \bibinfo {author}
  {\bibfnamefont {J.}~\bibnamefont {Wang}}, \bibinfo {author} {\bibfnamefont
  {B.}~\bibnamefont {Wilson}}, \ and\ \bibinfo {author} {\bibfnamefont
  {G.}~\bibnamefont {Rose}},\ }\href {\doibase 10.1038/nature10012} {\bibfield
  {journal} {\bibinfo  {journal} {Nature}\ }\textbf {\bibinfo {volume} {473}},\
  \bibinfo {pages} {194} (\bibinfo {year} {2011})}\BibitemShut {NoStop}%
\bibitem [{\citenamefont {Amin}\ \emph {et~al.}(2013)\citenamefont {Amin},
  \citenamefont {Dickson},\ and\ \citenamefont {Smith}}]{amin13}%
  \BibitemOpen
  \bibfield  {author} {\bibinfo {author} {\bibfnamefont {M.~H.~S.}\
  \bibnamefont {Amin}}, \bibinfo {author} {\bibfnamefont {N.~G.}\ \bibnamefont
  {Dickson}}, \ and\ \bibinfo {author} {\bibfnamefont {P.}~\bibnamefont
  {Smith}},\ }\href {\doibase 10.1007/s11128-012-0480-x} {\bibfield  {journal}
  {\bibinfo  {journal} {Quantum Inf. Process.}\ }\textbf {\bibinfo {volume}
  {12}},\ \bibinfo {pages} {1819} (\bibinfo {year} {2013})}\BibitemShut
  {NoStop}%
\bibitem [{\citenamefont {Zhao}\ \emph {et~al.}(2016)\citenamefont {Zhao},
  \citenamefont {{De Raedt}}, \citenamefont {Miyashita}, \citenamefont {Jin},\
  and\ \citenamefont {Michielsen}}]{ZHAO16}%
  \BibitemOpen
  \bibfield  {author} {\bibinfo {author} {\bibfnamefont {P.}~\bibnamefont
  {Zhao}}, \bibinfo {author} {\bibfnamefont {H.}~\bibnamefont {{De Raedt}}},
  \bibinfo {author} {\bibfnamefont {S.}~\bibnamefont {Miyashita}}, \bibinfo
  {author} {\bibfnamefont {F.}~\bibnamefont {Jin}}, \ and\ \bibinfo {author}
  {\bibfnamefont {K.}~\bibnamefont {Michielsen}},\ }\href {\doibase
  10.1103/PhysRevE.94.022126} {\bibfield  {journal} {\bibinfo  {journal} {Phys.
  Rev. E}\ }\textbf {\bibinfo {volume} {94}},\ \bibinfo {pages} {022126}
  (\bibinfo {year} {2016})}\BibitemShut {NoStop}%
\bibitem [{\citenamefont {{De Raedt}}\ \emph {et~al.}(2017)\citenamefont {{De
  Raedt}}, \citenamefont {Jin}, \citenamefont {Katsnelson},\ and\ \citenamefont
  {Michielsen}}]{RAED17b}%
  \BibitemOpen
  \bibfield  {author} {\bibinfo {author} {\bibfnamefont {H.}~\bibnamefont {{De
  Raedt}}}, \bibinfo {author} {\bibfnamefont {F.}~\bibnamefont {Jin}}, \bibinfo
  {author} {\bibfnamefont {M.}~\bibnamefont {Katsnelson}}, \ and\ \bibinfo
  {author} {\bibfnamefont {K.}~\bibnamefont {Michielsen}},\ }\href {\doibase
  10.1103/PhysRevE.96.053306} {\bibfield  {journal} {\bibinfo  {journal} {Phys.
  Rev. E}\ }\textbf {\bibinfo {volume} {96}},\ \bibinfo {pages} {053306}
  (\bibinfo {year} {2017})}\BibitemShut {NoStop}%
\bibitem [{\citenamefont {Shnirman}\ \emph {et~al.}(2005)\citenamefont
  {Shnirman}, \citenamefont {Sch\"on}, \citenamefont {Martin},\ and\
  \citenamefont {Makhlin}}]{shnirman05}%
  \BibitemOpen
  \bibfield  {author} {\bibinfo {author} {\bibfnamefont {A.}~\bibnamefont
  {Shnirman}}, \bibinfo {author} {\bibfnamefont {G.}~\bibnamefont {Sch\"on}},
  \bibinfo {author} {\bibfnamefont {I.}~\bibnamefont {Martin}}, \ and\ \bibinfo
  {author} {\bibfnamefont {Y.}~\bibnamefont {Makhlin}},\ }\href {\doibase
  10.1103/PhysRevLett.94.127002} {\bibfield  {journal} {\bibinfo  {journal}
  {Phys. Rev. Lett.}\ }\textbf {\bibinfo {volume} {94}},\ \bibinfo {pages}
  {127002} (\bibinfo {year} {2005})}\BibitemShut {NoStop}%
\bibitem [{\citenamefont {M\"uller}\ \emph {et~al.}(2009)\citenamefont
  {M\"uller}, \citenamefont {Shnirman},\ and\ \citenamefont
  {Makhlin}}]{mueller09}%
  \BibitemOpen
  \bibfield  {author} {\bibinfo {author} {\bibfnamefont {C.}~\bibnamefont
  {M\"uller}}, \bibinfo {author} {\bibfnamefont {A.}~\bibnamefont {Shnirman}},
  \ and\ \bibinfo {author} {\bibfnamefont {Y.}~\bibnamefont {Makhlin}},\ }\href
  {\doibase 10.1103/PhysRevB.80.134517} {\bibfield  {journal} {\bibinfo
  {journal} {Phys. Rev. B}\ }\textbf {\bibinfo {volume} {80}},\ \bibinfo
  {pages} {134517} (\bibinfo {year} {2009})}\BibitemShut {NoStop}%
\bibitem [{\citenamefont {Cole}\ \emph {et~al.}(2010)\citenamefont {Cole},
  \citenamefont {M\"uller}, \citenamefont {Bushev}, \citenamefont {Grabovskij},
  \citenamefont {Lisenfeld}, \citenamefont {Lukashenko}, \citenamefont
  {Ustinov},\ and\ \citenamefont {Shnirman}}]{cole10}%
  \BibitemOpen
  \bibfield  {author} {\bibinfo {author} {\bibfnamefont {J.~H.}\ \bibnamefont
  {Cole}}, \bibinfo {author} {\bibfnamefont {C.}~\bibnamefont {M\"uller}},
  \bibinfo {author} {\bibfnamefont {P.}~\bibnamefont {Bushev}}, \bibinfo
  {author} {\bibfnamefont {G.~J.}\ \bibnamefont {Grabovskij}}, \bibinfo
  {author} {\bibfnamefont {J.}~\bibnamefont {Lisenfeld}}, \bibinfo {author}
  {\bibfnamefont {A.}~\bibnamefont {Lukashenko}}, \bibinfo {author}
  {\bibfnamefont {A.~V.}\ \bibnamefont {Ustinov}}, \ and\ \bibinfo {author}
  {\bibfnamefont {A.}~\bibnamefont {Shnirman}},\ }\href {\doibase
  10.1063/1.3529457} {\bibfield  {journal} {\bibinfo  {journal} {Appl. Phys.
  Lett.}\ }\textbf {\bibinfo {volume} {97}},\ \bibinfo {pages} {252501}
  (\bibinfo {year} {2010})}\BibitemShut {NoStop}%
\bibitem [{\citenamefont {Born}\ and\ \citenamefont {Fock}(1928)}]{born28}%
  \BibitemOpen
  \bibfield  {author} {\bibinfo {author} {\bibfnamefont {M.}~\bibnamefont
  {Born}}\ and\ \bibinfo {author} {\bibfnamefont {V.}~\bibnamefont {Fock}},\
  }\href {\doibase 10.1007/BF01343193} {\bibfield  {journal} {\bibinfo
  {journal} {Z. Phys.}\ }\textbf {\bibinfo {volume} {51}},\ \bibinfo {pages}
  {165} (\bibinfo {year} {1928})}\BibitemShut {NoStop}%
\bibitem [{\citenamefont {Amin}(2009)}]{amin09_adiabatictheorem}%
  \BibitemOpen
  \bibfield  {author} {\bibinfo {author} {\bibfnamefont {M.~H.~S.}\
  \bibnamefont {Amin}},\ }\href {\doibase 10.1103/PhysRevLett.102.220401}
  {\bibfield  {journal} {\bibinfo  {journal} {Phys. Rev. Lett.}\ }\textbf
  {\bibinfo {volume} {102}},\ \bibinfo {pages} {220401} (\bibinfo {year}
  {2009})}\BibitemShut {NoStop}%
\bibitem [{\citenamefont {Albash}\ and\ \citenamefont
  {Lidar}(2018)}]{albash18}%
  \BibitemOpen
  \bibfield  {author} {\bibinfo {author} {\bibfnamefont {T.}~\bibnamefont
  {Albash}}\ and\ \bibinfo {author} {\bibfnamefont {D.~A.}\ \bibnamefont
  {Lidar}},\ }\href {\doibase 10.1103/RevModPhys.90.015002} {\bibfield
  {journal} {\bibinfo  {journal} {Rev. Mod. Phys.}\ }\textbf {\bibinfo {volume}
  {90}},\ \bibinfo {pages} {015002} (\bibinfo {year} {2018})}\BibitemShut
  {NoStop}%
\bibitem [{\citenamefont {Kadowaki}\ and\ \citenamefont
  {Nishimori}(1998)}]{kadowaki98}%
  \BibitemOpen
  \bibfield  {author} {\bibinfo {author} {\bibfnamefont {T.}~\bibnamefont
  {Kadowaki}}\ and\ \bibinfo {author} {\bibfnamefont {H.}~\bibnamefont
  {Nishimori}},\ }\href {\doibase 10.1103/PhysRevE.58.5355} {\bibfield
  {journal} {\bibinfo  {journal} {Phys. Rev. E}\ }\textbf {\bibinfo {volume}
  {58}},\ \bibinfo {pages} {5355} (\bibinfo {year} {1998})}\BibitemShut
  {NoStop}%
\bibitem [{\citenamefont {Harris}\ \emph
  {et~al.}(2010{\natexlab{b}})\citenamefont {Harris}, \citenamefont {Johnson},
  \citenamefont {Lanting}, \citenamefont {Berkley}, \citenamefont {Johansson},
  \citenamefont {Bunyk}, \citenamefont {Tolkacheva}, \citenamefont
  {Ladizinsky}, \citenamefont {Ladizinsky}, \citenamefont {Oh}, \citenamefont
  {Cioata}, \citenamefont {Perminov}, \citenamefont {Spear}, \citenamefont
  {Enderud}, \citenamefont {Rich}, \citenamefont {Uchaikin}, \citenamefont
  {Thom}, \citenamefont {Chapple}, \citenamefont {Wang}, \citenamefont
  {Wilson}, \citenamefont {Amin}, \citenamefont {Dickson}, \citenamefont
  {Karimi}, \citenamefont {Macready}, \citenamefont {Truncik},\ and\
  \citenamefont {Rose}}]{harris10_eightqubit}%
  \BibitemOpen
  \bibfield  {author} {\bibinfo {author} {\bibfnamefont {R.}~\bibnamefont
  {Harris}}, \bibinfo {author} {\bibfnamefont {M.~W.}\ \bibnamefont {Johnson}},
  \bibinfo {author} {\bibfnamefont {T.}~\bibnamefont {Lanting}}, \bibinfo
  {author} {\bibfnamefont {A.~J.}\ \bibnamefont {Berkley}}, \bibinfo {author}
  {\bibfnamefont {J.}~\bibnamefont {Johansson}}, \bibinfo {author}
  {\bibfnamefont {P.}~\bibnamefont {Bunyk}}, \bibinfo {author} {\bibfnamefont
  {E.}~\bibnamefont {Tolkacheva}}, \bibinfo {author} {\bibfnamefont
  {E.}~\bibnamefont {Ladizinsky}}, \bibinfo {author} {\bibfnamefont
  {N.}~\bibnamefont {Ladizinsky}}, \bibinfo {author} {\bibfnamefont
  {T.}~\bibnamefont {Oh}}, \bibinfo {author} {\bibfnamefont {F.}~\bibnamefont
  {Cioata}}, \bibinfo {author} {\bibfnamefont {I.}~\bibnamefont {Perminov}},
  \bibinfo {author} {\bibfnamefont {P.}~\bibnamefont {Spear}}, \bibinfo
  {author} {\bibfnamefont {C.}~\bibnamefont {Enderud}}, \bibinfo {author}
  {\bibfnamefont {C.}~\bibnamefont {Rich}}, \bibinfo {author} {\bibfnamefont
  {S.}~\bibnamefont {Uchaikin}}, \bibinfo {author} {\bibfnamefont {M.~C.}\
  \bibnamefont {Thom}}, \bibinfo {author} {\bibfnamefont {E.~M.}\ \bibnamefont
  {Chapple}}, \bibinfo {author} {\bibfnamefont {J.}~\bibnamefont {Wang}},
  \bibinfo {author} {\bibfnamefont {B.}~\bibnamefont {Wilson}}, \bibinfo
  {author} {\bibfnamefont {M.~H.~S.}\ \bibnamefont {Amin}}, \bibinfo {author}
  {\bibfnamefont {N.}~\bibnamefont {Dickson}}, \bibinfo {author} {\bibfnamefont
  {K.}~\bibnamefont {Karimi}}, \bibinfo {author} {\bibfnamefont
  {B.}~\bibnamefont {Macready}}, \bibinfo {author} {\bibfnamefont {C.~J.~S.}\
  \bibnamefont {Truncik}}, \ and\ \bibinfo {author} {\bibfnamefont
  {G.}~\bibnamefont {Rose}},\ }\href {\doibase 10.1103/PhysRevB.82.024511}
  {\bibfield  {journal} {\bibinfo  {journal} {Phys. Rev. B}\ }\textbf {\bibinfo
  {volume} {82}},\ \bibinfo {pages} {024511} (\bibinfo {year}
  {2010}{\natexlab{b}})}\BibitemShut {NoStop}%
\bibitem [{\citenamefont {Harris}\ \emph
  {et~al.}(2009{\natexlab{b}})\citenamefont {Harris}, \citenamefont {Lanting},
  \citenamefont {Berkley}, \citenamefont {Johansson}, \citenamefont {Johnson},
  \citenamefont {Bunyk}, \citenamefont {Ladizinsky}, \citenamefont
  {Ladizinsky}, \citenamefont {Oh},\ and\ \citenamefont {Han}}]{harris09}%
  \BibitemOpen
  \bibfield  {author} {\bibinfo {author} {\bibfnamefont {R.}~\bibnamefont
  {Harris}}, \bibinfo {author} {\bibfnamefont {T.}~\bibnamefont {Lanting}},
  \bibinfo {author} {\bibfnamefont {A.~J.}\ \bibnamefont {Berkley}}, \bibinfo
  {author} {\bibfnamefont {J.}~\bibnamefont {Johansson}}, \bibinfo {author}
  {\bibfnamefont {M.~W.}\ \bibnamefont {Johnson}}, \bibinfo {author}
  {\bibfnamefont {P.}~\bibnamefont {Bunyk}}, \bibinfo {author} {\bibfnamefont
  {E.}~\bibnamefont {Ladizinsky}}, \bibinfo {author} {\bibfnamefont
  {N.}~\bibnamefont {Ladizinsky}}, \bibinfo {author} {\bibfnamefont
  {T.}~\bibnamefont {Oh}}, \ and\ \bibinfo {author} {\bibfnamefont
  {S.}~\bibnamefont {Han}},\ }\href {\doibase 10.1103/PhysRevB.80.052506}
  {\bibfield  {journal} {\bibinfo  {journal} {Phys. Rev. B}\ }\textbf {\bibinfo
  {volume} {80}},\ \bibinfo {pages} {052506} (\bibinfo {year}
  {2009}{\natexlab{b}})}\BibitemShut {NoStop}%
\bibitem [{\citenamefont {Han}\ \emph {et~al.}(1989)\citenamefont {Han},
  \citenamefont {Lapointe},\ and\ \citenamefont {Lukens}}]{han89}%
  \BibitemOpen
  \bibfield  {author} {\bibinfo {author} {\bibfnamefont {S.}~\bibnamefont
  {Han}}, \bibinfo {author} {\bibfnamefont {J.}~\bibnamefont {Lapointe}}, \
  and\ \bibinfo {author} {\bibfnamefont {J.~E.}\ \bibnamefont {Lukens}},\
  }\href {\doibase 10.1103/PhysRevLett.63.1712} {\bibfield  {journal} {\bibinfo
   {journal} {Phys. Rev. Lett.}\ }\textbf {\bibinfo {volume} {63}},\ \bibinfo
  {pages} {1712} (\bibinfo {year} {1989})}\BibitemShut {NoStop}%
\bibitem [{\citenamefont {Boixo}\ \emph {et~al.}(2016)\citenamefont {Boixo},
  \citenamefont {Smelyanskiy}, \citenamefont {Shabani}, \citenamefont {Isakov},
  \citenamefont {Dykman}, \citenamefont {Denchev}, \citenamefont {Amin},
  \citenamefont {Smirnov}, \citenamefont {Mohseni},\ and\ \citenamefont
  {Neven}}]{boixo16}%
  \BibitemOpen
  \bibfield  {author} {\bibinfo {author} {\bibfnamefont {S.}~\bibnamefont
  {Boixo}}, \bibinfo {author} {\bibfnamefont {V.~N.}\ \bibnamefont
  {Smelyanskiy}}, \bibinfo {author} {\bibfnamefont {A.}~\bibnamefont
  {Shabani}}, \bibinfo {author} {\bibfnamefont {S.~V.}\ \bibnamefont {Isakov}},
  \bibinfo {author} {\bibfnamefont {M.}~\bibnamefont {Dykman}}, \bibinfo
  {author} {\bibfnamefont {V.~S.}\ \bibnamefont {Denchev}}, \bibinfo {author}
  {\bibfnamefont {M.~H.}\ \bibnamefont {Amin}}, \bibinfo {author}
  {\bibfnamefont {A.~Y.}\ \bibnamefont {Smirnov}}, \bibinfo {author}
  {\bibfnamefont {M.}~\bibnamefont {Mohseni}}, \ and\ \bibinfo {author}
  {\bibfnamefont {H.}~\bibnamefont {Neven}},\ }\href {\doibase
  10.1038/ncomms10327} {\bibfield  {journal} {\bibinfo  {journal} {Nat.
  Commun.}\ }\textbf {\bibinfo {volume} {7}},\ \bibinfo {pages} {10327}
  (\bibinfo {year} {2016})}\BibitemShut {NoStop}%
\bibitem [{\citenamefont {Chiarello}\ \emph {et~al.}(2005)\citenamefont
  {Chiarello}, \citenamefont {Carelli}, \citenamefont {Castellano},
  \citenamefont {Cosmelli}, \citenamefont {Gangemi}, \citenamefont {Leoni},
  \citenamefont {Poletto}, \citenamefont {Simeone},\ and\ \citenamefont
  {Torrioli}}]{Chiarello2005}%
  \BibitemOpen
  \bibfield  {author} {\bibinfo {author} {\bibfnamefont {F.}~\bibnamefont
  {Chiarello}}, \bibinfo {author} {\bibfnamefont {P.}~\bibnamefont {Carelli}},
  \bibinfo {author} {\bibfnamefont {M.~G.}\ \bibnamefont {Castellano}},
  \bibinfo {author} {\bibfnamefont {C.}~\bibnamefont {Cosmelli}}, \bibinfo
  {author} {\bibfnamefont {L.}~\bibnamefont {Gangemi}}, \bibinfo {author}
  {\bibfnamefont {R.}~\bibnamefont {Leoni}}, \bibinfo {author} {\bibfnamefont
  {S.}~\bibnamefont {Poletto}}, \bibinfo {author} {\bibfnamefont
  {D.}~\bibnamefont {Simeone}}, \ and\ \bibinfo {author} {\bibfnamefont
  {G.}~\bibnamefont {Torrioli}},\ }\href {\doibase 10.1088/0953-2048/18/10/021}
  {\bibfield  {journal} {\bibinfo  {journal} {Supercond. Sci. Technol.}\
  }\textbf {\bibinfo {volume} {18}},\ \bibinfo {pages} {1370} (\bibinfo {year}
  {2005})}\BibitemShut {NoStop}%
\bibitem [{\citenamefont {Lanting}\ \emph {et~al.}(2014)\citenamefont
  {Lanting}, \citenamefont {Przybysz}, \citenamefont {Smirnov}, \citenamefont
  {Spedalieri}, \citenamefont {Amin}, \citenamefont {Berkley}, \citenamefont
  {Harris}, \citenamefont {Altomare}, \citenamefont {Boixo}, \citenamefont
  {Bunyk}, \citenamefont {Dickson}, \citenamefont {Enderud}, \citenamefont
  {Hilton}, \citenamefont {Hoskinson}, \citenamefont {Johnson}, \citenamefont
  {Ladizinsky}, \citenamefont {Ladizinsky}, \citenamefont {Neufeld},
  \citenamefont {Oh}, \citenamefont {Perminov}, \citenamefont {Rich},
  \citenamefont {Thom}, \citenamefont {Tolkacheva}, \citenamefont {Uchaikin},
  \citenamefont {Wilson},\ and\ \citenamefont {Rose}}]{lanting14}%
  \BibitemOpen
  \bibfield  {author} {\bibinfo {author} {\bibfnamefont {T.}~\bibnamefont
  {Lanting}}, \bibinfo {author} {\bibfnamefont {A.~J.}\ \bibnamefont
  {Przybysz}}, \bibinfo {author} {\bibfnamefont {A.~Y.}\ \bibnamefont
  {Smirnov}}, \bibinfo {author} {\bibfnamefont {F.~M.}\ \bibnamefont
  {Spedalieri}}, \bibinfo {author} {\bibfnamefont {M.~H.}\ \bibnamefont
  {Amin}}, \bibinfo {author} {\bibfnamefont {A.~J.}\ \bibnamefont {Berkley}},
  \bibinfo {author} {\bibfnamefont {R.}~\bibnamefont {Harris}}, \bibinfo
  {author} {\bibfnamefont {F.}~\bibnamefont {Altomare}}, \bibinfo {author}
  {\bibfnamefont {S.}~\bibnamefont {Boixo}}, \bibinfo {author} {\bibfnamefont
  {P.}~\bibnamefont {Bunyk}}, \bibinfo {author} {\bibfnamefont
  {N.}~\bibnamefont {Dickson}}, \bibinfo {author} {\bibfnamefont
  {C.}~\bibnamefont {Enderud}}, \bibinfo {author} {\bibfnamefont {J.~P.}\
  \bibnamefont {Hilton}}, \bibinfo {author} {\bibfnamefont {E.}~\bibnamefont
  {Hoskinson}}, \bibinfo {author} {\bibfnamefont {M.~W.}\ \bibnamefont
  {Johnson}}, \bibinfo {author} {\bibfnamefont {E.}~\bibnamefont {Ladizinsky}},
  \bibinfo {author} {\bibfnamefont {N.}~\bibnamefont {Ladizinsky}}, \bibinfo
  {author} {\bibfnamefont {R.}~\bibnamefont {Neufeld}}, \bibinfo {author}
  {\bibfnamefont {T.}~\bibnamefont {Oh}}, \bibinfo {author} {\bibfnamefont
  {I.}~\bibnamefont {Perminov}}, \bibinfo {author} {\bibfnamefont
  {C.}~\bibnamefont {Rich}}, \bibinfo {author} {\bibfnamefont {M.~C.}\
  \bibnamefont {Thom}}, \bibinfo {author} {\bibfnamefont {E.}~\bibnamefont
  {Tolkacheva}}, \bibinfo {author} {\bibfnamefont {S.}~\bibnamefont
  {Uchaikin}}, \bibinfo {author} {\bibfnamefont {A.~B.}\ \bibnamefont
  {Wilson}}, \ and\ \bibinfo {author} {\bibfnamefont {G.}~\bibnamefont
  {Rose}},\ }\href {\doibase 10.1103/PhysRevX.4.021041} {\bibfield  {journal}
  {\bibinfo  {journal} {Phys. Rev. X}\ }\textbf {\bibinfo {volume} {4}},\
  \bibinfo {pages} {021041} (\bibinfo {year} {2014})}\BibitemShut {NoStop}%
\bibitem [{\citenamefont {Harris}\ \emph {et~al.}(2007)\citenamefont {Harris},
  \citenamefont {Berkley}, \citenamefont {Johnson}, \citenamefont {Bunyk},
  \citenamefont {Govorkov}, \citenamefont {Thom}, \citenamefont {Uchaikin},
  \citenamefont {Wilson}, \citenamefont {Chung}, \citenamefont {Holtham},
  \citenamefont {Biamonte}, \citenamefont {Smirnov}, \citenamefont {Amin},\
  and\ \citenamefont {{Maassen van den Brink}}}]{harris07}%
  \BibitemOpen
  \bibfield  {author} {\bibinfo {author} {\bibfnamefont {R.}~\bibnamefont
  {Harris}}, \bibinfo {author} {\bibfnamefont {A.~J.}\ \bibnamefont {Berkley}},
  \bibinfo {author} {\bibfnamefont {M.~W.}\ \bibnamefont {Johnson}}, \bibinfo
  {author} {\bibfnamefont {P.}~\bibnamefont {Bunyk}}, \bibinfo {author}
  {\bibfnamefont {S.}~\bibnamefont {Govorkov}}, \bibinfo {author}
  {\bibfnamefont {M.~C.}\ \bibnamefont {Thom}}, \bibinfo {author}
  {\bibfnamefont {S.}~\bibnamefont {Uchaikin}}, \bibinfo {author}
  {\bibfnamefont {A.~B.}\ \bibnamefont {Wilson}}, \bibinfo {author}
  {\bibfnamefont {J.}~\bibnamefont {Chung}}, \bibinfo {author} {\bibfnamefont
  {E.}~\bibnamefont {Holtham}}, \bibinfo {author} {\bibfnamefont {J.~D.}\
  \bibnamefont {Biamonte}}, \bibinfo {author} {\bibfnamefont {A.~Y.}\
  \bibnamefont {Smirnov}}, \bibinfo {author} {\bibfnamefont {M.~H.~S.}\
  \bibnamefont {Amin}}, \ and\ \bibinfo {author} {\bibfnamefont
  {A.}~\bibnamefont {{Maassen van den Brink}}},\ }\href {\doibase
  10.1103/PhysRevLett.98.177001} {\bibfield  {journal} {\bibinfo  {journal}
  {Phys. Rev. Lett.}\ }\textbf {\bibinfo {volume} {98}},\ \bibinfo {pages}
  {177001} (\bibinfo {year} {2007})}\BibitemShut {NoStop}%
\bibitem [{\citenamefont {van~den Brink}\ \emph {et~al.}(2005)\citenamefont
  {van~den Brink}, \citenamefont {Berkley},\ and\ \citenamefont
  {Yalowsky}}]{vandenbrink05}%
  \BibitemOpen
  \bibfield  {author} {\bibinfo {author} {\bibfnamefont {A.~M.}\ \bibnamefont
  {van~den Brink}}, \bibinfo {author} {\bibfnamefont {A.~J.}\ \bibnamefont
  {Berkley}}, \ and\ \bibinfo {author} {\bibfnamefont {M.}~\bibnamefont
  {Yalowsky}},\ }\href {http://stacks.iop.org/1367-2630/7/i=1/a=230} {\bibfield
   {journal} {\bibinfo  {journal} {New J. Phys.}\ }\textbf {\bibinfo {volume}
  {7}},\ \bibinfo {pages} {230} (\bibinfo {year} {2005})}\BibitemShut {NoStop}%
\bibitem [{\citenamefont {Suzuki}(1985)}]{suzuki84}%
  \BibitemOpen
  \bibfield  {author} {\bibinfo {author} {\bibfnamefont {M.}~\bibnamefont
  {Suzuki}},\ }\href {\doibase 10.1063/1.526596} {\bibfield  {journal}
  {\bibinfo  {journal} {J. Math. Phys.}\ }\textbf {\bibinfo {volume} {26}},\
  \bibinfo {pages} {601} (\bibinfo {year} {1985})}\BibitemShut {NoStop}%
\bibitem [{\citenamefont {{De Raedt}}(1987)}]{deraedt87}%
  \BibitemOpen
  \bibfield  {author} {\bibinfo {author} {\bibfnamefont {H.}~\bibnamefont {{De
  Raedt}}},\ }\href {\doibase 10.1016/0167-7977(87)90002-5} {\bibfield
  {journal} {\bibinfo  {journal} {Comp. Phys. Rep.}\ }\textbf {\bibinfo
  {volume} {7}},\ \bibinfo {pages} {1} (\bibinfo {year} {1987})}\BibitemShut
  {NoStop}%
\bibitem [{\citenamefont {{J\"ulich Supercomputing Centre}}(2018)}]{jureca}%
  \BibitemOpen
  \bibfield  {author} {\bibinfo {author} {\bibnamefont {{J\"ulich
  Supercomputing Centre}}},\ }\href {\doibase 10.17815/jlsrf-4-121-1}
  {\bibfield  {journal} {\bibinfo  {journal} {Journal of large-scale research
  facilities}\ }\textbf {\bibinfo {volume} {4}} (\bibinfo {year} {2018}),\
  10.17815/jlsrf-4-121-1}\BibitemShut {NoStop}%
\bibitem [{\citenamefont {{D-Wave Systems Inc.}}()}]{dwave_private}%
  \BibitemOpen
  \bibfield  {author} {\bibinfo {author} {\bibnamefont {{D-Wave Systems
  Inc.}}},\ }\href@noop {} {\emph {\bibinfo {title} {private
  communication}}}\BibitemShut {NoStop}%
\bibitem [{\citenamefont {Landau}(1932)}]{landau32}%
  \BibitemOpen
  \bibfield  {author} {\bibinfo {author} {\bibfnamefont {L.}~\bibnamefont
  {Landau}},\ }\href@noop {} {\bibfield  {journal} {\bibinfo  {journal} {Phys.
  Z. Sowjetunion}\ }\textbf {\bibinfo {volume} {2}},\ \bibinfo {pages} {46}
  (\bibinfo {year} {1932})}\BibitemShut {NoStop}%
\bibitem [{\citenamefont {Zener}(1932)}]{zener32}%
  \BibitemOpen
  \bibfield  {author} {\bibinfo {author} {\bibfnamefont {C.}~\bibnamefont
  {Zener}},\ }\href {\doibase 10.1098/rspa.1932.0165} {\bibfield  {journal}
  {\bibinfo  {journal} {Proc. R. Soc. London, Ser A}\ }\textbf {\bibinfo
  {volume} {137}},\ \bibinfo {pages} {696} (\bibinfo {year}
  {1932})}\BibitemShut {NoStop}%
\bibitem [{\citenamefont {Matsuda}\ \emph {et~al.}(2009)\citenamefont
  {Matsuda}, \citenamefont {Nishimori},\ and\ \citenamefont
  {Katzgraber}}]{Matsuda09}%
  \BibitemOpen
  \bibfield  {author} {\bibinfo {author} {\bibfnamefont {Y.}~\bibnamefont
  {Matsuda}}, \bibinfo {author} {\bibfnamefont {H.}~\bibnamefont {Nishimori}},
  \ and\ \bibinfo {author} {\bibfnamefont {H.~G.}\ \bibnamefont {Katzgraber}},\
  }\href {\doibase 10.1088/1742-6596/143/1/012003} {\bibfield  {journal}
  {\bibinfo  {journal} {J. Phys.: Conf. Ser.}\ }\textbf {\bibinfo {volume}
  {143}},\ \bibinfo {pages} {012003} (\bibinfo {year} {2009})}\BibitemShut
  {NoStop}%
\bibitem [{\citenamefont {Albash}\ \emph {et~al.}(2015)\citenamefont {Albash},
  \citenamefont {Vinci}, \citenamefont {Mishra}, \citenamefont {Warburton},\
  and\ \citenamefont {Lidar}}]{albash15}%
  \BibitemOpen
  \bibfield  {author} {\bibinfo {author} {\bibfnamefont {T.}~\bibnamefont
  {Albash}}, \bibinfo {author} {\bibfnamefont {W.}~\bibnamefont {Vinci}},
  \bibinfo {author} {\bibfnamefont {A.}~\bibnamefont {Mishra}}, \bibinfo
  {author} {\bibfnamefont {P.~A.}\ \bibnamefont {Warburton}}, \ and\ \bibinfo
  {author} {\bibfnamefont {D.~A.}\ \bibnamefont {Lidar}},\ }\href {\doibase
  10.1103/PhysRevA.91.042314} {\bibfield  {journal} {\bibinfo  {journal} {Phys.
  Rev. A}\ }\textbf {\bibinfo {volume} {91}},\ \bibinfo {pages} {042314}
  (\bibinfo {year} {2015})}\BibitemShut {NoStop}%
\bibitem [{\citenamefont {{D-Wave Systems Inc.}}(2019)}]{dwave_manual}%
  \BibitemOpen
  \bibfield  {author} {\bibinfo {author} {\bibnamefont {{D-Wave Systems
  Inc.}}},\ }\href {https://docs.dwavesys.com/docs/latest/doc_qpu.html} {\emph
  {\bibinfo {title} {Technical Description of the D-Wave Quantum Processing
  Unit}}} (\bibinfo {year} {2019})\BibitemShut {NoStop}%
\bibitem [{\citenamefont {Amin}\ \emph {et~al.}(2008)\citenamefont {Amin},
  \citenamefont {Love},\ and\ \citenamefont {Truncik}}]{amin08}%
  \BibitemOpen
  \bibfield  {author} {\bibinfo {author} {\bibfnamefont {M.~H.~S.}\
  \bibnamefont {Amin}}, \bibinfo {author} {\bibfnamefont {P.~J.}\ \bibnamefont
  {Love}}, \ and\ \bibinfo {author} {\bibfnamefont {C.~J.~S.}\ \bibnamefont
  {Truncik}},\ }\href {\doibase 10.1103/PhysRevLett.100.060503} {\bibfield
  {journal} {\bibinfo  {journal} {Phys. Rev. Lett.}\ }\textbf {\bibinfo
  {volume} {100}},\ \bibinfo {pages} {060503} (\bibinfo {year}
  {2008})}\BibitemShut {NoStop}%
\bibitem [{\citenamefont {Johansson}\ \emph {et~al.}(2009)\citenamefont
  {Johansson}, \citenamefont {Amin}, \citenamefont {Berkley}, \citenamefont
  {Bunyk}, \citenamefont {Choi}, \citenamefont {Harris}, \citenamefont
  {Johnson}, \citenamefont {Lanting}, \citenamefont {Lloyd},\ and\
  \citenamefont {Rose}}]{johansson09}%
  \BibitemOpen
  \bibfield  {author} {\bibinfo {author} {\bibfnamefont {J.}~\bibnamefont
  {Johansson}}, \bibinfo {author} {\bibfnamefont {M.~H.~S.}\ \bibnamefont
  {Amin}}, \bibinfo {author} {\bibfnamefont {A.~J.}\ \bibnamefont {Berkley}},
  \bibinfo {author} {\bibfnamefont {P.}~\bibnamefont {Bunyk}}, \bibinfo
  {author} {\bibfnamefont {V.}~\bibnamefont {Choi}}, \bibinfo {author}
  {\bibfnamefont {R.}~\bibnamefont {Harris}}, \bibinfo {author} {\bibfnamefont
  {M.~W.}\ \bibnamefont {Johnson}}, \bibinfo {author} {\bibfnamefont {T.~M.}\
  \bibnamefont {Lanting}}, \bibinfo {author} {\bibfnamefont {S.}~\bibnamefont
  {Lloyd}}, \ and\ \bibinfo {author} {\bibfnamefont {G.}~\bibnamefont {Rose}},\
  }\href {\doibase 10.1103/PhysRevB.80.012507} {\bibfield  {journal} {\bibinfo
  {journal} {Phys. Rev. B}\ }\textbf {\bibinfo {volume} {80}},\ \bibinfo
  {pages} {012507} (\bibinfo {year} {2009})}\BibitemShut {NoStop}%
\bibitem [{\citenamefont {Amin}\ \emph
  {et~al.}(2009{\natexlab{b}})\citenamefont {Amin}, \citenamefont {Averin},\
  and\ \citenamefont {Nesteroff}}]{amin09}%
  \BibitemOpen
  \bibfield  {author} {\bibinfo {author} {\bibfnamefont {M.~H.~S.}\
  \bibnamefont {Amin}}, \bibinfo {author} {\bibfnamefont {D.~V.}\ \bibnamefont
  {Averin}}, \ and\ \bibinfo {author} {\bibfnamefont {J.~A.}\ \bibnamefont
  {Nesteroff}},\ }\href {\doibase 10.1103/PhysRevA.79.022107} {\bibfield
  {journal} {\bibinfo  {journal} {Phys. Rev. A}\ }\textbf {\bibinfo {volume}
  {79}},\ \bibinfo {pages} {022107} (\bibinfo {year}
  {2009}{\natexlab{b}})}\BibitemShut {NoStop}%
\bibitem [{\citenamefont {Amin}\ and\ \citenamefont
  {Brito}(2009)}]{amin09_nonmarkovian}%
  \BibitemOpen
  \bibfield  {author} {\bibinfo {author} {\bibfnamefont {M.~H.~S.}\
  \bibnamefont {Amin}}\ and\ \bibinfo {author} {\bibfnamefont {F.}~\bibnamefont
  {Brito}},\ }\href {\doibase 10.1103/PhysRevB.80.214302} {\bibfield  {journal}
  {\bibinfo  {journal} {Phys. Rev. B}\ }\textbf {\bibinfo {volume} {80}},\
  \bibinfo {pages} {214302} (\bibinfo {year} {2009})}\BibitemShut {NoStop}%
\bibitem [{\citenamefont {Amin}(2015)}]{amin15}%
  \BibitemOpen
  \bibfield  {author} {\bibinfo {author} {\bibfnamefont {M.~H.}\ \bibnamefont
  {Amin}},\ }\href {\doibase 10.1103/PhysRevA.92.052323} {\bibfield  {journal}
  {\bibinfo  {journal} {Phys. Rev. A}\ }\textbf {\bibinfo {volume} {92}},\
  \bibinfo {pages} {052323} (\bibinfo {year} {2015})}\BibitemShut {NoStop}%
\bibitem [{\citenamefont {Phillips}(1972)}]{Phillips1972}%
  \BibitemOpen
  \bibfield  {author} {\bibinfo {author} {\bibfnamefont {W.~A.}\ \bibnamefont
  {Phillips}},\ }\href {\doibase 10.1007/BF00660072} {\bibfield  {journal}
  {\bibinfo  {journal} {J. Low Temp. Phys.}\ }\textbf {\bibinfo {volume} {7}},\
  \bibinfo {pages} {351} (\bibinfo {year} {1972})}\BibitemShut {NoStop}%
\bibitem [{\citenamefont {Anderson}\ \emph {et~al.}(1972)\citenamefont
  {Anderson}, \citenamefont {Halperin},\ and\ \citenamefont
  {Varma}}]{anderson72}%
  \BibitemOpen
  \bibfield  {author} {\bibinfo {author} {\bibfnamefont {P.~W.}\ \bibnamefont
  {Anderson}}, \bibinfo {author} {\bibfnamefont {B.~I.}\ \bibnamefont
  {Halperin}}, \ and\ \bibinfo {author} {\bibfnamefont {C.~M.}\ \bibnamefont
  {Varma}},\ }\href {\doibase 10.1080/14786437208229210} {\bibfield  {journal}
  {\bibinfo  {journal} {Phil. Mag.}\ }\textbf {\bibinfo {volume} {25}},\
  \bibinfo {pages} {1} (\bibinfo {year} {1972})}\BibitemShut {NoStop}%
\bibitem [{\citenamefont {Burnett}\ \emph {et~al.}(2014)\citenamefont
  {Burnett}, \citenamefont {Faoro}, \citenamefont {Wisby}, \citenamefont
  {Gurtovoi}, \citenamefont {Chernykh}, \citenamefont {Mikhailov},
  \citenamefont {Tulin}, \citenamefont {Shaikhaidarov}, \citenamefont
  {Antonov}, \citenamefont {Meeson}, \citenamefont {Tzalenchuk},\ and\
  \citenamefont {Lindstr{\"o}m}}]{burnett14}%
  \BibitemOpen
  \bibfield  {author} {\bibinfo {author} {\bibfnamefont {J.}~\bibnamefont
  {Burnett}}, \bibinfo {author} {\bibfnamefont {L.}~\bibnamefont {Faoro}},
  \bibinfo {author} {\bibfnamefont {I.}~\bibnamefont {Wisby}}, \bibinfo
  {author} {\bibfnamefont {V.~L.}\ \bibnamefont {Gurtovoi}}, \bibinfo {author}
  {\bibfnamefont {A.~V.}\ \bibnamefont {Chernykh}}, \bibinfo {author}
  {\bibfnamefont {G.~M.}\ \bibnamefont {Mikhailov}}, \bibinfo {author}
  {\bibfnamefont {V.~A.}\ \bibnamefont {Tulin}}, \bibinfo {author}
  {\bibfnamefont {R.}~\bibnamefont {Shaikhaidarov}}, \bibinfo {author}
  {\bibfnamefont {V.}~\bibnamefont {Antonov}}, \bibinfo {author} {\bibfnamefont
  {P.~J.}\ \bibnamefont {Meeson}}, \bibinfo {author} {\bibfnamefont {A.~Y.}\
  \bibnamefont {Tzalenchuk}}, \ and\ \bibinfo {author} {\bibfnamefont
  {T.}~\bibnamefont {Lindstr{\"o}m}},\ }\href {\doibase 10.1038/ncomms5119}
  {\bibfield  {journal} {\bibinfo  {journal} {Nat. Commun.}\ }\textbf {\bibinfo
  {volume} {5}},\ \bibinfo {pages} {4119} (\bibinfo {year} {2014})}\BibitemShut
  {NoStop}%
\bibitem [{\citenamefont {Faoro}\ and\ \citenamefont {Ioffe}(2015)}]{faoro15}%
  \BibitemOpen
  \bibfield  {author} {\bibinfo {author} {\bibfnamefont {L.}~\bibnamefont
  {Faoro}}\ and\ \bibinfo {author} {\bibfnamefont {L.~B.}\ \bibnamefont
  {Ioffe}},\ }\href {\doibase 10.1103/PhysRevB.91.014201} {\bibfield  {journal}
  {\bibinfo  {journal} {Phys. Rev. B}\ }\textbf {\bibinfo {volume} {91}},\
  \bibinfo {pages} {014201} (\bibinfo {year} {2015})}\BibitemShut {NoStop}%
\bibitem [{\citenamefont {Lisenfeld}\ \emph {et~al.}(2015)\citenamefont
  {Lisenfeld}, \citenamefont {Grabovskij}, \citenamefont {M{\"u}ller},
  \citenamefont {Cole}, \citenamefont {Weiss},\ and\ \citenamefont
  {Ustinov}}]{Lisenfeld15}%
  \BibitemOpen
  \bibfield  {author} {\bibinfo {author} {\bibfnamefont {J.}~\bibnamefont
  {Lisenfeld}}, \bibinfo {author} {\bibfnamefont {G.~J.}\ \bibnamefont
  {Grabovskij}}, \bibinfo {author} {\bibfnamefont {C.}~\bibnamefont
  {M{\"u}ller}}, \bibinfo {author} {\bibfnamefont {J.~H.}\ \bibnamefont
  {Cole}}, \bibinfo {author} {\bibfnamefont {G.}~\bibnamefont {Weiss}}, \ and\
  \bibinfo {author} {\bibfnamefont {A.~V.}\ \bibnamefont {Ustinov}},\ }\href
  {\doibase 10.1038/ncomms7182} {\bibfield  {journal} {\bibinfo  {journal}
  {Nat. Commun.}\ }\textbf {\bibinfo {volume} {6}},\ \bibinfo {pages} {6182}
  (\bibinfo {year} {2015})}\BibitemShut {NoStop}%
\bibitem [{\citenamefont {de~Graaf}\ \emph {et~al.}(2018)\citenamefont
  {de~Graaf}, \citenamefont {Faoro}, \citenamefont {Burnett}, \citenamefont
  {Adamyan}, \citenamefont {Tzalenchuk}, \citenamefont {Kubatkin},
  \citenamefont {Lindstr{\"o}m},\ and\ \citenamefont {Danilov}}]{deGraaf18}%
  \BibitemOpen
  \bibfield  {author} {\bibinfo {author} {\bibfnamefont {S.~E.}\ \bibnamefont
  {de~Graaf}}, \bibinfo {author} {\bibfnamefont {L.}~\bibnamefont {Faoro}},
  \bibinfo {author} {\bibfnamefont {J.}~\bibnamefont {Burnett}}, \bibinfo
  {author} {\bibfnamefont {A.~A.}\ \bibnamefont {Adamyan}}, \bibinfo {author}
  {\bibfnamefont {A.~Y.}\ \bibnamefont {Tzalenchuk}}, \bibinfo {author}
  {\bibfnamefont {S.~E.}\ \bibnamefont {Kubatkin}}, \bibinfo {author}
  {\bibfnamefont {T.}~\bibnamefont {Lindstr{\"o}m}}, \ and\ \bibinfo {author}
  {\bibfnamefont {A.~V.}\ \bibnamefont {Danilov}},\ }\href {\doibase
  10.1038/s41467-018-03577-2} {\bibfield  {journal} {\bibinfo  {journal} {Nat.
  Commun.}\ }\textbf {\bibinfo {volume} {9}},\ \bibinfo {pages} {1143}
  (\bibinfo {year} {2018})}\BibitemShut {NoStop}%
\bibitem [{\citenamefont {M\"uller}\ \emph {et~al.}(2019)\citenamefont
  {M\"uller}, \citenamefont {Cole},\ and\ \citenamefont
  {Lisenfeld}}]{mueller19}%
  \BibitemOpen
  \bibfield  {author} {\bibinfo {author} {\bibfnamefont {C.}~\bibnamefont
  {M\"uller}}, \bibinfo {author} {\bibfnamefont {J.~H.}\ \bibnamefont {Cole}},
  \ and\ \bibinfo {author} {\bibfnamefont {J.}~\bibnamefont {Lisenfeld}},\
  }\href {\doibase 10.1088/1361-6633/ab3a7e} {\bibfield  {journal} {\bibinfo
  {journal} {Rep. Prog. Phys.}\ }\textbf {\bibinfo {volume} {82}},\ \bibinfo
  {pages} {124501} (\bibinfo {year} {2019})}\BibitemShut {NoStop}%
\bibitem [{\citenamefont {{Jin}}\ \emph {et~al.}(2013)\citenamefont {{Jin}},
  \citenamefont {{Michielsen}}, \citenamefont {Novotny}, \citenamefont
  {{Miyashita}}, \citenamefont {{Yuan}},\ and\ \citenamefont {{De
  Raedt}}}]{JIN13a}%
  \BibitemOpen
  \bibfield  {author} {\bibinfo {author} {\bibfnamefont {F.}~\bibnamefont
  {{Jin}}}, \bibinfo {author} {\bibfnamefont {K.}~\bibnamefont {{Michielsen}}},
  \bibinfo {author} {\bibfnamefont {M.~A.}\ \bibnamefont {Novotny}}, \bibinfo
  {author} {\bibfnamefont {S.}~\bibnamefont {{Miyashita}}}, \bibinfo {author}
  {\bibfnamefont {S.}~\bibnamefont {{Yuan}}}, \ and\ \bibinfo {author}
  {\bibfnamefont {H.}~\bibnamefont {{De Raedt}}},\ }\href {\doibase
  10.1103/PhysRevA.87.022117} {\bibfield  {journal} {\bibinfo  {journal} {Phys.
  Rev. A}\ }\textbf {\bibinfo {volume} {87}},\ \bibinfo {pages} {022117}
  (\bibinfo {year} {2013})}\BibitemShut {NoStop}%
\bibitem [{\citenamefont {{J\"ulich Supercomputing Centre}}(2019)}]{JUWELS}%
  \BibitemOpen
  \bibfield  {author} {\bibinfo {author} {\bibnamefont {{J\"ulich
  Supercomputing Centre}}},\ }\href {\doibase 10.17815/jlsrf-5-171} {\bibfield
  {journal} {\bibinfo  {journal} {Journal of large-scale research facilities}\
  }\textbf {\bibinfo {volume} {5}} (\bibinfo {year} {2019}),\
  10.17815/jlsrf-5-171}\BibitemShut {NoStop}%
\bibitem [{\citenamefont {Saito}\ and\ \citenamefont
  {Miyashita}(2001)}]{saito01}%
  \BibitemOpen
  \bibfield  {author} {\bibinfo {author} {\bibfnamefont {K.}~\bibnamefont
  {Saito}}\ and\ \bibinfo {author} {\bibfnamefont {S.}~\bibnamefont
  {Miyashita}},\ }\href {\doibase 10.1143/JPSJ.70.3385} {\bibfield  {journal}
  {\bibinfo  {journal} {J. Phys. Soc. Jpn.}\ }\textbf {\bibinfo {volume}
  {70}},\ \bibinfo {pages} {3385} (\bibinfo {year} {2001})}\BibitemShut
  {NoStop}%
\bibitem [{\citenamefont {Hams}\ and\ \citenamefont {{De
  Raedt}}(2000)}]{HAMS00}%
  \BibitemOpen
  \bibfield  {author} {\bibinfo {author} {\bibfnamefont {A.}~\bibnamefont
  {Hams}}\ and\ \bibinfo {author} {\bibfnamefont {H.}~\bibnamefont {{De
  Raedt}}},\ }\href {\doibase 10.1103/PhysRevE.62.4365} {\bibfield  {journal}
  {\bibinfo  {journal} {Phys. Rev. E}\ }\textbf {\bibinfo {volume} {62}},\
  \bibinfo {pages} {4365} (\bibinfo {year} {2000})}\BibitemShut {NoStop}%
\bibitem [{\citenamefont {{De Raedt}}\ and\ \citenamefont
  {Michielsen}(2006)}]{RAED06}%
  \BibitemOpen
  \bibfield  {author} {\bibinfo {author} {\bibfnamefont {H.}~\bibnamefont {{De
  Raedt}}}\ and\ \bibinfo {author} {\bibfnamefont {K.}~\bibnamefont
  {Michielsen}},\ }in\ \href@noop {} {\emph {\bibinfo {booktitle} {Handbook of
  Theoretical and Computational Nanotechnology}}},\ \bibinfo {editor} {edited
  by\ \bibinfo {editor} {\bibfnamefont {M.}~\bibnamefont {Rieth}}\ and\
  \bibinfo {editor} {\bibfnamefont {W.}~\bibnamefont {Schommers}}}\ (\bibinfo
  {publisher} {American Scientific Publishers},\ \bibinfo {address} {Los
  Angeles},\ \bibinfo {year} {2006})\ pp.\ \bibinfo {pages} {2 --
  48}\BibitemShut {NoStop}%
\bibitem [{\citenamefont {{De Raedt}}\ \emph {et~al.}(2019)\citenamefont {{De
  Raedt}}, \citenamefont {Jin}, \citenamefont {Willsch}, \citenamefont
  {Willsch}, \citenamefont {Yoshioka}, \citenamefont {Ito}, \citenamefont
  {Yuan},\ and\ \citenamefont {Michielsen}}]{deraedt18}%
  \BibitemOpen
  \bibfield  {author} {\bibinfo {author} {\bibfnamefont {H.}~\bibnamefont {{De
  Raedt}}}, \bibinfo {author} {\bibfnamefont {F.}~\bibnamefont {Jin}}, \bibinfo
  {author} {\bibfnamefont {D.}~\bibnamefont {Willsch}}, \bibinfo {author}
  {\bibfnamefont {M.}~\bibnamefont {Willsch}}, \bibinfo {author} {\bibfnamefont
  {N.}~\bibnamefont {Yoshioka}}, \bibinfo {author} {\bibfnamefont
  {N.}~\bibnamefont {Ito}}, \bibinfo {author} {\bibfnamefont {S.}~\bibnamefont
  {Yuan}}, \ and\ \bibinfo {author} {\bibfnamefont {K.}~\bibnamefont
  {Michielsen}},\ }\href {\doibase 10.1016/j.cpc.2018.11.005} {\bibfield
  {journal} {\bibinfo  {journal} {Comput. Phys. Commun.}\ }\textbf {\bibinfo
  {volume} {237}},\ \bibinfo {pages} {47} (\bibinfo {year} {2019})}\BibitemShut
  {NoStop}%
\end{thebibliography}%

\end{document}